\documentclass[aps,pra,twocolumn,superscriptaddress,showpacs,reprint]{revtex4-1}

\usepackage{amsmath}
\usepackage{amsthm}
\usepackage{graphicx}
\usepackage{amsfonts}
\usepackage{amssymb}
\usepackage{times,txfonts}
\usepackage{hyperref}
\usepackage{relsize}
\usepackage{braket}
\usepackage{mathrsfs}
\usepackage{array,multirow}
\usepackage{color}
\usepackage{booktabs}

\definecolor{darkblue}{RGB}{54,59,149}

\def\M3N{$\mathcal{M}_{N}^{3}$}

\newtheorem{lemma}{Lemma}[section]
\newtheorem{theorem}{Theorem}[section]
\newtheorem{corollary}{Corollary}[section]

\begin{document}

\title{\sf \bfseries Accessible quantification of multiparticle entanglement}

\author{Marco Cianciaruso}
\email{cianciaruso.marco@gmail.com}
\affiliation{Centre for the Mathematics and Theoretical Physics of Quantum Non-Equilibrium Systems, School of Mathematical Sciences, The University of Nottingham, University Park, Nottingham NG7 2RD, United Kingdom}

\author{Thomas R. Bromley}
\email{thomas.r.bromley@gmail.com}
\affiliation{Centre for the Mathematics and Theoretical Physics of Quantum Non-Equilibrium Systems, School of Mathematical Sciences, The University of Nottingham, University Park, Nottingham NG7 2RD, United Kingdom}

\author{Gerardo Adesso}
\email{gerardo.adesso@nottingham.ac.uk}
\affiliation{Centre for the Mathematics and Theoretical Physics of Quantum Non-Equilibrium Systems, School of Mathematical Sciences, The University of Nottingham, University Park, Nottingham NG7 2RD, United Kingdom}

\begin{abstract}
{Entanglement is a key ingredient for quantum technologies and a fundamental signature of quantumness in a broad range of phenomena encompassing  many-body physics, thermodynamics, cosmology, and life sciences. For arbitrary multiparticle systems, entanglement quantification typically involves nontrivial optimisation problems, and may require demanding tomographical techniques. Here we develop an experimentally feasible approach to the evaluation of geometric measures of multiparticle entanglement. Our framework provides analytical results for particular classes of mixed states of {\em N} qubits, and computable lower bounds to global, partial, or genuine multiparticle entanglement of any general state.  For global and partial entanglement, useful bounds are obtained with minimum effort, requiring local measurements in just three settings for any {\em N}.  For genuine entanglement, a number of measurements scaling linearly with {\em N} is required. We demonstrate the power of our approach to estimate and quantify different types of multiparticle entanglement in a variety of {\em N}-qubit states useful for quantum information processing and recently engineered in laboratories with quantum optics and trapped ion setups.}
\end{abstract}
%

\maketitle


\section{\sf \bfseries Introduction}

The fascination with quantum entanglement has evolved over the last eight decades, from the realm of philosophical debate \cite{EPR1935} to a very concrete recognition  of its resource role in a range of applied sciences \cite{Vedral2014,Horodecki2009}. While considerable progress has been achieved in the detection of entanglement \cite{GuhneToth2009,Guhne2010,Gao2014,Levi2013,Huber2010,DurCirac2000,Gabriel2010,Klockl2015,ExpFriendly}, its experimentally accessible quantification remains an open problem for any real implementation of an entangled system \cite{Guhne2015,Hofmann2014,SiewertPRL,SiewertSciRep,Wu2012,Hashemi2012,Audenart2006,Wunderlich2009,Ma2011,QWitness2,QWitness1}. Quantifying entanglement is yet necessary to gauge precisely the quantum enhancement in information processing and computation \cite{Vedral2014,PlenioVirmani2007,Horodecki2009}, and to pin down exactly how much a physical or biological system under observation departs from an essentially classical behaviour \cite{Whaley2010}. This is especially relevant in the case of complex, multiparticle systems, for which only quite recently have notable advances been reported on the control of entanglement \cite{Cramer2011,Cramer2013,Marty2014,Marty2015}.

\begin{figure}[tbh]
\centering \includegraphics[width=7.5cm]{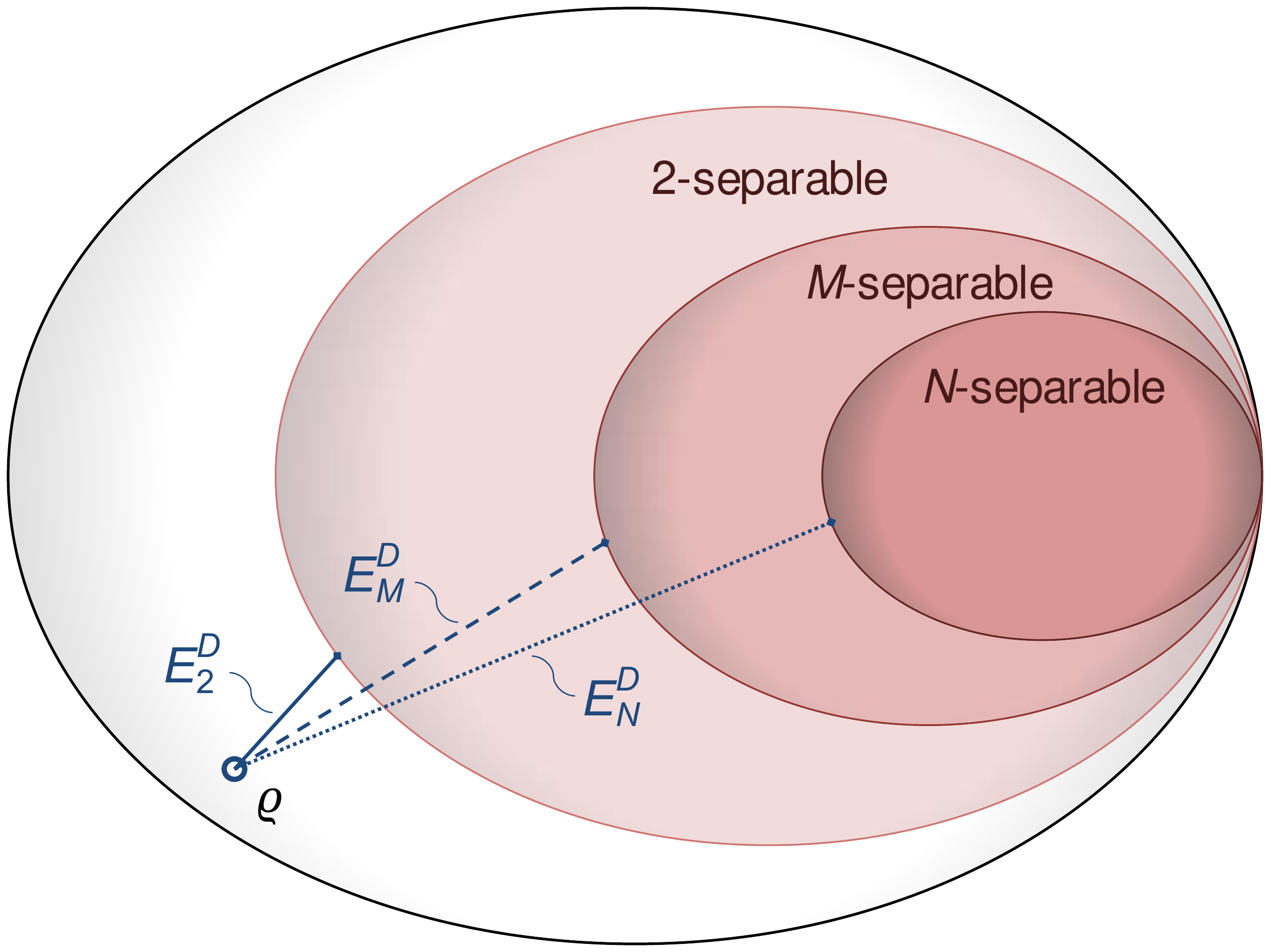}
\caption{Geometric picture of multiparticle entanglement in a quantum system of $N$ particles. Each red-shaded convex set contains $M$-separable states (for $2 \leq M \leq N$), which can be defined as follows. The set of pure $M$-separable states is given by the union of the sets of tensor products $\ket{\Psi_M}=\bigotimes_{k=1}^M \ket{\psi^{(k)}}$ of pure states $\ket{\psi^{(k)}}$, with respect to any partition of the $N$ particles into $M$ subsystems $k=1,\ldots,M$; the set of general (mixed)  $M$-separable states is then formed by all convex mixtures of pure $M$-separable states, where each term in the mixture may be factorised with respect to a different multipartition. This provides a partition-independent classification of separability (see also the Supplementary Material). For any $M$, the multiparticle entanglement measure $E^D_M$ of a state $\varrho$ is defined as the minimum distance, with respect to a contractive and jointly convex distance functional $D$, from the set of $M$-separable states. We refer to the case $M=N$ (dotted line) as global entanglement, the case $M=2$ (solid line) as genuine entanglement, and any intermediate case (dashed line) as partial entanglement, as detailed in the main text.
}
\label{figMSE}
\end{figure}

An intuitive framework for quantifying the degree of multiparticle entanglement relies on a geometric perspective \cite{Plenio1997,Wei2003,Blasone2008}. Within this approach, one first identifies a hierarchy of non-entangled multiparticle states, also referred to as $M$-separable states for $2 \leq M \leq N$, where $N$ is the number of particles composing the quantum system of interest; see Figure~\ref{figMSE}.
Introducing then a distance functional $D$ respecting natural properties of contractivity under quantum operations and joint convexity (see Methods) \cite{Bengtsson2006}, the quantity $E^D_{M}$ defined as
\begin{equation}\label{geome}
E^D_{M}(\varrho)=\inf_{\mbox{$\varsigma$ $M$-separable}} D(\varrho, \varsigma)\,,
\end{equation}
is a valid geometric measure of ($M$-inseparable) {\em multiparticle entanglement} in the state $\varrho$.

Some special cases are prominent in this hierarchy.
For $M=N$, the distance from $N$-separable (also known as fully separable) states defines the {\em global} multiparticle entanglement $E^D_N$, which accounts for any form of entanglement distributed among two or more of the $N$ particles. Geometric measures of global entanglement have been successfully employed to characterise quantum phase transitions in many-body systems \cite{Wei2005} and directly assess the usefulness of initial states for Grover's search algoritms \cite{Groverian}.
On the other extreme of the hierarchy, for $M=2$, the distance from $2$-separable (also known as biseparable) states defines instead the {\em genuine} multiparticle entanglement $E^D_2$, which quantifies the entanglement shared by all the $N$ particles, that is the highest degree of inseparability.
Genuine multiparticle entanglement is an essential ingredient for quantum technologies including multiuser quantum cryptography \cite{Markham2014}, quantum metrology \cite{Giovannetti2011}, and measurement-based quantum computation \cite{BriegelNP}. 
Finally, for any intermediate $M$, we can refer to $E_M^D$ as {\em partial} multiparticle entanglement. The presence  of partial entanglement is relevant in quantum informational tasks such as quantum secret sharing \cite{Gabriel2010} and may play a relevant role in  biological phenomena \cite{Tiersch2012,Whaley2010}. Probing and quantifying different types of entanglement can shed light on which nonclassical features of a mixed multiparticle state are necessary for quantum-enhanced performance in specific tasks \cite{Levi2013} and can guide the understanding of the emergence of classicality in multiparticle quantum systems of increasing complexity \cite{Blatt2010}.

The quantitative amount of multiparticle entanglement, be it global or genuine (or any intermediate type), has an intuitive operational meaning when adopting the geometric approach. Namely, $E^D_{M}$ measures how distinguishable a given state $\varrho$ is from the closest $M$-separable state. Given some widely adopted metrics, such a distinguishability is directly connected to the usefulness of $\varrho$ for quantum information protocols relying on multiparticle entanglement. For instance,  the trace distance of entanglement is operationally related to the minimum probability of error in the discrimination between $\varrho$ and any $M$-separable state with a single measurement \cite{Bengtsson2006}.  Furthermore, the geometric entanglement with respect to relative entropy or Bures distance sets quantitative bounds on the number of orthogonal states that can be discriminated by local operations and classical communication (LOCC) \cite{Hayashi2006}. The geometric entanglement based on infidelity \cite{Wei2003} (monotonically related to Bures distance) has also a dual interpretation based on the convex roof construction \cite{Streltsov2010}, that is, it quantifies the minimum price (in units of pure-state entanglement) that has to be spent on average to create a given density matrix $\varrho$ as a statistical mixture of pure states.

It is therefore clear that finding the minimum in Eq.~(\ref{geome}), and hence evaluating geometric measures of multiparticle entanglement defined by meaningful distances, is a central challenge to benchmark quantum technologies. However, obtaining such a solution for general multiparticle states is in principle a formidable problem. Even if possible, there would remain major challenges for experimental evaluation, which would in general require a complete reconstruction of the state through full tomography. For multiparticle states of any reasonable number of qubits, full state tomography places significant demands on experimental resources, and it is thus highly desirable to provide quantitative guarantees on the geometric multiparticle entanglement present in a state, via non-trivial lower bounds, in an experimentally accessible way \cite{Guhne2015,Hofmann2014,SiewertPRL,SiewertSciRep,Wu2012,Hashemi2012,Audenart2006,Wunderlich2009,Ma2011,QWitness2,QWitness1}.


Here we provide substantial advances towards addressing this problem in a general fashion. We identify a {\em general framework} for the provision of experimentally friendly quantitative guarantees on the geometric multiparticle entanglement present in a state. This approach consists of:
\begin{description}
\item[\em (1) Choosing a set of reference states]{
Find a restricted family of $N$-qubit states   with the property that \emph{any} state may be mapped into this family through a fixed procedure of single-qubit LOCC. This reference family should be simple to characterise, and can be chosen from experimental or theoretical considerations.
} 
\item[\em (2) Identifying $M$-separable reference states]{
Apply the fixed LOCC procedure to the general set of $M$-separable states, hence identifying the subset of $M$-separable states within the reference family.
}
\item[\em (3) Calculating $E_{M}^{D}$ for the reference states]{
Solve the optimisation problem for the geometric entanglement of reference states. This is dramatically simplified by using the properties of contractivity and joint convexity, that hold for any distance functional $D$ defining a valid entanglement measure, and imply in particular that one of the closest $M$-separable states to any reference state is to be found itself within the reference family.
}
\item[\em (4) Deriving optimised lower bounds for any state]{
Exploit the freedom to apply single-qubit unitaries to any $N$-qubit state $\varrho$ in order to find the corresponding reference state with the highest geometric entanglement, providing an optimised lower bound to $E_{M}^{D}(\varrho)$.
}
\end{description}
This process presents a versatile and comprehensive approach to obtain lower bounds on  geometric multiparticle entanglement measures according to any valid distance. While building on some previously utilised methods for steps (1) \cite{Hofmann2014,Guhne2015,SiewertPRL,SiewertSciRep} and (4) \cite{Guhne2015,SiewertPRL,SiewertSciRep}, it introduces novel techniques in steps (2) and most importantly (3), which are crucial for completing the framework and making it effective in practice (see e.g.~Appendix C, D, E, and F of the Supplementary Material).

To illustrate the power of our approach, we focus initially on a reference family of mixed states $\varpi$ of $N$ qubits, that we label \M3N states, which form a subset of the class of states having all maximally mixed marginals.
This family includes maximally entangled Bell states of two qubits and their mixtures, as well as multiparticle bound entangled states \cite{BeatrizSmolin,PianiSmolin,BoundNature,GeneSmolin}.
For any $N$, these states are completely specified by three easily measurable quantities, given by the correlation functions $c_j=\langle \sigma_j^{\otimes N}\rangle$, where $\{\sigma_j\}_{j=1,2,3}$ are the Pauli matrices.
In the following we show how every entanglement monotone $E^D_{M}$ can be evaluated exactly for any even $N$ on these states, by revealing an intuitive geometric picture common to all valid distances $D$. For odd $N$, the results are distance-dependent; we show nonetheless that $E^D_{M}$ can still be evaluated exactly if $D$ denotes the trace distance. The results are nontrivial for all $M > \left \lceil{N/2}\right \rceil$ in the hierarchy of Figure~\ref{figMSE}. A central observation, in line with the general framework, is that an arbitrary state of $N$ qubits can be transformed into an \M3N state by a LOCC procedure, which cannot increase entanglement by definition. This implies that our exact  formulae readily provide practical lower bounds to the degree of global and partial multiparticle entanglement in completely general states. Importantly, the bounds are obtained by measuring only the three correlation functions $\{c_j\}$ for any number of qubits, and can be further improved by adjusting the local measurement basis (see Figure~\ref{figTB} for an illustration).


Furthermore, we discuss how our results can be extended to allow for the quantitative estimation of genuine multiparticle entanglement as well, at the cost of performing extra measurements. Since \M3N states are always biseparable, we must consider a different reference family. We focus on the class of $N$-qubit states obtained as mixtures of Greenberger-Horne-Zeilinger (GHZ) states \cite{GHZ90,Guhne2010}, the latter being central resources for quantum communication and estimation; this class of states depends on $2^N-1$ real parameters. We calculate exactly distance-based measures of genuine multiparticle entanglement $E_2^D$ for these states, for every valid  $D$. Once more, these analytical results provide lower bounds to geometric measures of genuine entanglement for any general state of $N$ qubits, obtainable experimentally in this case by performing at least $N+1$ local measurements  \cite{Guhne2007}.

We demonstrate that our results provide overall accessible quantitative assessments of global, partial, and genuine multiparticle entanglement in a variety of noisy states produced in recent experiments \cite{Blatt2005,PianiSmolin,Blatt2010,BlattPRL,Prevedel2009,Wieczorek2009}, going beyond mere detection
\cite{GuhneToth2009,Guhne2010,Gao2014,Levi2013,Huber2010,DurCirac2000,Gabriel2010,Klockl2015,ExpFriendly}, yet with a signficantly reduced experimental overhead. Compared with some recent complementary approaches to the quantification of multiparticle entanglement \cite{Guhne2015,Hofmann2014,SiewertPRL,SiewertSciRep,Wu2012,Hashemi2012,Audenart2006,Wunderlich2009,Ma2011,QWitness2,QWitness1}, we find that our results, obtained via the general quantitative framework discussed above, fare surprisingly well in their efficiency and versatility despite the minimal experimental requirements (see Table \ref{TabComp} for an in-depth comparison).

 \begin{table*}[bht]
 \centering
\begin{tabular}{cccccc}
\hline \hline
 \multicolumn{6}{c}{Comparison Table: Experimentally friendly methods to quantify multiparticle entanglement} \\
\hline \hline
 Ref. & \qquad Experimental \qquad \qquad  & Computational & \qquad  $M$-inseparability \qquad  & \qquad  Entanglement  \qquad \qquad   \\
  & \qquad friendliness \qquad \qquad  & friendliness & \qquad quantified \qquad  & \qquad  measure  \qquad \qquad \\ \hline
\cite{QWitness2,QWitness1} & Variable & Optimisation required  & \qquad $2\leq M \leq N$ & Any convex and continuous measure \\

\cite{Ma2011} & $O(2^N)$ & Optimisation required   & \qquad 2 & Genuine multiparticle concurrence &  \\

\cite{Audenart2006,Wunderlich2009} & $O(N)$ & Closed formula & \qquad  2 & Robustness of entanglement \\

\cite{Hashemi2012} & $O(N)$ & Closed formula & \qquad 2 & Genuine multiparticle concurrence &  \\

\cite{Wu2012} & $O(N)$ & Closed formula & \qquad 2 & Genuine multiparticle concurrence  &  \\

\cite{SiewertPRL,SiewertSciRep} & $O(N)$ & Closed formula & \qquad 2 &  Polynomial invariant (three-tangle)  &  \\

\cite{Hofmann2014} & $O(N)$ & Closed formula & \qquad 2 & Genuine multiparticle negativity   \\

\cite{Guhne2015} & $O(N)$ & Closed formula & \qquad $2\leq M \leq N$ & Infidelity-based geometric measure  \\

[$\ast$] (GHZ-diagonal set) & $O(N)$ & Closed formula & \qquad 2 & All distance-based measures \\

[$\ast$] (\M3N set) & $3$ & Closed formula & \qquad $\left \lceil{N/2}\right \rceil \leq M \leq N$ & All distance-based  measures\\

\hline \hline
\end{tabular}

\caption{\label{TabComp} A comparison of  relevant literature on experimentally friendly quantification of multiparticle entanglement (based on accessible lower bounds). For each reference, we give the experimental friendliness, in terms of the number of local measurement settings required, and also the computational friendliness. The levels of $M$-inseparability quantified are given, along with the entanglement measures to which each work applies; [$\ast$] refers to this paper.
}
\end{table*}

\section{\sf \bfseries Results}

\subsection{\sf \bfseries Global and partial multiparticle entanglement}

\begin{figure*}[tbh]
\centering \includegraphics[width=15cm]{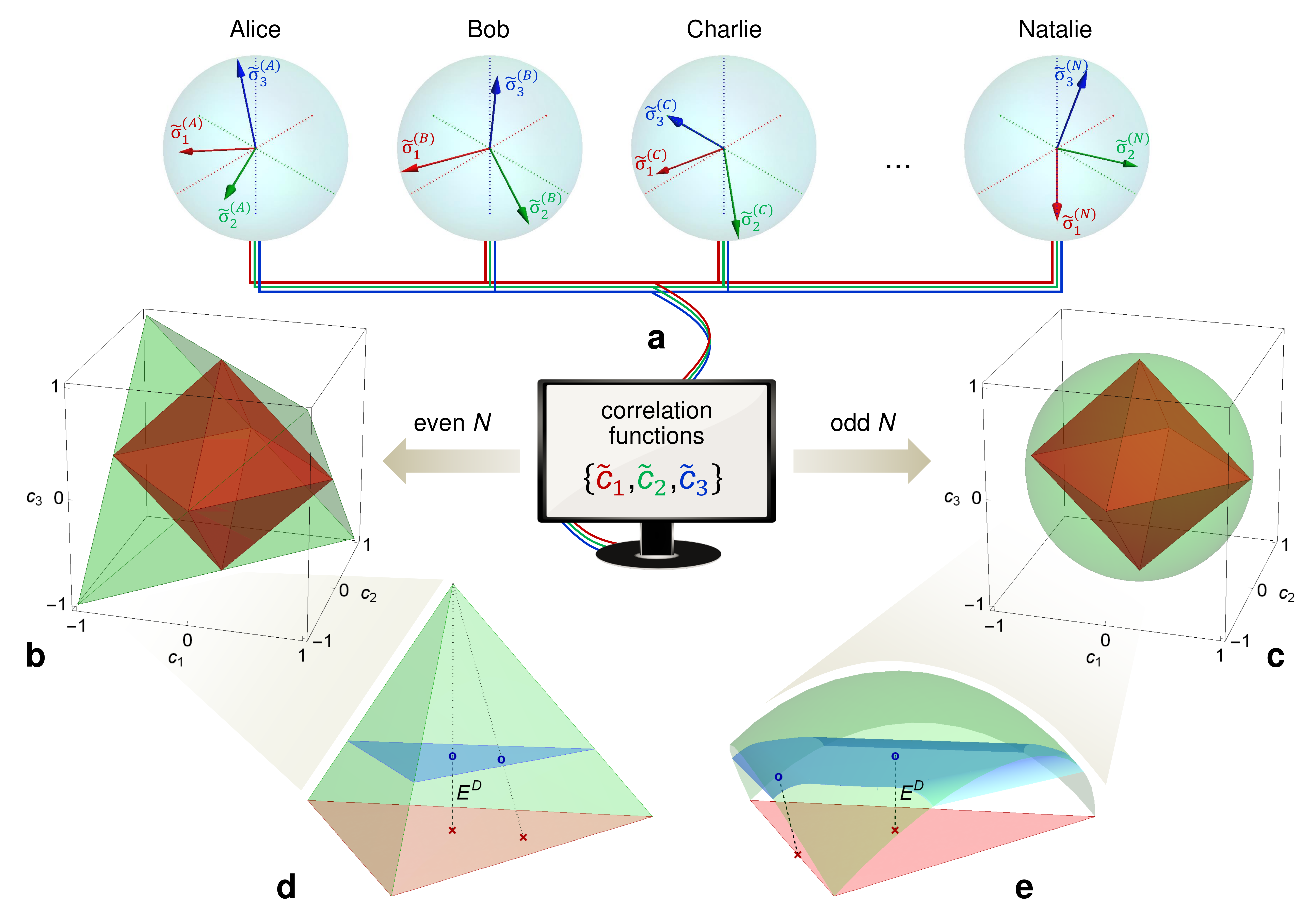}\\
\caption{{Experimentally friendly protocol to quantify global and partial $N$-particle entanglement.} {\em Top row:} (a) A state $\varrho$ of $N$ qubits is shared by $N$ parties, named Alice, Bob, Charlie, ..., Natalie. Each party, labelled by $\alpha=A,\ldots,N$, locally measures her or his qubit in three orthogonal directions $\{\tilde{\sigma}_j^{(\alpha)}\}$, with $j=1,2,3$, indicated by the solid arrows. If the shared state $\varrho$ is completely unknown, a standard choice can be to measure the three canonical Pauli operators for all the qubits (corresponding to the directions of the dashed axes); if instead some partial information on $\varrho$ is available, the measurement directions can be optimised a priori. Once all the data are collected, the $N$ parties communicate classically to construct the three correlation functions $\{\tilde{c}_j\}$, with $\tilde{c}_j = \langle \bigotimes_\alpha \tilde{\sigma}_j^{(\alpha)}\rangle$.  {\em Middle row}: For any $N$, one can define a reference subset of $N$-qubit states with all maximally mixed marginals (\M3N states), which are completely specified by a triple of orthogonal correlation functions $\{{c}_j\}$. These states enjoy a convenient representation in the space of $\{{c}_1, {c}_2, {c}_3\}$. (b) For even $N$, \M3N states fill the tetrahedron with vertices $\{1,(-1)^{N/2},1\}$, $\{-1,-(-1)^{N/2},1\}$, $\{1,-(-1)^{N/2},-1\}$ and $\{-1,(-1)^{N/2},-1\}$. (c) For odd $N$, they are instead contained in the  unit Bloch ball. For any $M > \left \lceil{N/2}\right \rceil$, $M$-separable \M3N states are confined to the octahedron
 with vertices $\{\pm 1, 0, 0\}$, $\{0, \pm 1, 0\}$ and $\{0, 0, \pm 1\}$, illustrated in red in both panels; conversely, for $M \leq \lceil{N/2}\rceil$, all \M3N states are $M$-separable.
{\em Bottom row}: Geometric analysis of multiparticle entanglement. The bottom panels depict zooms of (d) a corner of the tetrahedron for even $N$ and (e) a sector of the unit sphere for odd $N$, opposing a face of the octahedron of $M$-separable \M3N states (for $M > \left \lceil{N/2}\right \rceil$). Instances of inseparable \M3N states are indicated by blue circles, and their closest $M$-separable states by red crosses. The cyan surfaces in each of the two bottom panels contain states with equal global and partial multiparticle entanglement $E^D_{M}$, which we compute exactly. The results are valid for any contractive and jointly convex distance $D$ in the even $N$ case, and for the trace distance in the odd $N$ case. The  entanglement of an \M3N state with correlation functions $\{\tilde{c}_1,\tilde{c}_2,\tilde{c}_3\}$ provides an analytical lower bound for the  entanglement of any $N$-qubit state with the same correlation functions, such as the state $\varrho$ initially shared by the $N$ parties in (a). The  bound is effective for the most relevant families of $N$-qubit states in theoretical and experimental investigations of quantum information processing, as we show in this article.
}
\label{figTB}
\end{figure*}

We begin by choosing as our reference family the set of $N$-qubit \M3N states. An \M3N state is defined as $\varpi = \frac{1}{2^N}\left(\mathbb{I}^{\otimes N} + \sum_{j=1}^{3} c_j \sigma_j^{\otimes N}\right)$, where $\mathbb{I}$ is the $2 \times 2$ identity matrix. These states are invariant under permutations of any pair of qubits and enjoy a nice geometrical representation in the space of the three correlation functions $c_j$, corresponding to a tetrahedron for even $N$ and to the unit ball for odd $N$, as depicted in Figure~\ref{figTB}. We can then characterise the subset of $M$-separable $\mathcal{M}_{N}^{3}$ states for any $N$. We find that, if $M > \left \lceil{N/2}\right \rceil$, then the $M$-separable $\mathcal{M}_{N}^{3}$ states fill a subset corresponding to an octahedron in the space of the correlation functions (see Figure~\ref{figTB}). When $M \leq \left \lceil{N/2}\right \rceil$, all $\mathcal{M}_{N}^{3}$ states are instead $M$-separable. The proofs are deferred to the Supplementary Material.

We can now tackle the quantification of global and partial multiparticle entanglement in these states; for the latter, we will always focus on the nontrivial case  $M > \left \lceil{N/2}\right \rceil$ throughout this section.
First, we observe that the closest $M$-separable state $\varsigma_\varpi$ to an $\mathcal{M}_{N}^{3}$ state $\varpi$, which solves the optimisation in Eq.~(\ref{geome}), can always be found within the subset of  $M$-separable $\mathcal{M}_{N}^{3}$ states, yielding a considerable simplification of the general problem. To find the exact form of $\varsigma_\varpi$, and consequently of $E^D_{M}(\varpi)$, we approach the cases of even and odd $N$ separately.

For even $N$, we prove that there exists a unique solution to the minimisation in Eq.~(\ref{geome}), independent of the specific choice of contractive and jointly convex distance $D$. Namely, the closest $M$-separable state $\varsigma_\varpi$ is on the face of the octahedron bounding the corner of the tetrahedron in which $\varpi$ is located, and is identified by the intersection of such octahedron face with the line connecting $\varpi$ to the vertex of the tetrahedron corner, as depicted in Figure~\ref{figTB}(d).
It follows that, 
for any nontrivial $M$,  valid $D$, and even $N$, the multiparticle entanglement $E^D_{M}(\varpi^{\{c_j\}})$ of an \M3N state $\varpi^{\{c_j\}}$ with correlation functions $\{c_j\}$ is only a monotonically increasing function of the Euclidean distance between the point of coordinates $\{c_j\}$ and the closest octahedron face, which is in turn proportional to $h_\varpi = \frac12( \sum_{j=1}^3 |c_j|-1)$ (notice that $h_\varpi$ equals the bipartite measure known as concurrence for $N=2$ \cite{PlenioVirmani2007,Carnio2015}). We have then a closed formula for any valid geometric measure of global and partial multiparticle entanglement on an arbitrary  $\mathcal{M}_{N}^{3}$ state $\varpi^{\{c_j\}}$ with even $N$, given by
\begin{equation}\label{evN}
E^D_{M}(\varpi^{\{c_j\}}) = \left\{
                           \begin{array}{ll}
                             0 \,, & \hbox{$h_\varpi \leq 0$ (or $M \leq N/2$);} \\
                             f_D(h_\varpi)\,, & \hbox{otherwise,}
                           \end{array}
                         \right.
\end{equation}
where  $f_D$ denotes a monotonically increasing function whose explicit form is specific to each distance $D$. In Table~\ref{tableD} we present the expression of $f_D$ for relevant distances in quantum information theory.

\begin{table}[bht]
\begin{tabular}{ccc}
\hline \hline
Distance $D$ & $D(\varrho,\varsigma)$ & $f_{D}(h_{\varpi})$ \\ \hline
$\begin{array}{c}\mbox{\!Relative\!\!}\\ \mbox{\!entropy $D_{\text{RE}}$\!\!\!\!}\end{array}$ & $\text{Tr}\left[ \varrho \left( \log_2 \varrho - \log_2 \varsigma \right)\right]$ &
$\begin{array}{l} \frac{1}{2}\big[\! \left(1-h_{\varpi}\right) \log_2\! \left(1-h_{\varpi}\right) \\
\ \  +\left(1+h_{\varpi}\right) \log_2\! \left(1+h_{\varpi}\right)\! \big] \end{array}$ \\[9pt]
\!Trace $D_{\text{Tr}}$  & $\frac{1}{2}\text{Tr} \left| \varrho - \varsigma \right|$ & $\frac{1}{2} h_{\varpi}$ \\[9pt]
\!Infidelity $D_{\text{F}}$ & $1-\left[\text{Tr}\left( \sqrt{\sqrt{\varsigma} \varrho \sqrt{\varsigma}} \right)\right]^2$  & $\frac{1}{2} \left(1-\sqrt{1-h_{\varpi}^2}\right)$ \\[9pt]
$\begin{array}{c}\mbox{\!Squared\!}\\ \mbox{\!Bures $D_{\text{B}}$\!\!\!\!}\end{array}$& $2\left[1-\text{Tr}\left( \sqrt{\sqrt{\varsigma} \varrho \sqrt{\varsigma}} \right)\right]$ & $2 - \sqrt{1-h_{\varpi}}-\sqrt{1+h_{\varpi}}$ \\[9pt]
$\begin{array}{c}\mbox{\!Squared\!}\\ \mbox{\!Hellinger $D_{\text{H}}$\!\!\!\!}\end{array}$ & $2\left[ 1-\text{Tr}\left(\sqrt{\varrho}\sqrt{\varsigma}\right) \right]$ & $2 - \sqrt{1-h_{\varpi}}-\sqrt{1+h_{\varpi}}$ \\[3pt] \hline \hline
\end{tabular}
\caption{\label{tableD}  {Analytical expression of global and partial multiparticle entanglement} $E^D_{M}$  for \M3N states of an even number $N$ of qubits as defined by Eq.~(\ref{evN}), for representative choices of the distance  $D$.}
\end{table}

For odd $N$, the closest $M$-separable state $\varsigma_{\varpi}$ to any $\mathcal{M}_{N}^{3}$ state $\varpi$ is still independent of (any nontrivial) $M$. However, different choices of $D$ in Eq.~(\ref{geome}) are minimised by different states $\varsigma_\varpi$. We focus on the important but notoriously hard-to-evaluate case of the trace distance $D_{\text{Tr}}(\varpi,\varsigma)$ (see Table~\ref{tableD}).
In the representation of Figure~\ref{figTB}(c), the trace distance amounts to half the Euclidean distance on the unit ball. It follows that the closest $M$-separable state $\varsigma_\varpi$ to $\varpi$ is the Euclidean orthogonal projection onto the boundary of the octahedron, see Figure~\ref{figTB}(e). We can then get a closed formula for the trace distance measure of global and partial multiparticle entanglement $E^{D_{\text{Tr}}}_{M}(\varpi^{\{c_j\}})$ of an arbitrary  $\mathcal{M}_{N}^{3}$ state $\varpi^{\{c_j\}}$ with odd $N$ as well, given by
\begin{equation}\label{odN}
E^{D_{\text{Tr}}}_{M}(\varpi^{\{c_j\}}) = \left\{
                           \begin{array}{l}
                             0 \,, \,\, \, \,\,\, \hbox{$h_\varpi\leq 0$ (or $M\leq \left \lceil{N/2}\right \rceil$);} \\
                             \frac{h_\varpi}{\sqrt{3}}\,, \,\, \hbox{$0 < h_{\varpi} \leq 3 |c_{j}|/2$ $\forall j$;} \\
\underset{j}{\min}\, \frac12\! \sqrt{|c_j|^2+\frac12(2 h_{\varpi}-|c_{j}|)^{2}} \,, \,\, \hbox{otherwise.}
                           \end{array}
                         \right.
\end{equation}



The usefulness of the just derived analytical results for multiparticle entanglement is not limited to the  \M3N states. In accordance with our general framework, a crucial observation  is that the \M3N states are {\it extremal} among all quantum states with given correlation functions $\{c_j\}$.
Specifically, any general state $\varrho$ of $N$ qubits can be transformed into an \M3N state with the same $\{c_j\}$ by means of a procedure that we name \M3N-fication, involving only LOCC (see Methods). This immediately implies that, for any $M > \left \lceil{N/2}\right \rceil$, the multiparticle entanglement $E_M^D$ of $\varrho$
can have a nontrivial exact lower bound given by the corresponding multiparticle entanglement of the \M3N state $\varpi$ with the same $\{c_j\}$,
\begin{equation}\label{low}
E^D_{M}(\varpi^{\{c_j\}}) \leq E^D_{M}(\varrho)\,, \mbox{  $\forall\,\varrho$ : $\text{Tr}\big(\varrho\, \sigma_j^{\otimes N}\big)=c_j$ ($j=1,2,3$)}.
\end{equation}
From a practical point of view, one needs only to measure the three correlation functions $\{c_j\}$, as routinely done in optical, atomic, and spin systems \cite{PianiSmolin,Blatt2005,BlattPRL,GuhneToth2009}, to obtain an estimate of the global and partial multiparticle entanglement content of an unknown state $\varrho$ with no need for a full state reconstruction.

 \begin{table*}[bth]
\centering
    \begin{minipage}{.52\textwidth}
\begin{flushleft}
\begin{tabular}{ccccc}
 \hline \hline
$N$ & \multicolumn{1}{c}{State} & \multicolumn{1}{c}{$\{\tilde c_1,\tilde c_2,\tilde c_3\}$} & \multicolumn{1}{c}{$\sum_{j=1}^{3} |\tilde{c}_j|$} & \multicolumn{1}{c}{$\{\theta,\psi,\phi\}$} \\
 \hline
 \parbox[t]{2mm}{\multirow{2}{*}{\rotatebox[origin=c]{90}{$N=3$}}} &
$|\text{GHZ}^{(3)}\rangle$ & $\left\lbrace-\sqrt{\frac{8}{27}},\sqrt{\frac{8}{27}}, -\sqrt{\frac{8}{27}} \right\rbrace$  &$2\sqrt{\frac{2}{3}}$                                       & $\left\lbrace \cos^{-1}(\frac{1}{\sqrt{3}}),\frac{5\pi}{30}, \frac{\pi}{4} \right\rbrace$   \\ &
$|\text{W}^{(3)}\rangle $       & $\left\lbrace\frac{1}{\sqrt{3}},-\frac{1}{\sqrt{3}} ,\frac{1}{\sqrt{3}} \right\rbrace$                                                               & $\sqrt{3} $                                                       & $\left\lbrace \cos^{-1}(\frac{1}{\sqrt{3}}),0 ,\frac{\pi}{4} \right\rbrace$ \\
 \hline
 \parbox[t]{2mm}{\multirow{8}{*}{\rotatebox[origin=c]{90}{$N=4$}}} &
$|\text{GHZ}^{(4)}\rangle$ & $\left\lbrace 1, 1,  1\right\rbrace$                                                                                                                                                 & $3$                                                                     & $\left\lbrace 0,0, 0 \right\rbrace$   \\ &
$|\text{W}^{(4)}\rangle $       & $\left\lbrace\frac{5}{9},\frac{5}{9} ,\frac{5}{9} \right\rbrace$                                                                                           & $\frac{5}{3}$                                                 & $\left\lbrace \cos^{-1}(\frac{1}{\sqrt{3}}),0 ,\frac{\pi}{4} \right\rbrace$ \\ &
$\varrho^{(4)}_{\text{Wei}}(x) $ & $\left\lbrace x, x ,2x-1 \right\rbrace$                                                                                                                                           & $2x+|2x-1|$                                                     & $\left\lbrace0,0 ,0 \right\rbrace$ \\ &
$|\text{C}^{(4)}_{1}\rangle $        & $\left\lbrace1,1 ,1 \right\rbrace$                                                                                                                                                  & $3$                                                                   & * \\ &
$|\text{C}^{(4)}_{2}\rangle $        & $\left\lbrace1,1 ,1 \right\rbrace$                                                                                                                                                  & $3$                                                                   & $\left\lbrace \frac{\pi}{4},0, 0 \right\rbrace$ \\ &
$|\text{D}_{2}^{(4)}\rangle $        & $\left\lbrace1,1 ,1 \right\rbrace$                                                                                                                                                  & $3$                                                                   & $\left\lbrace 0,0, 0 \right\rbrace$ \\ &
$| \Psi^{(4)}\rangle $        & $\left\lbrace1,1 ,1 \right\rbrace$                                                                                                                                                  & $3$                                                                   & $\left\lbrace 0,0, 0 \right\rbrace$ \\ &
$\varrho_{\text{S}}^{(4)}$        & $\left\lbrace 1,1 ,1 \right\rbrace$                                                                                                                                                  & $3$                                                                   & $\left\lbrace 0,0, 0 \right\rbrace$ \\
\hline
 \parbox[t]{2mm}{\multirow{4}{*}{\rotatebox[origin=c]{90}{$N=5$}}} &
$|\text{GHZ}^{(5)}\rangle$ & $\left\lbrace \frac{1}{\sqrt{2}}, \frac{1}{\sqrt{2}},  0\right\rbrace$                                                                                       & $\sqrt{2}$                                                         & $\left\lbrace 0, \frac{\pi}{40}, \frac{\pi}{40} \right\rbrace$   \\ &
$|\text{W}^{(5)}\rangle $       & $\left\lbrace\frac{7}{9\sqrt{3}},-\frac{7}{9\sqrt{3}} ,\frac{7}{9\sqrt{3}} \right\rbrace$                                                        & $\frac{7}{3\sqrt{3}}$                                         & $\left\lbrace \cos^{-1}(\frac{1}{\sqrt{3}}),0 ,\frac{\pi}{4} \right\rbrace$ \\ &
$\varrho^{(5)}_{\text{Wei}}(x) $ & $\left\lbrace \frac{x}{\sqrt{2}}, \frac{x}{\sqrt{2}},  0\right\rbrace$                                                                                                                                           & $\sqrt{2}x$                                                     & $\left\lbrace0, \frac{\pi}{40} , \frac{\pi}{40} \right\rbrace$ \\
 &
$|\text{C}^{(5)}_{1}\rangle $        & $\left\lbrace \frac{1}{2},\frac{1}{2} ,\frac{1}{2} \right\rbrace$                                                                                                                                                  & $\frac{3}{2}$                                                                   & * \\
 \hline
 \parbox[t]{2mm}{\multirow{6}{*}{\rotatebox[origin=c]{90}{$N=6$}}} &
$|\text{GHZ}^{(6)}\rangle$ & $\left\lbrace 1, -1,  1\right\rbrace$                                                                                                                                                 & $3$                                                                     & $\left\lbrace 0,0, 0 \right\rbrace$   \\ 
&$\varrho^{(6)}_{\text{Wei}}(x) $ & $\left\lbrace x, -x ,2x-1 \right\rbrace$                                                                                                                                           & $2x+|2x-1|$                                                     & $\left\lbrace0,0 ,0 \right\rbrace$ \\ &
$|\text{C}^{(6)}_{1}\rangle $        & $\left\lbrace1,-1 ,1 \right\rbrace$                                                                                                                                                  & $3$                                                                   & * \\ &
$|\text{C}^{(6)}_{2}\rangle $        & $\left\lbrace1,-1 ,1 \right\rbrace$                                                                                                                                                  & $3$                                                                   & * \\ &
$|\text{D}_{3}^{(6)}\rangle $        & $\left\lbrace1,1 ,-1 \right\rbrace$                                                                                                                                                  & $3$                                                                   & $\left\lbrace 0,0, 0 \right\rbrace$ \\ &
$\varrho_{\text{S}}^{(6)}$        & $\left\lbrace -1,-1 ,-1 \right\rbrace$                                                                                                                                                  & $3$                                                                   & $\left\lbrace 0,0, 0 \right\rbrace$ \\
 \hline
 \parbox[t]{2mm}{\multirow{3}{*}{\rotatebox[origin=c]{90}{$N=7$}}} &
$|\text{GHZ}^{(7)}\rangle$ & $\left\lbrace \frac{1}{\sqrt{2}}, -\frac{1}{\sqrt{2}},  0\right\rbrace$                                                                                       & $\sqrt{2}$                                                         & $\left\lbrace 0, \frac{\pi}{56}, \frac{\pi}{56} \right\rbrace$   \\
&
$\varrho^{(7)}_{\text{Wei}}(x) $ & $\left\lbrace \frac{x}{\sqrt{2}}, -\frac{x}{\sqrt{2}},  0\right\rbrace$                                                                                      & $\sqrt{2}x$                                                     & $\left\lbrace0, \frac{\pi}{56} , \frac{\pi}{56} \right\rbrace$ \\
 &
$|\text{C}^{(7)}_{1}\rangle $        & $\left\lbrace \frac{1}{2},-\frac{1}{2} ,\frac{1}{2} \right\rbrace$                                                                                                                                                  & $\frac{3}{2}$                                                                   & * \\
 \hline
\parbox[t]{2mm}{\multirow{5}{*}{\rotatebox[origin=c]{90}{$N=8$}}} &
$|\text{GHZ}^{(8)}\rangle$ & $\left\lbrace 1, 1,  1\right\rbrace$                                                                                                                                                 & $3$                                                                     & $\left\lbrace 0,0, 0 \right\rbrace$   \\ 
&$\varrho^{(8)}_{\text{Wei}}(x) $ & $\left\lbrace x, x ,2x-1 \right\rbrace$                                                                                                                                           & $2x+|2x-1|$                                                     & $\left\lbrace0,0 ,0 \right\rbrace$ \\
 &
$|\text{C}^{(8)}_{1}\rangle $        & $\left\lbrace1,1 ,1 \right\rbrace$                                                                                                                                                  & $3$                                                                   & * \\ &
$|\text{D}_{4}^{(8)}\rangle $        & $\left\lbrace1,1 ,1 \right\rbrace$                                                                                                                                                  & $3$                                                                   & $\left\lbrace 0,0, 0 \right\rbrace$ \\ &
$\varrho_{\text{S}}^{(8)}$        & $\left\lbrace 1,1 ,1 \right\rbrace$                                                                                                                                                  & $3$                                                                   & $\left\lbrace 0,0, 0 \right\rbrace$ \\
 \hline
 \hline
 \end{tabular}
 \end{flushleft}
    \end{minipage}%
    \begin{minipage}{0.48\textwidth}
    \centering
\caption{\label{TabTh} Applications of our framework to construct accessible lower bounds on global and partial ($M$-inseparable) multiparticle entanglement (which are nonzero for any $M > \left \lceil{N/2}\right \rceil$ when $\sum_j |\tilde{c}_j|>1$), for the families of $N$-qubit states listed as follows.
(i) $N$-qubit $GHZ$ states \cite{GHZ90}
$|\text{GHZ}^{(N)}\rangle = \frac{1}{\sqrt{2}} \left(|00\cdots 00 \rangle + |11\cdots 11 \rangle\right)$
with $N\geq 3$. 
(ii) $N$-qubit $\text{W}$ states \cite{Dur2000}
$|\text{W}^{(N)}\rangle = \frac{1}{\sqrt{N}} \left(|00\cdots 01 \rangle + |00\cdots 10 \rangle + \cdots + |10\cdots 00\rangle \right)
$
with $N\geq 3$.
(iii) $N$-qubit Wei states  \cite{Wei2004,Wei2008}
$\varrho^{(N)}_{\text{Wei}}(x) = x |\text{GHZ}^{(N)}\rangle\langle \text{GHZ}^{(N)}| + \frac{(1-x)}{2N} \sum_{k=1}^N \left(P_k + \overline{P}_k\right),
$
where $N\geq 4$, $x\in[0,1]$ and $P_k$ is the projector onto the binary $N$-qubit representation of $2^{k-1}$ whereas $\bar{P}_{i} = \sigma_{1}^{\otimes N} P_{i} \sigma_{1}^{\otimes N}$. 
(iv) $N$-qubit linear cluster states $|\text{C}^{(N)}_{1}\rangle$ corresponding to the $N$-vertex linear graph $\footnotesize{\begin{array}{ccccccc}
\!\!\bullet \!\!&\!\! \mbox{---} \!\!&\!\! \bullet \!\!&\!\! \mbox{---} \!\!&\!\! \cdots \!\!&\!\! \mbox{---} \!\!&\!\! \bullet
\end{array}}$ \cite{Hofmann2014,BriegelNP}.
(v) $N$-qubit rectangular cluster states $|\text{C}^{(N)}_{2}\rangle$ corresponding to the $N$-vertex ladder-type graph $\footnotesize{\begin{array}{ccccccc}
\!\!\bullet \!\!&\!\! \mbox{---} \!\!&\!\! \bullet \!\!&\!\! \mbox{---} \!\!&\!\! \cdots \!\!&\!\! \mbox{---} \!\!&\!\! \bullet \\[-.1cm]
\!\!\vert \!\!&\!\! &\!\! \vert \!\!&\!\! \!\!&\!\! \!\!&\!\! \!\!&\!\! \vert \\[-.1cm]
\!\!\bullet \!\!&\!\! \mbox{---} \!\!&\!\! \bullet \!\!&\!\! \mbox{---} \!\!&\!\! \cdots \!\!&\!\! \mbox{---} \!\!&\!\! \bullet
\end{array}}$ \cite{Raussendorf2001,BriegelNP}.
(vi) $N$-qubit (symmetric) Dicke states $|{\text{D}}_{k}^{(N)}\rangle = \frac{1}{\sqrt{Z}} \sum_{i} \Pi_{i}(| 0 \rangle^{\otimes N-k} \otimes | 1 \rangle^{\otimes k})$, which are superpositions of all states with  $k$ qubits in the excited state $|1\rangle$ and  $N-k$ qubits in the ground state $|0\rangle$, with the symbol $\{{\Pi}_{i}(\cdot)\}_{i=1}^{Z}$ denoting all the $Z \equiv {N \choose k}$ distinct permutations of 0's and 1's; we focus on half-excited Dicke states, given by $k=N/2$ for any even $N$ \cite{Dicke1954,AcinScience,Wieczorek2009,Prevedel2009}.
(vii) $4$-qubit singlet state \cite{weinfurter2001four} $| \Psi^{(4)}\rangle = \frac{1}{\sqrt{3}}\big[|0011 \rangle+|1100 \rangle- (|0101 \rangle+|0110 \rangle+|1001 \rangle+|1010 \rangle)/2\big]$.
(viii) $N$-qubit generalised Smolin states \cite{Smolin2001,BeatrizSmolin,GeneSmolin} $\varrho_{\text{S}}^{(N)}$ for even $N \geq 4$, which are instances of $\mathcal{M}^3_N$ states with correlation triple $\{(-1)^{N/2},(-1)^{N/2},(-1)^{N/2}\}$, hence their entanglement quantification is exact. The asterisk * indicates non-permutationally invariant states for which the optimisation of the bounds requires different angles for each qubit (not reported here).
Notice that in the table we listed mostly pure states. In general, if the triple  $\{\tilde{c}_{j}\}$ is optimal for a pure $N$-qubit state $|\Phi^{(N)}\rangle$, then for the mixed state  $\varrho^{(N)}(q)=q |\Phi^{(N)}\rangle\langle\Phi^{(N)}| + \frac{1-q}{2^N} \mathbb{I}^{\otimes N}$, obtained by mixing $|\Phi^{(N)}\rangle$  with white noise, one still gets nonzero lower bounds to global and partial entanglement for all $q > 1/\sum_{j=1}^{3} |\tilde{c}_j|$, as shown in Figure~\ref{figWei} for some representative examples. \\}
    \end{minipage}
\end{table*}

Furthermore, the lower bound can be improved if a partial knowledge of the state $\varrho$ is assumed, as is usually the case for experiments aiming to produce specific families of states for applications in quantum information processing \cite{Blatt2005,PianiSmolin,BoundNature}. In those realisations, one typically aims to detect entanglement by constructing optimised entanglement witnesses tailored on the target states \cite{GuhneToth2009}. By exploiting similar ideas, we can optimise the quantitative lower bound in Eq.~(\ref{low}) over all possible single-qubit local unitaries applied to the state $\varrho$ before the \M3N-fication,
\begin{eqnarray}\label{lowopt}
&&\sup_{U_\otimes} E^D_{M}(\varpi^{\{\tilde{c}_j\}}) \leq  E^D_{M}(U_\otimes \varrho U_\otimes^{\dagger}) = E^D_{M}(\varrho)\,,
\end{eqnarray}
where $\text{Tr}\big(U_\otimes \varrho U_\otimes^{\dagger}\, \sigma_j^{\otimes N}\big)=\tilde{c}_j$ and $U_\otimes = \bigotimes_{\alpha=1}^{N} U^{(\alpha)}$ denotes a single-qubit local unitary operation. Experimentally, the optimised bound can then be still accessed by measuring a triple of correlations functions $\{\tilde{c}_j\}$ given by the expectation values of correspondingly rotated Pauli operators on each qubit, $\tilde{c}_j = \langle U_\otimes^\dagger \sigma_j^{\otimes N} U_\otimes \rangle$, as illustrated in Figure~\ref{figTB}(a).
Optimality in Eq.~(\ref{lowopt}) can be achieved by the choice of $U_\otimes$ such that $\tilde{h}_\varpi = \frac12( \sum_{j=1}^3 |\tilde{c}_j|-1)$ is maximum.
The optimisation procedure can be significantly simplified when considering a state $\varrho$ which is invariant under permutations of any pair of qubits. In such a case, one may need to optimise only over three angles $\{\theta, \psi, \phi\}$ parameterising a generic unitary applied to each single qubit; the optimisation can be equivalently performed over an orthogonal matrix acting on the Bloch vector of each qubit (see Methods).



We can now investigate how useful our results are on concrete examples.
Table~\ref{TabTh} presents a compendium of optimised analytical lower bounds on the global and partial multiparticle entanglement of several relevant families of $N$-qubit states \cite{GHZ90,Dur2000,Cabello2003,Raussendorf2001,Kiesel2005,Dicke1954,Kiesel2007,Prevedel2009,Wieczorek2009,Wei2004,Wei2008,GeneSmolin}, up to $N = 8$. All the bounds are experimentally accessible by measuring the three correlation functions $\{\tilde{c}_j\}$, corresponding to optimally rotated Pauli operators (see also Figure~\ref{figTB}).

\begin{figure*}[tbh]
\centering \includegraphics[width=16.5cm]{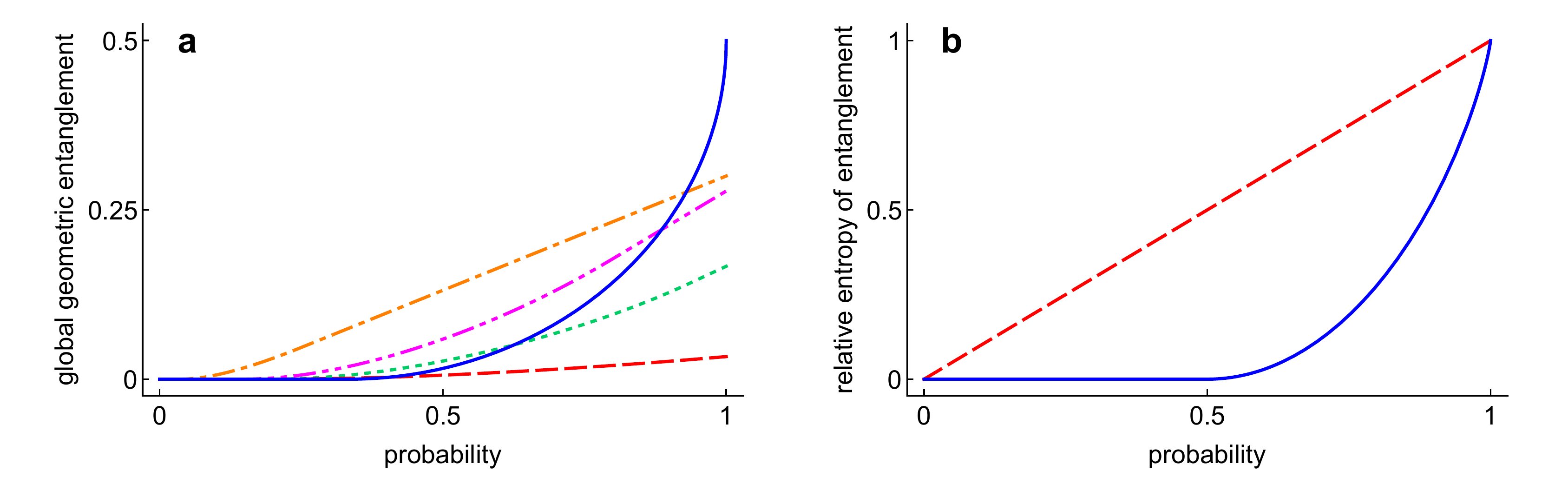}
\caption{(a) Lower bounds to the global geometric entanglement $E^{D_{\text{F}}}_N$ based on infidelity for noisy versions of some $N$-qubit states (defined in the caption of Table~\ref{TabTh}), as functions of the probability $q$ of obtaining the corresponding pure states. The non-solid lines refer to bounds obtained by the method of Ref.~\cite{Guhne2015} for: $4$-qubit linear cluster state  (green dotted), $6$-qubit rectangular cluster state (red dashed), $6$-qubit half-excited Dicke state (orange dot-dashed), $4$-qubit singlet state (magenta dot-dot-dashed). The solid blue line corresponds to our bound based on \M3N-fication for all the considered states, which is accessible by measuring only the three correlation functions $\tilde{c}_j =\langle \bigotimes_\alpha \tilde{\sigma}_j^{(\alpha)}\rangle$. (b) {Relative entropy of multiparticle entanglement of $N$-qubit Wei states} $\varrho^{(N)}_{\text{Wei}}$ defined in Table~\ref{TabTh}, as a function of the probability $x$ of obtaining a GHZ state. The dashed red line $E^{D_{\text{RE}}}_{N}(\varrho^{(N)}_{\text{Wei}})=x$ denotes the exact value of the global relative entropy of entanglement as computed in \cite{Wei2008}. The solid blue line denotes our accessible lower bound, obtained by combining Eqs.~(\ref{evN}) and (\ref{lowopt}) with the expressions in Tables~\ref{tableD} and \ref{TabTh}, and given explicitly by $E_{N,\text{low}}^{D_{\text{RE}}}(\varrho^{(N)}_{\text{Wei}}) =  \log_{2}(2-2x) + x \big(\log_{2}(x)- \log_{2}(1-x)\big)$  for $\frac12<x \leq 1$, while it vanishes for  $0\leq x \leq \frac12$.  The bound becomes tight for $x=1$, thus quantifying exactly the global multiparticle entanglement of pure GHZ states. We further show that our lower bound to global entanglement coincides with the exact genuine multiparticle entanglement of Wei states, $E_{N,\text{low}}^{D_{\text{RE}}}(\varrho^{(N)}_{\text{Wei}})=E_{2}^{D_{\text{RE}}}(\varrho^{(N)}_{\text{Wei}})$, that is computed in the next section of this paper. The results are scale-invariant and hold for any even $N$.
}
\label{figWei}
\end{figure*}

Let us comment on some cases where our analysis is particularly effective. For GHZ states, cluster states, and half-excited Dicke states, which constitute primary resources for quantum computation and metrology \cite{BriegelNP,Giovannetti2011}, we get the maximum $h_\varpi=1$ for any even $N$. This means that our bounds remain robust to estimate global and partial entanglement in noisy versions of these states (i.e. when one considers mixtures of any of these states with probability $q$ and the maximally mixed state with probability $1-q$) for all $q>1/3$. Notably, for values of $q$ sufficiently close to $1$, our bounds to global entanglement can be tighter than the (more experimentally demanding) ones derived very recently in Ref.~\cite{Guhne2015}, as shown in Figure~\ref{figWei}(a). Focusing on noisy GHZ states, we observe however that our scale-invariant threshold $q>1/3$, obtained by measuring the three canonical Pauli operators for each qubit, is weaker than the well-established inseparability threshold $q>1/(1+2^{N-1})$ \cite{DurCirac2000}. Nevertheless, we note that our simple quantitative bound given by Eq.~(\ref{low}) becomes tight in the paradigmatic limit of pure GHZ states ($q=1$) of any even number $N$ of qubits, thus returning the {\it exact} value of their global multiparticle entanglement via Eq.~(\ref{evN}), despite the fact that such states are not (and are very different from) \M3N states.  Eq.~(\ref{low}) also provides a useful nonvanishing lower bound to the global (and partial) $N$-particle entanglement of Wei states in the interval $x \in \big(\frac12,1\big]$, for any even $N$. A comparison between such a bound (with $D$ denoting the relative entropy), which requires only three local measurements, and the true value of the relative entropy of global $N$-particle entanglement for these states \cite{Wei2008}, whose experimental evaluation would conventionally require a complete state tomography, is presented in Figure~\ref{figWei}(b).



\subsection{\sf \bfseries Genuine multiparticle entanglement}
We now show how general analytical results for geometric measures of genuine multiparticle entanglement can be obtained as well within our approach.
The results from the previous section, while quite versatile, cannot provide useful bounds for the complete hierarchy of multiparticle entanglement, because \M3N states are $M$-separable for all $M \leq \left \lceil{N/2}\right \rceil$, and thus in particular biseparable for any number $N$ of qubits.
Therefore, to investigate genuine entanglement we consider a different reference set of states, specifically formed by mixtures of GHZ states, hence incarnating  archetypical representatives of full inseparability \cite{GHZ90,Guhne2010}. Any such state $\xi$, which will be referred to as a GHZ-diagonal (in short, ${\cal G}_N$) state, can be written as
$\xi = \sum_{i,\pm} p_i^{\pm} |\beta_i^{\pm}\rangle\langle\beta_i^{\pm}|\,,$
in terms of its eigenvalues $p_i^{\pm}$, with the eigenvectors $|\beta_i^{\pm}\rangle=\frac{1}{\sqrt{2}} \big(\mathbb{I}^{\otimes N} \pm \sigma_1^{\otimes N}\big) |i\rangle$  forming a basis of $N$-qubit GHZ states (where  $\{|i\rangle\}_{i=0}^{2^N-1}$ denotes the binary ordered $N$-qubit computational basis). The  ${\cal G}_N$ states have been studied in recent years as testbeds for multiparticle entanglement detection \cite{Guhne2007,Guhne2010}, and specific algebraic measures of genuine multiparticle entanglement such as the $N$-particle concurrence \cite{Ma2011,Hashemi2012} and negativity \cite{Hofmann2014} have been computed for these states. Here, we calculate exactly the whole class of geometric measures of genuine multiparticle entanglement $E^D_{2}$ defined by Eq.~(\ref{geome}), with respect to any contractive and jointly convex distance $D$, for ${\cal G}_N$ states of an arbitrary number $N$ of qubits.

By applying our general framework, we can prove that, for every valid $D$, the closest biseparable state to any ${\cal G}_N$ state can be found within the subset of biseparable ${\cal G}_N$ states (see Supplementary Material for detailed derivations). The latter subset is well characterised \cite{Guhne2010}, and is formed by all, and only, the ${\cal G}_N$ states with eigenvalues such that $p_{\max}\equiv\max_{i,\pm} p_i^\pm \leq 1/2$. We can then show that the closest biseparable ${\cal G}_N$ state to an arbitrary ${\cal G}_N$ state has maximum eigenvalue equal to $1/2$, which allows us to solve the optimisation in the definition of $E^D_2$, with respect to every valid $D$. We have then a closed formula for the geometric multiparticle entanglement of any ${\cal G}_N$ state $\xi$ with maximum eigenvalue $p_{\max}$, given by
\begin{equation}\label{egN}
E^D_{2}(\xi^{\{p_i^{\pm}\}}) = \left\{
                           \begin{array}{ll}
                             0 \,, & \hbox{$p_{\max} \leq 1/2$;} \\
                             g_D(p_{\max})\,, & \hbox{otherwise,}
                           \end{array}
                         \right.
\end{equation}
where $g_D$ denotes a monotonically increasing function whose explicit form is specific to each distance $D$, as reported in Table~\ref{tableGD} for typical instances.

\begin{table}[bht]
\begin{tabular}{cc}
\hline \hline
Distance $D$ & $g_{D}({p_{\max}})$ \\ \hline
$\begin{array}{c}\mbox{Relative}\\ \mbox{entropy $D_{\text{RE}}$}\end{array}$  &
$\begin{array}{l} 1 + p_{\max}\log_2 p_{\max}  \\
\ \ \ +\ (1-p_{\max})\log_2 (1-p_{\max}) \end{array}$ \\[9pt]
Trace $D_{\text{Tr}}$  & $p_{\max}-\frac{1}{2}$ \\[9pt]
Infidelity   $D_{\text{F}}$ & $\frac{1}{2} - \sqrt{p_{\max}(1-p_{\max})}$ \\[9pt]
$\begin{array}{c}\mbox{\!Squared\!}\\ \mbox{\!Bures $D_{\text{B}}$\!\!\!\!}\end{array}$ & $2-\sqrt{2} \left(\sqrt{1-p_{\max}}+\sqrt{p_{\max}}\right)$ \\[9pt]
$\begin{array}{c}\mbox{\!Squared\!}\\ \mbox{\!Hellinger $D_{\text{H}}$\!\!\!\!}\end{array}$ & $2-\sqrt{2} \left(\sqrt{1-p_{\max}}+\sqrt{p_{\max}}\right)$ \\[3pt] \hline \hline
\end{tabular}
\caption{\label{tableGD}  {Analytical expression of genuine multiparticle entanglement} $E^D_{2}$  for GHZ-diagonal states of any number $N$ of qubits as defined by Eq.~(\ref{egN}), for representative choices of the distance function $D$ (introduced in Table~\ref{tableD}).}
\end{table}

Let us comment on some particular results. The genuine multiparticle trace distance of entanglement $E_2^{D_{\text{Tr}}}$ is found to coincide with the genuine multiparticle negativity \cite{Hofmann2014} and with half the genuine multiparticle concurrence \cite{Hashemi2012} for all ${\cal G}_N$ states, thus providing the latter 
entanglement measures with an insightful geometrical interpretation on this important set of states. Examples of ${\cal G}_N$ states include several resources for quantum information processing, such as the noisy GHZ states and Wei states introduced in the previous section.
In particular, for noisy GHZ states (described by a pure-state probability $q$ as detailed in Table~\ref{TabTh}), we recover that every geometric measure of genuine multiparticle entanglement is nonzero if and only if $q>\big(1+(1-2^N)^{-1}\big)/2 {}_{\overrightarrow{\ N \gg 1\ }} 1/2$ \cite{Guhne2010} and monotonically increasing with $q$, as expected; for $q=1$ (pure GHZ states), genuine and global entanglement coincide, i.e.~the hierarchy of Figure~\ref{figMSE} collapses, meaning that all the entanglement of $N$-qubit GHZ states is genuinely shared among all the $N$ particles \cite{Blasone2008}.
On the other hand, the relative entropy of genuine multiparticle entanglement of Wei states \cite{Wei2004,Wei2008} can be calculated exactly via Eq.~(\ref{egN}); interestingly, for even $N$ it is found to coincide with the lower bound to their global entanglement that we had obtained by \M3N-fication, plotted as a solid line in Figure~\ref{figWei}(b). This means that for these states also the genuine multiparticle entanglement can be quantified entirely by measuring the three canonical correlation functions $\{c_j\}$, for any $N$. More generally, for arbitrary ${\cal G}_N$ states, all the genuine entanglement measures given by Eq.~(\ref{egN}) can be obtained by measuring the maximum GHZ overlap $p_{\max}$, which requires $N+1$ local measurement settings given explicitly in Ref.~\cite{Guhne2007}. This is remarkable, since with the same experimental effort needed to detect full inseparability \cite{Guhne2010} we have now a complete quantitative picture of genuine entanglement in these states based on any geometric measure, agreeing with and extending the findings of \cite{Hashemi2012,Hofmann2014}. Furthermore, as evident from Eq.~(\ref{egN}), all the geometric measures are monotonic functions of each other: our analysis thus reveals that there is a unique ordering of genuinely entangled  ${\cal G}_N$ states within the distance-based approach of Fig.~\ref{figMSE}.

In the same spirit as the previous section, and in compliance with our general framework, we note that the exact results obtained for the particular reference family of ${\cal G}_N$ states provide quantitative lower bounds to the genuine entanglement of general $N$-qubit states. This follows from the observation that any $N$-qubit state $\varrho$ can be transformed into a ${\cal G}_N$  state with eigenvalues $p_i^\pm=\langle\beta_i^\pm|\varrho|\beta_i^\pm\rangle$ by a LOCC procedure that we may call GHZ-diagonalisation \cite{Hofmann2014}. Therefore, given a completely general state $\varrho$, one only needs to measure its overlap with a suitable reference GHZ state; if this overlap is found larger than $1/2$, then by using Eq.~(\ref{egN}) with $p_{\max}$ equal to the measured overlap one obtains analytical lower bounds to the genuine multiparticle entanglement $E^D_2$ of $\varrho$ with respect to any desired distance $D$. As before, the bounds can be optimised in situations of partial prior knowledge, e.g.~by applying local unitaries on each qubit before the GHZ-diagonalisation, which has the effect of maximising the overlap with a chosen particular GHZ vector in the basis $\{|\beta_i^{\pm}\rangle\}$. The bounds then remain accessible for any state $\varrho$ by  $N+1$ local measurements \cite{Guhne2007}, with exactly the same demand as for just witnessing entanglement \cite{Guhne2010}.

For instance, for the singlet state $|\Psi^{(4)}\rangle$ \cite{weinfurter2001four}, which is a relevant resource in a number of quantum protocols including multiuser secret sharing \cite{bourennane2004decoherence,murao1999quantum,gaertner2007experimental}, one has $p_{\max} = \langle \beta_3^+|\Psi^{(4)}\rangle \langle \Psi^{(4)}|\beta_3^+\rangle = 2/3>1/2$, obtainable by measuring the overlap with the GHZ basis state $|\beta_3^+\rangle =  (|0011\rangle + |1100\rangle)/\sqrt{2}$. Optimised bounds to the genuine multiparticle entanglement of half-excited Dicke states $|\text{D}^{(N)}_{N/2}\rangle$ (for even $N \geq 4$), defined in Table~\ref{TabTh} \cite{Dicke1954,AcinScience}, can be found as well based on GHZ-diagonalisation, and are expressed by $p_{\max}^{(N)} = {N \choose {N/2}}\ 2^{1-N}$, meaning that they become looser with increasing $N$ and stay nonzero only up to $N=8$. In this respect, we note that alternative methods to detect full inseparability of Dicke states for any $N$ are available \cite{Prevedel2009,Wieczorek2009,GuhneToth2009}, but quantitative results are lacking in general. Nevertheless, applying our general approach to an alternative reference family more tailored to the Dicke states could yield tighter lower bounds that do not vanish beyond $N=8$.

Finally, notice that a lower bound to a distance-based measure of genuine multiparticle entanglement, as derived in this section, is automatically also a lower bound to corresponding measures of global and any form of partial entanglement, as evident by looking at the geometric picture in Figure~\ref{figMSE}. However, for states which are entangled yet not genuinely entangled, the simple bound from the previous section remains instrumental to assess their inseparability with minimum effort. \M3N states are themselves instances of such states (in fact, for even $N$, \M3N states are also ${\cal G}_N$ states, but with $p_{\max} \leq 1/2$ for $N>2$).



\subsection{\sf \bfseries Applications to experimental states}

 \begin{table*}[bht]
 \centering
\begin{tabular}{ccccccc}
\hline \hline
{\sf \bfseries a} & \multicolumn{6}{c}{Global and partial multiparticle entanglement} \\
\hline \hline
State & Ref. & \qquad Fidelity (\%) \qquad \qquad  & \multicolumn{2}{c}{$\{\tilde{c}_1,\tilde{c}_2,\tilde{c}_3\}$} & \qquad  $\sum_{j=1}^3|\tilde{c}_j|$ \qquad \qquad & \qquad \qquad $E^{D_{\text{Tr}}}_M$  \qquad \qquad \qquad \\ \hline
$\varrho_{\text{S}}^{(4)}$& \cite{PianiSmolin} & $96.83 \pm 0.05$ &  \multicolumn{2}{c}{$\{0.401 \pm 0.004,0.362 \pm 0.004, 0.397 \pm 0.008\}$} & $1.16 \pm 0.01$ &  $0.040 \pm 0.002$ \\
$\varrho_{\text{D}_3}^{(6)}$& \cite{Prevedel2009} & $56 \pm 2$ & \multicolumn{2}{c}{$\{0.8 \pm 0.2,0.5 \pm 0.2, -0.3 \pm 0.1\}$} & $1.6 \pm 0.3$ & $0.15 \pm 0.08$ \\
$\varrho_{\text{D}_3}^{(6)}$& \cite{Wieczorek2009} & $65 \pm 2$ &  \multicolumn{2}{c}{$\{0.63 \pm 0.02,0.63 \pm 0.02, -0.42 \pm 0.02\}$} & $1.69 \pm 0.04$ & $0.17 \pm 0.01$ \\
$\varrho_{\text{GHZ}}^{(3)}$& \cite{MonzPhD} & $87.9$ & \multicolumn{2}{c}{$\{-0.497, 0.515, -0.341\}$} & $1.35$ & $0.102$ \\
$\varrho_{\text{GHZ}}^{(4)}$& \cite{MonzPhD} & $80.3$ & \multicolumn{2}{c}{$\{0.663,0.683,0.901\}$} & $2.25$  & $0.312$ \\
$\varrho_{\text{W}_{A}}^{(4)}$ & \cite{MonzPhD} & $19.4$ & \multicolumn{2}{c}{$\{-0.404 , 0.454 , -0.378\}$} & $1.24$ & $0.0589$ \\
$\varrho_{\text{W}_{B}}^{(4)}$& \cite{MonzPhD} & $31.4$ & \multicolumn{2}{c}{$\{0.472 ,- 0.468 , -0.446\}$} & $1.39$ & $0.0963$ \\
\hline \hline
{\sf \bfseries b} & \multicolumn{6}{c}{Genuine multiparticle entanglement} \\
\hline \hline
State & Ref. & \qquad Fidelity (\%) \qquad \ & $E^{D_{\text{RE}}}_2$ & $E^{D_{\text{Tr}}}_2$ & $E^{D_{\text{F}}}_2$ & $E^{D_{\text{B}}}_2$  \\ \hline
$\varrho_\text{GHZ}^{(3)}$ & \cite{BlattPRL} & $97.0 \pm 0.3$ &  $0.81 \pm 0.02$ &  $0.470 \pm 0.003$ &  $0.329 \pm 0.008$ &  {$0.36 \pm 0.01$} \\
$\varrho_\text{GHZ}^{(4)}$ & \cite{BlattPRL}& $95.7 \pm 0.3$ &  $0.74 \pm 0.01$ &  $0.457 \pm 0.003$ &  $0.297 \pm 0.007$ &  {$0.323 \pm 0.008$}  \\
$\varrho_\text{GHZ}^{(5)}$ & \cite{BlattPRL}& $94.4 \pm 0.5$ &  $0.69 \pm 0.02$ &  $0.444 \pm 0.005$ &  $0.27 \pm 0.01$ &  {$0.29 \pm 0.01$} \\
$\varrho_\text{GHZ}^{(6)}$ & \cite{BlattPRL}& $89.2 \pm 0.4$ &  $0.51 \pm 0.01$ &  $0.392 \pm 0.004$ &  $0.190 \pm 0.005$ &  {$0.200 \pm 0.006$} \\
$\varrho_\text{GHZ}^{(8)}$ & \cite{BlattPRL}& $81.7 \pm 0.4$ &  $0.313 \pm 0.009$ &  $0.317 \pm 0.004$ &  $0.113 \pm 0.003$ &  {$0.117 \pm 0.003$} \\
$\varrho_\text{GHZ}^{(10)}$ & \cite{BlattPRL}& $62.6 \pm 0.6$ &  $0.046 \pm 0.004$ &  $0.126 \pm 0.006$ &  $0.016 \pm 0.002$ &  {$0.016 \pm 0.002$} \\
$\varrho_\text{GHZ}^{(14)}$ & \cite{BlattPRL}& $50.8 \pm 0.9$ &  $0.0002 \pm 0.0004$ &  $0.008 \pm 0.009$ &  $0.0001 \pm 0.0001$ &  {$0.0001 \pm 0.0001$} \\
 \hline \hline
\end{tabular}

\caption{\label{TabEx} (a) {Accessible lower bounds to global and partial multiparticle entanglement of some experimentally prepared states}, given by Eq.~(\ref{lowopt}) and evaluated in particular for the trace distance of entanglement $E^{D_{\text{Tr}}}$ by using Eq.~(\ref{evN}) for even $N$ and Eq.~(\ref{odN}) for odd $N$. Following the theoretical analysis of Table~\ref{TabTh}, data obtained by direct measurements of the canonical correlation functions were used to construct bounds for a noisy Smolin state of 4 photons \cite{PianiSmolin}, noisy Dicke states of 6 photons \cite{Prevedel2009,Wieczorek2009}, and noisy GHZ states of 4 ions \cite{MonzPhD}. For noisy GHZ states of 3 ions and noisy $\text{W}$ states of 4 ions (two implementations labelled as $A$ and $B$) \cite{MonzPhD}, full datasets were used to extract the optimised correlation functions $\{\tilde{c}_j\}$ required for the bounds. For all the presented experimental states (whose fidelities with the ideal target states are reported for reference), we are able to provide a reliable estimate of the multiparticle entanglement $E^D_M$ for any $M>\left \lceil{N/2}\right \rceil$.
(b) Lower bounds to genuine multiparticle entanglement of experimental noisy GHZ states of up to 14 ions \cite{BlattPRL}, as quantified in terms of all the distance-based entanglement measures $E_2^D$ reported in Table~\ref{tableGD}, obtained by Eq.~(\ref{egN}) with $p_{\max}$ given in each case by the measured fidelity with the pure reference GHZ state. All the reported entanglement estimates are obtained from the same data needed to witness full inseparability, which for general $N$-qubit states can be accessed by $N+1$ local measurements without the need for a full tomography.
}
\end{table*}

In this section, we benchmark the applicability of our results to {\it real data} from recent experiments  \cite{Blatt2005,BlattPRL,MonzPhD,PianiSmolin,Prevedel2009,Wieczorek2009}.

In Refs.~\cite{PianiSmolin,BoundNature}, the authors used quantum optical setups to prepare an instance of a bound entangled four-qubit state, known as Smolin state \cite{Smolin2001}. Such a state cannot be written as a convex mixture of product states of the four qubits, yet no  entanglement can be distilled out of it, thus incarnating the irreversibility in entanglement manipulation while still representing a useful resource for
information locking and quantum secret sharing \cite{Horodecki2009,GeneSmolin}.  It turns out that noisy Smolin states are particular types of \M3N states (for any even $N$) \cite{BeatrizSmolin,GeneSmolin}, that in the representation of Figure~\ref{figTB}(b) are located along the segment connecting the tetrahedron vertex $\{(-1)^{N/2},(-1)^{N/2},(-1)^{N/2}\}$ with the origin. Therefore, this work provides exact analytical formulae for all the nontrivial hierarchy of their global and partial entanglement, as mentioned in Table~\ref{TabTh}. In the specific experimental implementation of Ref.~\cite{PianiSmolin} for $N=4$, the global entanglement was detected (but not quantified) via a witness constructed  by measuring precisely the three correlation functions $\{c_j\}$.   Based on the existing data alone (and without assuming that the produced state is within the \M3N family), we can then provide a quantitative estimate to the multiparticle entanglement of this experimental bound entangled state in terms of any geometric measure $E^D_{M}$, by using Table~\ref{tableD}. The results are reported in Table~\ref{TabEx}(a) for the illustrative case of the trace distance.

Remaining within the domain of quantum optics, recently two laboratories reported the creation of six-photon Dicke states $|\text{D}^{(6)}_3\rangle$ \cite{Prevedel2009,Wieczorek2009}. Dicke states \cite{Dicke1954} are valuable resources for quantum metrology, computation, and networked communication, and emerge naturally in many-body systems as ground states of the isotropic Lipkin-Meshkov-Glick  model \cite{AcinScience}. Based on the values of the three correlation functions $\{c_j\}$, which were measured in Refs.~\cite{Prevedel2009,Wieczorek2009} to construct some entanglement witnesses, we can provide quantitative bounds to their global and partial geometric entanglement $E^D_M$ (for $4 \leq M \leq 6$) from Eq.~(\ref{low}); see Table~\ref{TabEx}(a).

A  series of  experiments at Innsbruck \cite{Blatt2005,MonzPhD,BlattPRL,Blatt2010} resulted in the generation of a variety of relevant multi-qubit states with trapped ion setups, for explorations of fundamental science and for the implementation of quantum protocols. In those realisations, data acquisition and processing for the purpose of entanglement verification was often a more demanding task than running the experiment itself \cite{Blatt2005}. Focusing first on global and partial entanglement, we obtained full datasets for experimental density matrices corresponding to particularly noisy GHZ and $\text{W}$ states of up to four qubits, produced during laboratory test runs \cite{MonzPhD}. Despite the relatively low fidelity with their ideal target states, we still obtain meaningful quantitative bounds from  Eq.~(\ref{lowopt}). The results are compactly presented in Table~\ref{TabEx}(a).

Regarding now genuine multiparticle entanglement, the authors of Ref.~\cite{BlattPRL} reported the creation of (noisy) GHZ states of up to $N=14$ trapped ions. In each of these states, full inseparability was witnessed by measuring precisely the maximum overlap $p_{\max}$ with a reference pure GHZ state, without the need for complete state tomography. Thanks to Eq.~(\ref{egN}), we can now use the same data to obtain a full quantification of the genuine $N$-particle  entanglement of these realistic states, according to any measure $E^D_2$, at no extra cost in terms of experimental or computational resources. The results are in Table~\ref{TabEx}(b), for all the representative choices of distances enumerated in Table~\ref{tableGD}. Notice that we do not need to assume that the experimentally produced states are in the ${\cal G}_N$ set: the obtained results can be still safely regarded as lower bounds.

\bigskip

\section{\sf \bfseries Discussion}

We have introduced a general framework for estimating and quantifying geometric entanglement monotones. This enabled us to achieve a compendium of exact results on the quantification of general distance-based  measures of (global, partial, and genuine) multiparticle entanglement in some pivotal reference families of $N$-qubit mixed states. In turn, these results allowed us to establish faithful lower bounds to various forms of  multiparticle entanglement for  arbitrary states, accessible by few local measurements and effective on prominent resource states for quantum information processing.

Our results can be regarded as realising simple yet particularly convenient instances of quantitative entanglement witnesses \cite{QWitness1,QWitness2}, with the crucial advance that our lower bounds are analytical (in contrast to conventional numerical approaches requiring semidefinite programming) and hold for {\it all} valid geometric measures of entanglement, which are endowed with meaningful operational interpretations yet have been traditionally hard to evaluate \cite{Chen2013,Guhne2015}.

A key aspect of our analysis lies in fact in the generality of the adopted techniques, which rely on natural information-theoretic requirements of contractivity and joint convexity of any valid distance $D$ entering Eq.~(\ref{geome}). We can expect our general framework to be applicable to other reference families of states (for example, states diagonal in a basis of cluster states  \cite{Chen2013,Hofmann2014}, or more general states with X-shaped density matrices \cite{Hashemi2012}), thereby leading to alternative entanglement bounds for arbitrary states, which might be more tailored to different classes, or to specific measurement settings in laboratory.

Furthermore, our framework lends itself to numerous other applications. These include the obtention of accessible analytical results for the geometric quantification of other useful forms of multiparticle quantum correlations, such as Einstein-Podolsky-Rosen steering \cite{WisemanPRL,SteeringNPhys}, and Bell nonlocality in many-body systems \cite{AcinScience}. This can eventually lead to a unifying characterisation, resting on the structure of information geometry, of the whole spectrum of genuine signatures of quantumness in cooperative phenomena. We plan to extend our approach in this sense in subsequent works.


Another key feature of our results is the experimental accessibility. Having tested our entanglement bounds on a selection of very different families of theoretical and experimentally produced states with high levels of noise, we can certify their usefulness in realistic scenarios. We recall that, for instance, three canonical local measurements suffice to quantify exactly the global entanglement of GHZ states of any even number $N$ of qubits, while  $N+1$ local measurements provide their exact genuine entanglement, according to every geometric measure for any $N$, when such states are realistically mixed with white noise.
Compared to other complementary studies of accessible quantification of multiparticle entanglement \cite{Guhne2015,Hofmann2014,SiewertPRL,SiewertSciRep,Wu2012,Hashemi2012,Audenart2006,Wunderlich2009,Ma2011,QWitness2,QWitness1},  our study retains not only a comparably low resource demand but also crucial aspects such as efficiency and versatility, as shown in Table \ref{TabComp}.
This can lead to a considerable simplification of quantitative resource assessment in future experiments based on large-scale entangled registers, involving e.g.~two quantum bytes (16 qubits) and beyond \cite{BlattPRL,MonzPhD}.


\section{\sf \bfseries Methods}

\noindent {\bfseries Distance-based measures of multiparticle entanglement}.
A general distance-based measure of multiparticle entanglement $E^D_{M}$ is defined in Eq.~(\ref{geome}). In this work, the distance $D$ is required to satisfy the following two physical constraints:\\
\begin{tabular}{lp{0.88\columnwidth}}
(D.i) & Contractivity under quantum channels, i.e. $D(\Omega(\varrho),\Omega(\varrho')) \leq D(\varrho,\varrho')$, for any states $\varrho$, $\varrho'$, and any completely positive trace preserving map $\Omega$;\\
(D.ii) & Joint convexity, i.e.
$
D(q \varrho + (1-q) \varrho',q \chi + (1-q)\chi') \leq q D(\varrho,\chi)+ (1-q) D(\varrho',\chi')
$, for any states $\varrho$, $\varrho'$, $\chi$, and $\chi'$, and any $q \in [0,1]$.
\end{tabular}
Constraint (D.i) implies that $E^D_{M}$ is invariant under local unitaries and monotonically nonincreasing under LOCC (i.e., it is an entanglement monotone \cite{PlenioVirmani2007}). Constraint (D.ii) implies that $E^D_{M}$ is also convex. A selection of distance functionals respecting these properties is given in Table~\ref{tableD}.

\medskip

\noindent {\bfseries ${\boldsymbol{\mathcal{M}^3_N}}$-fication}.
\noindent \textit{Theorem}. Any $N$-qubit state $\varrho$ can be transformed into a corresponding $\mathcal{M}^{3}_{N}$ state $\varpi$ through a fixed transformation, ${\Theta}$, consisting of single-qubit LOCC, such that
$
{\Theta}(\varrho) = \varpi = \frac{1}{2^{N}}\left( \mathbb{I}^{\otimes N} + \sum_{i=1}^{3} c_{i} \sigma_{i}^{\otimes N}\right),$
where $c_{i} = \text{Tr}(\varrho \sigma_{i}^{\otimes N})$.

\noindent \textit{Proof}.
Here we sketch the form of the \M3N-fication channel ${\Theta}$. We begin by setting $2(N-1)$ single-qubit local unitaries $\{U_{j}\}_{j=1}^{2(N-1)}=\big\{(\sigma_{1} \otimes \sigma_{1} \otimes \mathbb{I}^{\otimes N-2}),\,
(\mathbb{I} \otimes \sigma_{1} \otimes \sigma_{1} \otimes \mathbb{I}^{\otimes N-3}), \,
\ldots,\, (\mathbb{I}^{\otimes N-3} \otimes \sigma_{1} \otimes \sigma_{1} \otimes \mathbb{I}),\,
(\mathbb{I}^{\otimes N-2} \otimes \sigma_{1} \otimes \sigma_{1}),\,
(\sigma_{2} \otimes \sigma_{2} \otimes \mathbb{I}^{\otimes N-2}),\,
(\mathbb{I} \otimes \sigma_{2} \otimes \sigma_{2} \otimes \mathbb{I}^{\otimes N-3}),\,
\ldots,\,
(\mathbb{I}^{\otimes N-3} \otimes \sigma_{2} \otimes \sigma_{2} \otimes \mathbb{I}),\,
(\mathbb{I}^{\otimes N-2} \otimes \sigma_{2} \otimes \sigma_{2})\big\}.$
Then, we fix a sequence of states $\{\varrho_{0},\varrho_{1}, \ldots \varrho_{2(N-1)}\}$ defined by
$\varrho_{j} = \frac{1}{2}\left( \varrho_{j-1}+U_{j}\varrho_{j-1}U_{j}^{\dagger} \right),
$
for $j \in \{1,2, \ldots 2(N-1)\}$. By setting $\varrho_{0} = \varrho$ and $\varrho_{2(N-1)}={\Theta}(\varrho)$, we define the required channel: ${\Theta}(\varrho)= \frac{1}{2^{2(N-1)}}\sum_{i=1}^{2^{2(N-1)}} U_{i}' \varrho U_{i}'^{\dagger}$, where: $\{U_{i}'\}_{i=1}^{2^{2(N-1)}}=\Big\{
  \mathbb{I}^{\otimes N},\,
  \{U_{i_{1}}\}_{i_{1}=1}^{2(N-1)},\,
  \{U_{i_{2}}U_{i_{1}}\}_{i_{2}>i_{1}=1}^{2(N-1)},\,
  \ldots,\,$ $\{U_{i_{2(N-1)}} \ldots U_{i_{2}}U_{i_{1}}\}_{i_{2(N-1)}>\ldots>i_{2}>i_{1}=1}^{2(N-1)}\Big\}.
$
Notice that  $\{U_{i}'\}_{i=1}^{2^{2(N-1)}}$ is still a sequence of single-qubit local unitaries. Since  ${\Theta}$ is a convex mixture of such local unitaries, it belongs to the class of single-qubit LOCC, mapping any $M$-separable set into itself. In the Supplementary Material, we show that ${\Theta}(\varrho) = \varpi$, concluding the proof.

\medskip

\noindent{\bfseries Lower bound optimisation}.
For any valid distance-based measure of global and partial multiparticle entanglement $E^D_{M}$, the maximisation in Eq.~(\ref{lowopt}) is equivalent (for even $N$) to maximising $|\tilde{c}_1|+|\tilde{c}_2|+|\tilde{c}_3|$, where $\tilde{c}_j = \text{Tr}[U_\otimes \varrho U_\otimes ^\dagger \sigma_j^{\otimes N}]$, over local single-qubit unitaries $U_\otimes=\bigotimes_\alpha U^{(\alpha)}$ ($\alpha = 1,\ldots,N$). By using the well known correspondence between the special unitary group ${\sf SU}(2)$ and special orthogonal group ${\sf SO}(3)$, we have that to any one-qubit unitary $U^{(\alpha)}$ corresponds the orthogonal $3\times 3$ matrix $O^{(\alpha)}$ such that
$U^{(\alpha)} \vec{n}\cdot\vec{\sigma}U^{(\alpha)\dagger} = (O^{(\alpha)} \vec{n})\cdot\vec{\sigma}$,
where $\vec{n}=\{n_1,n_3,n_3\}\in\mathbb{R}^3$ and $\vec{\sigma} = \{\sigma_1,\sigma_2,\sigma_3\}$ is the vector of Pauli matrices. We have then that $\sup_{\{U^{(\alpha)}\}} (|\tilde{c}_1|+|\tilde{c}_2|+|\tilde{c}_3|)=\sup_{\{O^{(\alpha)}\}} (|\tilde{T}_{11\cdots 1}|+|\tilde{T}_{22\cdots 2}|+|\tilde{T}_{33\cdots 3}|)$,
where
$\tilde{T}_{i_1i_2\cdots i_N} =  \sum_{j_1j_2\cdots j_N} T_{j_1j_2\cdots j_N}O^{(1)}_{i_1j_1}O^{(2)}_{i_2j_2}\cdots O^{(N)}_{i_Nj_N}$, and ${T}_{i_1i_2\cdots i_N}= \text{Tr}\left[\varrho \left(\sigma_{i_1}\otimes\sigma_{i_2}\otimes\cdots\otimes\sigma_{i_N}\right) \right]$.
In the case of permutationally invariant states $\varrho$, the $3\times 3\times\cdots\times 3$ tensor ${T}_{i_1i_2\cdots i_N}$ is fully symmetric, i.e. ${T}_{i_1i_2\cdots i_N}={T}_{\vartheta(i_1i_2\cdots i_N)}$ for any permutation $\vartheta$ of the indices, so that the optimisation can be achieved when $O^{(1)}=O^{(2)}=\cdots= O^{(N)}$ \cite{Comon2007}. As indicated in the main text, we then need to perform the maximisation over just the three angles $\{\theta, \psi, \phi\}$ which determine the orthogonal matrix $O^{(\alpha)}$ corresponding to an arbitrary single-qubit unitary $$U^{(\alpha)} =  \left( \begin{array}{cc}
\cos\frac{\theta}{2} e^{-i\frac{\psi+\phi}{2}} & -i \sin\frac{\theta}{2} e^{-i\frac{\phi-\psi}{2}}  \\
-i \sin\frac{\theta}{2} e^{i\frac{\phi-\psi}{2}} & \cos\frac{\theta}{2} e^{i\frac{\psi+\phi}{2}}   \\
 \end{array} \right).$$
As a special case, for a two-qubit state ($N=2$) the optimal local operation is the one which diagonalises the correlation matrix $(T_{i_1 i_2})$.


\section*{\sf \bfseries Acknowledgments}
We warmly thank Thomas Monz, Mauro Paternostro, Christian Schwemmer, and Witlef Wieczorek for providing experimental data, and we acknowledge fruitful discussions with (in alphabetical order) I.~Almeida~Silva, M.~Blasone, D.~Cavalcanti, E.~Carnio, T.~Chanda, M.~Christandl, P.~Comon, M.~Cramer, M.~Gessner, T.~Ginestra, D.~Gross, O.~G\"uhne, M.~Gu{\c t}{\u a}, M.~Huber, F.~Illuminati, I.~Kogias, T.~Kypraios, P.~Liuzzo-Scorpo, C.~Macchiavello, A.~Milne, T.~Monz, P.~Ott, A.~K.~Pal, M.~Piani, M.~Prater, S.~Rat, A.~Sanpera, P.~Skrzypczyk, A.~Streltsov, G.~T{\' o}th, A.~Winter. This work was supported by the European Research Council (ERC) Starting Grant GQCOP, Grant Agreement No.~637352.

\vspace*{.3cm}

\section*{\sf \bfseries Author contributions}
M. C. and T. R. B. contributed equally to this work.
All the authors conceived the idea, derived the technical results, discussed all stages of the project, and prepared the manuscript and figures.

\section*{\newline \sf \bfseries Competing financial interests}
The authors declare that they have no competing financial interests.

\section*{\sf \bfseries Corresponding author}
Correspondence to: \\ Gerardo Adesso  (gerardo.adesso@nottingham.ac.uk)




\begin{thebibliography}{0}%
\makeatletter
\providecommand \@ifxundefined [1]{%
 \@ifx{#1\undefined}
}%
\providecommand \@ifnum [1]{%
 \ifnum #1\expandafter \@firstoftwo
 \else \expandafter \@secondoftwo
 \fi
}%
\providecommand \@ifx [1]{%
 \ifx #1\expandafter \@firstoftwo
 \else \expandafter \@secondoftwo
 \fi
}%
\providecommand \natexlab [1]{#1}%
\providecommand \enquote  [1]{``#1''}%
\providecommand \bibnamefont  [1]{#1}%
\providecommand \bibfnamefont [1]{#1}%
\providecommand \citenamefont [1]{#1}%
\providecommand \href@noop [0]{\@secondoftwo}%
\providecommand \href [0]{\begingroup \@sanitize@url \@href}%
\providecommand \@href[1]{\@@startlink{#1}\@@href}%
\providecommand \@@href[1]{\endgroup#1\@@endlink}%
\providecommand \@sanitize@url [0]{\catcode `\\12\catcode `\$12\catcode
  `\&12\catcode `\#12\catcode `\^12\catcode `\_12\catcode `\%12\relax}%
\providecommand \@@startlink[1]{}%
\providecommand \@@endlink[0]{}%
\providecommand \url  [0]{\begingroup\@sanitize@url \@url }%
\providecommand \@url [1]{\endgroup\@href {#1}{\urlprefix }}%
\providecommand \urlprefix  [0]{URL }%
\providecommand \Eprint [0]{\href }%
\providecommand \doibase [0]{http://dx.doi.org/}%
\providecommand \selectlanguage [0]{\@gobble}%
\providecommand \bibinfo  [0]{\@secondoftwo}%
\providecommand \bibfield  [0]{\@secondoftwo}%
\providecommand \translation [1]{[#1]}%
\providecommand \BibitemOpen [0]{}%
\providecommand \bibitemStop [0]{}%
\providecommand \bibitemNoStop [0]{.\EOS\space}%
\providecommand \EOS [0]{\spacefactor3000\relax}%
\providecommand \BibitemShut  [1]{\csname bibitem#1\endcsname}%
\let\auto@bib@innerbib\@empty
\end{thebibliography}%


\begin{thebibliography}{10}
\expandafter\ifx\csname url\endcsname\relax
  \def\url#1{\texttt{#1}}\fi
\expandafter\ifx\csname urlprefix\endcsname\relax\def\urlprefix{URL }\fi
\providecommand{\bibinfo}[2]{#2}
\providecommand{\eprint}[2][]{\url{#2}}

\bibitem{EPR1935}
\bibinfo{author}{Einstein, A.}, \bibinfo{author}{Podolsky, B.} \&
  \bibinfo{author}{Rosen, N.}
\newblock \bibinfo{title}{Can quantum-mechanical description of physical
  reality be considered complete?}
\newblock \emph{\bibinfo{journal}{Phys. Rev.}} \textbf{\bibinfo{volume}{47}},
  \bibinfo{pages}{777--780} (\bibinfo{year}{1935}).

\bibitem{Vedral2014}
\bibinfo{author}{Vedral, V.}
\newblock \bibinfo{title}{Quantum entanglement}.
\newblock \emph{\bibinfo{journal}{Nat. Phys.}} \textbf{\bibinfo{volume}{10}},
  \bibinfo{pages}{256--258} (\bibinfo{year}{2014}).

\bibitem{Horodecki2009}
\bibinfo{author}{Horodecki, R.}, \bibinfo{author}{Horodecki, P.},
  \bibinfo{author}{Horodecki, M.} \& \bibinfo{author}{Horodecki, K.}
\newblock \bibinfo{title}{Quantum entanglement}.
\newblock \emph{\bibinfo{journal}{Rev. Mod. Phys.}}
  \textbf{\bibinfo{volume}{81}}, \bibinfo{pages}{865--942}
  (\bibinfo{year}{2009}).

\bibitem{GuhneToth2009}
\bibinfo{author}{G{\"{u}}hne, O.} \& \bibinfo{author}{T{\'{o}}th, G.}
\newblock \bibinfo{title}{Entanglement detection}.
\newblock \emph{\bibinfo{journal}{Phys. Rep.}} \textbf{\bibinfo{volume}{474}},
  \bibinfo{pages}{1--75} (\bibinfo{year}{2009}).

\bibitem{Guhne2010}
\bibinfo{author}{G\"uhne, O.} \& \bibinfo{author}{Seevinck, M.}
\newblock \bibinfo{title}{Separability criteria for genuine multiparticle
  entanglement}.
\newblock \emph{\bibinfo{journal}{New J. Phys.}} \textbf{\bibinfo{volume}{12}},
  \bibinfo{pages}{053002} (\bibinfo{year}{2010}).

\bibitem{Gao2014}
\bibinfo{author}{Gao, T.}, \bibinfo{author}{Yan, F.} \& \bibinfo{author}{van
  Enk, S.~J.}
\newblock \bibinfo{title}{Permutationally invariant part of a density matrix
  and nonseparability of $n$-qubit states}.
\newblock \emph{\bibinfo{journal}{Phys. Rev. Lett.}}
  \textbf{\bibinfo{volume}{112}}, \bibinfo{pages}{180501}
  (\bibinfo{year}{2014}).

\bibitem{Levi2013}
\bibinfo{author}{Levi, F.} \& \bibinfo{author}{Mintert, F.}
\newblock \bibinfo{title}{Hierarchies of multipartite entanglement}.
\newblock \emph{\bibinfo{journal}{Phys. Rev. Lett.}}
  \textbf{\bibinfo{volume}{110}}, \bibinfo{pages}{150402}
  (\bibinfo{year}{2013}).

\bibitem{Huber2010}
\bibinfo{author}{Huber, M.}, \bibinfo{author}{Mintert, F.},
  \bibinfo{author}{Gabriel, A.} \& \bibinfo{author}{Hiesmayr, B.~C.}
\newblock \bibinfo{title}{Detection of high-dimensional genuine multipartite
  entanglement of mixed states}.
\newblock \emph{\bibinfo{journal}{Phys. Rev. Lett.}}
  \textbf{\bibinfo{volume}{104}}, \bibinfo{pages}{210501}
  (\bibinfo{year}{2010}).

\bibitem{DurCirac2000}
\bibinfo{author}{D\"ur, W.} \& \bibinfo{author}{Cirac, J.~I.}
\newblock \bibinfo{title}{Classification of multiqubit mixed states:
  Separability and distillability properties}.
\newblock \emph{\bibinfo{journal}{Phys. Rev. A}} \textbf{\bibinfo{volume}{61}},
  \bibinfo{pages}{042314} (\bibinfo{year}{2000}).

\bibitem{Gabriel2010}
\bibinfo{author}{Gabriel, A.}, \bibinfo{author}{Hiesmayr, B.~C.} \&
  \bibinfo{author}{Huber, M.}
\newblock \bibinfo{title}{Criterion for k-separability in mixed multipartite
  systems}.
\newblock \emph{\bibinfo{journal}{Quantum Information and Computation}}
  \textbf{\bibinfo{volume}{10}}, \bibinfo{pages}{829--836}
  (\bibinfo{year}{2010}).

\bibitem{Klockl2015}
\bibinfo{author}{Kl\"ockl, C.} \& \bibinfo{author}{Huber, M.}
\newblock \bibinfo{title}{Characterizing multipartite entanglement without
  shared reference frames}.
\newblock \emph{\bibinfo{journal}{Phys. Rev. A}} \textbf{\bibinfo{volume}{91}},
  \bibinfo{pages}{042339} (\bibinfo{year}{2015}).

\bibitem{ExpFriendly}
\bibinfo{author}{Badzi{\c{a}}g, P.}, \bibinfo{author}{Brukner, {\v{C}}.},
  \bibinfo{author}{Laskowski, W.}, \bibinfo{author}{Paterek, T.} \&
  \bibinfo{author}{{\.Z}ukowski, M.}
\newblock \bibinfo{title}{Experimentally friendly geometrical criteria for
  entanglement}.
\newblock \emph{\bibinfo{journal}{Phys. Rev. Lett.}}
  \textbf{\bibinfo{volume}{100}}, \bibinfo{pages}{140403}
  (\bibinfo{year}{2008}).

\bibitem{Guhne2015}
\bibinfo{author}{Buchholz, L.~E.}, \bibinfo{author}{Moroder, T.} \&
  \bibinfo{author}{Gühne, O.}
\newblock \bibinfo{title}{Evaluating the geometric measure of multiparticle
  entanglement}.
\newblock \emph{\bibinfo{journal}{Ann. Phys. (Berlin)}} \bibinfo{pages}{1--10
  doi:10.1002/andp.201500293} (\bibinfo{year}{2015}).

\bibitem{Hofmann2014}
\bibinfo{author}{Hofmann, M.}, \bibinfo{author}{Moroder, T.} \&
  \bibinfo{author}{G\"uhne, O.}
\newblock \bibinfo{title}{Analytical characterization of the genuine
  multiparticle negativity}.
\newblock \emph{\bibinfo{journal}{J. Phys. A: Math. Theor.}}
  \textbf{\bibinfo{volume}{47}}, \bibinfo{pages}{155301}
  (\bibinfo{year}{2014}).

\bibitem{SiewertPRL}
\bibinfo{author}{Siewert, J.} \& \bibinfo{author}{Eltschka, C.}
\newblock \bibinfo{title}{Quantifying tripartite entanglement of three-qubit
  generalized werner states}.
\newblock \emph{\bibinfo{journal}{Physical review letters}}
  \textbf{\bibinfo{volume}{108}}, \bibinfo{pages}{230502}
  (\bibinfo{year}{2012}).

\bibitem{SiewertSciRep}
\bibinfo{author}{Eltschka, C.} \& \bibinfo{author}{Siewert, J.}
\newblock \bibinfo{title}{A quantitative witness for
  greenberger-horne-zeilinger entanglement}.
\newblock \emph{\bibinfo{journal}{Sci. Rep.}} \textbf{\bibinfo{volume}{2}}
  (\bibinfo{year}{2012}).

\bibitem{Wu2012}
\bibinfo{author}{Wu, J.-Y.}, \bibinfo{author}{Kampermann, H.},
  \bibinfo{author}{Bru\ss{}, D.}, \bibinfo{author}{Kl\"ockl, C.} \&
  \bibinfo{author}{Huber, M.}
\newblock \bibinfo{title}{Determining lower bounds on a measure of multipartite
  entanglement from few local observables}.
\newblock \emph{\bibinfo{journal}{Phys. Rev. A}} \textbf{\bibinfo{volume}{86}},
  \bibinfo{pages}{022319} (\bibinfo{year}{2012}).

\bibitem{Hashemi2012}
\bibinfo{author}{Hashemi~Rafsanjani, S.~M.}, \bibinfo{author}{Huber, M.},
  \bibinfo{author}{Broadbent, C.~J.} \& \bibinfo{author}{Eberly, J.~H.}
\newblock \bibinfo{title}{Genuinely multipartite concurrence of $n$-qubit $x$
  matrices}.
\newblock \emph{\bibinfo{journal}{Phys. Rev. A}} \textbf{\bibinfo{volume}{86}},
  \bibinfo{pages}{062303} (\bibinfo{year}{2012}).

\bibitem{Audenart2006}
\bibinfo{author}{Audenaert, K.} \& \bibinfo{author}{Plenio, M.}
\newblock \bibinfo{title}{When are correlations quantum?--verification and
  quantification of entanglement by simple measurements}.
\newblock \emph{\bibinfo{journal}{New Journal of Physics}}
  \textbf{\bibinfo{volume}{8}}, \bibinfo{pages}{266} (\bibinfo{year}{2006}).

\bibitem{Wunderlich2009}
\bibinfo{author}{Wunderlich, H.} \& \bibinfo{author}{Plenio, M.~B.}
\newblock \bibinfo{title}{Quantitative verification of entanglement and
  fidelities from incomplete measurement data}.
\newblock \emph{\bibinfo{journal}{Journal of Modern Optics}}
  \textbf{\bibinfo{volume}{56}}, \bibinfo{pages}{2100--2105}
  (\bibinfo{year}{2009}).

\bibitem{Ma2011}
\bibinfo{author}{Ma, Z.-H.} \emph{et~al.}
\newblock \bibinfo{title}{Measure of genuine multipartite entanglement with
  computable lower bounds}.
\newblock \emph{\bibinfo{journal}{Phys. Rev. A}} \textbf{\bibinfo{volume}{83}},
  \bibinfo{pages}{062325} (\bibinfo{year}{2011}).

\bibitem{QWitness2}
\bibinfo{author}{G\"uhne, O.}, \bibinfo{author}{Reimpell, M.} \&
  \bibinfo{author}{Werner, R.~F.}
\newblock \bibinfo{title}{Estimating entanglement measures in experiments}.
\newblock \emph{\bibinfo{journal}{Phys. Rev. Lett.}}
  \textbf{\bibinfo{volume}{98}}, \bibinfo{pages}{110502}
  (\bibinfo{year}{2007}).

\bibitem{QWitness1}
\bibinfo{author}{Eisert, J.}, \bibinfo{author}{Brand{\~a}o, F.~G.} \&
  \bibinfo{author}{Audenaert, K.~M.}
\newblock \bibinfo{title}{Quantitative entanglement witnesses}.
\newblock \emph{\bibinfo{journal}{New J. Phys.}} \textbf{\bibinfo{volume}{9}},
  \bibinfo{pages}{46} (\bibinfo{year}{2007}).

\bibitem{PlenioVirmani2007}
\bibinfo{author}{Plenio, M.~B.} \& \bibinfo{author}{Virmani, S.}
\newblock \bibinfo{title}{An introduction to entanglement measures}.
\newblock \emph{\bibinfo{journal}{Quant. Inf. Comput.}}
  \textbf{\bibinfo{volume}{7}}, \bibinfo{pages}{1} (\bibinfo{year}{2007}).

\bibitem{Whaley2010}
\bibinfo{author}{Sarovar, M.}, \bibinfo{author}{Ishizaki, A.},
  \bibinfo{author}{Fleming, G.~R.} \& \bibinfo{author}{Whaley, K.~B.}
\newblock \bibinfo{title}{Quantum entanglement in photosynthetic
  light-harvesting complexes}.
\newblock \emph{\bibinfo{journal}{Nat. Phys.}} \textbf{\bibinfo{volume}{6}},
  \bibinfo{pages}{462--467} (\bibinfo{year}{2010}).

\bibitem{Cramer2011}
\bibinfo{author}{Cramer, M.}, \bibinfo{author}{Plenio, M.~B.} \&
  \bibinfo{author}{Wunderlich, H.}
\newblock \bibinfo{title}{Measuring entanglement in condensed matter systems}.
\newblock \emph{\bibinfo{journal}{Phys. Rev. Lett.}}
  \textbf{\bibinfo{volume}{106}}, \bibinfo{pages}{020401}
  (\bibinfo{year}{2011}).

\bibitem{Cramer2013}
\bibinfo{author}{Cramer, M.} \emph{et~al.}
\newblock \bibinfo{title}{Spatial entanglement of bosons in optical lattices}.
\newblock \emph{\bibinfo{journal}{Nat. Commun.}} \textbf{\bibinfo{volume}{4}},
  \bibinfo{pages}{2161} (\bibinfo{year}{2013}).

\bibitem{Marty2014}
\bibinfo{author}{Marty, O.} \emph{et~al.}
\newblock \bibinfo{title}{Quantifying entanglement with scattering
  experiments}.
\newblock \emph{\bibinfo{journal}{Phys. Rev. B}} \textbf{\bibinfo{volume}{89}},
  \bibinfo{pages}{125117} (\bibinfo{year}{2014}).

\bibitem{Marty2015}
\bibinfo{author}{Marty, O.}, \bibinfo{author}{Cramer, M.} \&
  \bibinfo{author}{Plenio, M.~B.}
\newblock \bibinfo{title}{Practical entanglement estimation for spin-system
  quantum simulators}.
\newblock \emph{\bibinfo{journal}{Phys. Rev. Lett.}}
  \textbf{\bibinfo{volume}{116}}, \bibinfo{pages}{105301}
  (\bibinfo{year}{2016}).

\bibitem{Plenio1997}
\bibinfo{author}{Vedral, V.}, \bibinfo{author}{Plenio, M.~B.},
  \bibinfo{author}{Rippin, M.~A.} \& \bibinfo{author}{Knight, P.~L.}
\newblock \bibinfo{title}{Quantifying entanglement}.
\newblock \emph{\bibinfo{journal}{Phys. Rev. Lett.}}
  \textbf{\bibinfo{volume}{78}}, \bibinfo{pages}{2275--2279}
  (\bibinfo{year}{1997}).

\bibitem{Wei2003}
\bibinfo{author}{Wei, T.-C.} \& \bibinfo{author}{Goldbart, P.~M.}
\newblock \bibinfo{title}{Geometric measure of entanglement and applications to
  bipartite and multipartite quantum states}.
\newblock \emph{\bibinfo{journal}{Phys. Rev. A}} \textbf{\bibinfo{volume}{68}},
  \bibinfo{pages}{042307} (\bibinfo{year}{2003}).

\bibitem{Blasone2008}
\bibinfo{author}{Blasone, M.}, \bibinfo{author}{Dell'Anno, F.},
  \bibinfo{author}{De~Siena, S.} \& \bibinfo{author}{Illuminati, F.}
\newblock \bibinfo{title}{Hierarchies of geometric entanglement}.
\newblock \emph{\bibinfo{journal}{Phys. Rev. A}} \textbf{\bibinfo{volume}{77}},
  \bibinfo{pages}{062304} (\bibinfo{year}{2008}).

\bibitem{Bengtsson2006}
\bibinfo{author}{Bengtsson, I.} \& \bibinfo{author}{Zyczkowski, K.}
\newblock \emph{\bibinfo{title}{Geometry of Quantum States: An Introduction to
  Quantum Entanglement}} (\bibinfo{publisher}{Cambridge University Press},
  \bibinfo{year}{2006}).

\bibitem{Wei2005}
\bibinfo{author}{Wei, T.-C.}, \bibinfo{author}{Das, D.},
  \bibinfo{author}{Mukhopadyay, S.}, \bibinfo{author}{Vishveshwara, S.} \&
  \bibinfo{author}{Goldbart, P.~M.}
\newblock \bibinfo{title}{Global entanglement and quantum criticality in spin
  chains}.
\newblock \emph{\bibinfo{journal}{Phys. Rev. A}} \textbf{\bibinfo{volume}{71}},
  \bibinfo{pages}{060305} (\bibinfo{year}{2005}).

\bibitem{Groverian}
\bibinfo{author}{Biham, O.}, \bibinfo{author}{Nielsen, M.~A.} \&
  \bibinfo{author}{Osborne, T.~J.}
\newblock \bibinfo{title}{Entanglement monotone derived from grover's
  algorithm}.
\newblock \emph{\bibinfo{journal}{Phys. Rev. A}} \textbf{\bibinfo{volume}{65}},
  \bibinfo{pages}{062312} (\bibinfo{year}{2002}).

\bibitem{Markham2014}
\bibinfo{author}{Bell, B.~A.} \emph{et~al.}
\newblock \bibinfo{title}{Experimental demonstration of graph-state quantum
  secret sharing}.
\newblock \emph{\bibinfo{journal}{Nat. Commun.}} \textbf{\bibinfo{volume}{5}},
  \bibinfo{pages}{5480} (\bibinfo{year}{2014}).

\bibitem{Giovannetti2011}
\bibinfo{author}{Giovannetti, V.}, \bibinfo{author}{Lloyd, S.} \&
  \bibinfo{author}{Maccone, L.}
\newblock \bibinfo{title}{Advances in quantum metrology}.
\newblock \emph{\bibinfo{journal}{Nat. Photon.}} \textbf{\bibinfo{volume}{5}},
  \bibinfo{pages}{222} (\bibinfo{year}{2011}).

\bibitem{BriegelNP}
\bibinfo{author}{Briegel, H.~J.}, \bibinfo{author}{Browne, D.~E.},
  \bibinfo{author}{D\"ur, W.}, \bibinfo{author}{Raussendorf, R.} \&
  \bibinfo{author}{den Nest, M.~V.}
\newblock \emph{\bibinfo{journal}{Nat. Phys.}} \textbf{\bibinfo{volume}{5}},
  \bibinfo{pages}{19--26} (\bibinfo{year}{2009}).

\bibitem{Tiersch2012}
\bibinfo{author}{Tiersch, M.}, \bibinfo{author}{Popescu, S.} \&
  \bibinfo{author}{Briegel, H.~J.}
\newblock \bibinfo{title}{A critical view on transport and entanglement in
  models of photosynthesis}.
\newblock \emph{\bibinfo{journal}{Phil. Trans. Roy. Soc. A}}
  \textbf{\bibinfo{volume}{370}}, \bibinfo{pages}{3771--3786}
  (\bibinfo{year}{2012}).

\bibitem{Blatt2010}
\bibinfo{author}{Barreiro, J.~T.} \emph{et~al.}
\newblock \bibinfo{title}{Experimental multiparticle entanglement dynamics
  induced by decoherence}.
\newblock \emph{\bibinfo{journal}{Nat. Phys.}} \textbf{\bibinfo{volume}{6}},
  \bibinfo{pages}{943} (\bibinfo{year}{2010}).

\bibitem{Hayashi2006}
\bibinfo{author}{Hayashi, M.}, \bibinfo{author}{Markham, D.},
  \bibinfo{author}{Murao, M.}, \bibinfo{author}{Owari, M.} \&
  \bibinfo{author}{Virmani, S.}
\newblock \bibinfo{title}{Bounds on multipartite entangled orthogonal state
  discrimination using local operations and classical communication}.
\newblock \emph{\bibinfo{journal}{Phys. Rev. Lett.}}
  \textbf{\bibinfo{volume}{96}}, \bibinfo{pages}{040501}
  (\bibinfo{year}{2006}).

\bibitem{Streltsov2010}
\bibinfo{author}{Streltsov, A.}, \bibinfo{author}{Kampermann, H.} \&
  \bibinfo{author}{Bru\ss{}, D.}
\newblock \bibinfo{title}{Linking a distance measure of entanglement to its
  convex roof}.
\newblock \emph{\bibinfo{journal}{New J. Phys.}} \textbf{\bibinfo{volume}{12}},
  \bibinfo{pages}{123004} (\bibinfo{year}{2010}).

\bibitem{BeatrizSmolin}
\bibinfo{author}{Hiesmayr, B.~C.}, \bibinfo{author}{Hipp, F.},
  \bibinfo{author}{Huber, M.}, \bibinfo{author}{Krammer, P.} \&
  \bibinfo{author}{Spengler, C.}
\newblock \bibinfo{title}{Simplex of bound entangled multipartite qubit
  states}.
\newblock \emph{\bibinfo{journal}{Phys. Rev. A}} \textbf{\bibinfo{volume}{78}},
  \bibinfo{pages}{042327} (\bibinfo{year}{2008}).

\bibitem{PianiSmolin}
\bibinfo{author}{Lavoie, J.}, \bibinfo{author}{Kaltenbaek, R.},
  \bibinfo{author}{Piani, M.} \& \bibinfo{author}{Resch, K.~J.}
\newblock \bibinfo{title}{Experimental bound entanglement in a four-photon
  state}.
\newblock \emph{\bibinfo{journal}{Phys. Rev. Lett.}}
  \textbf{\bibinfo{volume}{105}}, \bibinfo{pages}{130501}
  (\bibinfo{year}{2010}).

\bibitem{BoundNature}
\bibinfo{author}{Amselem, E.} \& \bibinfo{author}{Bourennane, M.}
\newblock \bibinfo{title}{Experimental four-qubit bound entanglement}.
\newblock \emph{\bibinfo{journal}{Nat. Phys.}} \textbf{\bibinfo{volume}{5}},
  \bibinfo{pages}{748} (\bibinfo{year}{2009}).

\bibitem{GeneSmolin}
\bibinfo{author}{Augusiak, R.} \& \bibinfo{author}{Horodecki, P.}
\newblock \bibinfo{title}{Generalized smolin states and their properties}.
\newblock \emph{\bibinfo{journal}{Physical Review A}}
  \textbf{\bibinfo{volume}{73}}, \bibinfo{pages}{012318}
  (\bibinfo{year}{2006}).

\bibitem{GHZ90}
\bibinfo{author}{Greenberger, D.~M.}, \bibinfo{author}{Horne, M.} \&
  \bibinfo{author}{Zeilinger, A.}
\newblock \bibinfo{title}{{Going Beyond Bell's Theorem}}
  (\bibinfo{year}{1989}).
\newblock \bibinfo{note}{{pp.} 69--72 {in} {\it Bell's Theorem, Quantum Theory,
  and Conceptions of the Universe}, Kafatos, M. (Ed.), Kluwer, Dordrecht},
  \eprint{arXiv:0712.0921}.

\bibitem{Guhne2007}
\bibinfo{author}{G\"uhne, O.}, \bibinfo{author}{Lu, C.-Y.},
  \bibinfo{author}{Gao, W.-B.} \& \bibinfo{author}{Pan, J.-W.}
\newblock \bibinfo{title}{Toolbox for entanglement detection and fidelity
  estimation}.
\newblock \emph{\bibinfo{journal}{Phys. Rev. A}} \textbf{\bibinfo{volume}{76}},
  \bibinfo{pages}{030305} (\bibinfo{year}{2007}).

\bibitem{Blatt2005}
\bibinfo{author}{H{\"a}ffner, H.} \emph{et~al.}
\newblock \bibinfo{title}{Scalable multiparticle entanglement of trapped ions}.
\newblock \emph{\bibinfo{journal}{Nature}} \textbf{\bibinfo{volume}{438}},
  \bibinfo{pages}{643--646} (\bibinfo{year}{2005}).

\bibitem{BlattPRL}
\bibinfo{author}{Monz, T.} \emph{et~al.}
\newblock \bibinfo{title}{14-qubit entanglement: Creation and coherence}.
\newblock \emph{\bibinfo{journal}{Phys. Rev. Lett.}}
  \textbf{\bibinfo{volume}{106}}, \bibinfo{pages}{130506}
  (\bibinfo{year}{2011}).

\bibitem{Prevedel2009}
\bibinfo{author}{Prevedel, R.} \emph{et~al.}
\newblock \bibinfo{title}{Experimental realization of dicke states of up to six
  qubits for multiparty quantum networking}.
\newblock \emph{\bibinfo{journal}{Phys. Rev. Lett.}}
  \textbf{\bibinfo{volume}{103}}, \bibinfo{pages}{020503}
  (\bibinfo{year}{2009}).

\bibitem{Wieczorek2009}
\bibinfo{author}{Wieczorek, W.} \emph{et~al.}
\newblock \bibinfo{title}{Experimental entanglement of a six-photon symmetric
  dicke state}.
\newblock \emph{\bibinfo{journal}{Phys. Rev. Lett.}}
  \textbf{\bibinfo{volume}{103}}, \bibinfo{pages}{020504}
  (\bibinfo{year}{2009}).

\bibitem{Carnio2015}
\bibinfo{author}{Carnio, E.~G.}, \bibinfo{author}{Buchleitner, A.} \&
  \bibinfo{author}{Gessner, M.}
\newblock \bibinfo{title}{Robust asymptotic entanglement under multipartite
  collective dephasing}.
\newblock \emph{\bibinfo{journal}{Phys. Rev. Lett.}}
  \textbf{\bibinfo{volume}{115}}, \bibinfo{pages}{010404}
  (\bibinfo{year}{2015}).

\bibitem{Dur2000}
\bibinfo{author}{D\"ur, W.}, \bibinfo{author}{Vidal, G.} \&
  \bibinfo{author}{Cirac, J.~I.}
\newblock \bibinfo{title}{Three qubits can be entangled in two inequivalent
  ways}.
\newblock \emph{\bibinfo{journal}{Phys. Rev. A}} \textbf{\bibinfo{volume}{62}},
  \bibinfo{pages}{062314} (\bibinfo{year}{2000}).

\bibitem{Wei2004}
\bibinfo{author}{Wei, T.~C.}, \bibinfo{author}{Altepeter, J.~B.},
  \bibinfo{author}{Goldbart, P.~M.} \& \bibinfo{author}{Munro, W.~J.}
\newblock \bibinfo{title}{Measures of entanglement in multipartite bound
  entangled states}.
\newblock \emph{\bibinfo{journal}{Phys. Rev. A}} \textbf{\bibinfo{volume}{70}},
  \bibinfo{pages}{022322} (\bibinfo{year}{2004}).

\bibitem{Wei2008}
\bibinfo{author}{Wei, T.-C.}
\newblock \bibinfo{title}{Relative entropy of entanglement for multipartite
  mixed states: Permutation-invariant states and d\"ur states}.
\newblock \emph{\bibinfo{journal}{Phys. Rev. A}} \textbf{\bibinfo{volume}{78}},
  \bibinfo{pages}{012327} (\bibinfo{year}{2008}).

\bibitem{Raussendorf2001}
\bibinfo{author}{Raussendorf, R.} \& \bibinfo{author}{Briegel, H.~J.}
\newblock \bibinfo{title}{A one-way quantum computer}.
\newblock \emph{\bibinfo{journal}{Phys. Rev. Lett.}}
  \textbf{\bibinfo{volume}{86}}, \bibinfo{pages}{5188--5191}
  (\bibinfo{year}{2001}).

\bibitem{Dicke1954}
\bibinfo{author}{Dicke, R.~H.}
\newblock \bibinfo{title}{Coherence in spontaneous radiation processes}.
\newblock \emph{\bibinfo{journal}{Phys. Rev.}} \textbf{\bibinfo{volume}{93}},
  \bibinfo{pages}{99--110} (\bibinfo{year}{1954}).

\bibitem{AcinScience}
\bibinfo{author}{Tura, J.} \emph{et~al.}
\newblock \bibinfo{title}{Detecting nonlocality in many-body quantum states}.
\newblock \emph{\bibinfo{journal}{Science}} \textbf{\bibinfo{volume}{344}},
  \bibinfo{pages}{1256--1258} (\bibinfo{year}{2014}).

\bibitem{weinfurter2001four}
\bibinfo{author}{Weinfurter, H.} \& \bibinfo{author}{{\.Z}ukowski, M.}
\newblock \bibinfo{title}{Four-photon entanglement from down-conversion}.
\newblock \emph{\bibinfo{journal}{Physical Review A}}
  \textbf{\bibinfo{volume}{64}}, \bibinfo{pages}{010102}
  (\bibinfo{year}{2001}).

\bibitem{Smolin2001}
\bibinfo{author}{Smolin, J.~A.}
\newblock \bibinfo{title}{Four-party unlockable bound entangled state}.
\newblock \emph{\bibinfo{journal}{Phys. Rev. A}} \textbf{\bibinfo{volume}{63}},
  \bibinfo{pages}{032306} (\bibinfo{year}{2001}).

\bibitem{Cabello2003}
\bibinfo{author}{Cabello, A.}
\newblock \bibinfo{title}{Solving the liar detection problem using the
  four-qubit singlet state}.
\newblock \emph{\bibinfo{journal}{Phys. Rev. A}} \textbf{\bibinfo{volume}{68}},
  \bibinfo{pages}{012304} (\bibinfo{year}{2003}).

\bibitem{Kiesel2005}
\bibinfo{author}{Kiesel, N.} \emph{et~al.}
\newblock \bibinfo{title}{Experimental analysis of a four-qubit photon cluster
  state}.
\newblock \emph{\bibinfo{journal}{Phys. Rev. Lett.}}
  \textbf{\bibinfo{volume}{95}}, \bibinfo{pages}{210502}
  (\bibinfo{year}{2005}).

\bibitem{Kiesel2007}
\bibinfo{author}{Kiesel, N.}, \bibinfo{author}{Schmid, C.},
  \bibinfo{author}{T\'oth, G.}, \bibinfo{author}{Solano, E.} \&
  \bibinfo{author}{Weinfurter, H.}
\newblock \bibinfo{title}{Experimental observation of four-photon entangled
  dicke state with high fidelity}.
\newblock \emph{\bibinfo{journal}{Phys. Rev. Lett.}}
  \textbf{\bibinfo{volume}{98}}, \bibinfo{pages}{063604}
  (\bibinfo{year}{2007}).

\bibitem{bourennane2004decoherence}
\bibinfo{author}{Bourennane, M.} \emph{et~al.}
\newblock \bibinfo{title}{Decoherence-free quantum information processing with
  four-photon entangled states}.
\newblock \emph{\bibinfo{journal}{Physical review letters}}
  \textbf{\bibinfo{volume}{92}}, \bibinfo{pages}{107901}
  (\bibinfo{year}{2004}).

\bibitem{murao1999quantum}
\bibinfo{author}{Murao, M.}, \bibinfo{author}{Jonathan, D.},
  \bibinfo{author}{Plenio, M.} \& \bibinfo{author}{Vedral, V.}
\newblock \bibinfo{title}{Quantum telecloning and multiparticle entanglement}.
\newblock \emph{\bibinfo{journal}{Physical Review A}}
  \textbf{\bibinfo{volume}{59}}, \bibinfo{pages}{156} (\bibinfo{year}{1999}).

\bibitem{gaertner2007experimental}
\bibinfo{author}{Gaertner, S.}, \bibinfo{author}{Kurtsiefer, C.},
  \bibinfo{author}{Bourennane, M.} \& \bibinfo{author}{Weinfurter, H.}
\newblock \bibinfo{title}{Experimental demonstration of four-party quantum
  secret sharing}.
\newblock \emph{\bibinfo{journal}{Physical Review Letters}}
  \textbf{\bibinfo{volume}{98}}, \bibinfo{pages}{020503}
  (\bibinfo{year}{2007}).

\bibitem{MonzPhD}
\bibinfo{author}{Monz, T.}
\newblock \emph{\bibinfo{title}{Quantum information processing beyond ten
  ion-qubits}}.
\newblock Ph.D. thesis, \bibinfo{school}{Institute for Experimental Physics,
  University of Innsbruck} (\bibinfo{year}{2011}).

\bibitem{Chen2013}
\bibinfo{author}{Chen, X.-y.}, \bibinfo{author}{Yu, P.},
  \bibinfo{author}{Jiang, L.-z.} \& \bibinfo{author}{Tian, M.}
\newblock \bibinfo{title}{Genuine entanglement of four-qubit cluster diagonal
  states}.
\newblock \emph{\bibinfo{journal}{Phys. Rev. A}} \textbf{\bibinfo{volume}{87}},
  \bibinfo{pages}{012322} (\bibinfo{year}{2013}).

\bibitem{WisemanPRL}
\bibinfo{author}{Wiseman, H.~M.}, \bibinfo{author}{Jones, S.~J.} \&
  \bibinfo{author}{Doherty, A.~C.}
\newblock \bibinfo{title}{Steering, entanglement, nonlocality, and the
  einstein-podolsky-rosen paradox}.
\newblock \emph{\bibinfo{journal}{Phys. Rev. Lett.}}
  \textbf{\bibinfo{volume}{98}}, \bibinfo{pages}{140402}
  (\bibinfo{year}{2007}).

\bibitem{SteeringNPhys}
\bibinfo{author}{Armstrong, S.} \emph{et~al.}
\newblock \bibinfo{title}{Multipartite einstein-podolsky-rosen steering and
  genuine tripartite entanglement with optical networks}.
\newblock \emph{\bibinfo{journal}{Nat. Phys.}} \textbf{\bibinfo{volume}{11}},
  \bibinfo{pages}{167} (\bibinfo{year}{2015}).

\bibitem{Comon2007}
\bibinfo{author}{Comon, P.} \& \bibinfo{author}{Sorensen, M.}
\newblock \bibinfo{title}{Tensor diagonalization by orthogonal transforms}.
\newblock \emph{\bibinfo{journal}{Report ISRN I3S-RR-2007-06-FR}}
  (\bibinfo{year}{2007}).

\end{thebibliography}


\section*{\sf \bfseries Supplementary Material}
\appendix

\section{\sf \bfseries Multiparticle entanglement}

When considering a multiparticle quantum system, there exist two different approaches to entanglement, one referring to a particular partition of the composite system under consideration (partition-dependent setting), and another which considers indiscriminately all the partitions with a set number of parties (partition-independent setting).

In order to characterise the possible partitions of an $N$-qubit system, we will employ the following notation \cite{Blasone2008} :
\begin{itemize}
\item the positive integer $M$, $1< M \leq N$, representing the number of subsystems;
\item the sequence of positive integers $\{K_\alpha\}_{\alpha=1}^M:=\{K_1, K_2,\cdots, K_M\}$, where a given $K_\alpha$ represents the number of qubits belonging to the $\alpha$-th subsystem;
\item the sequence of sequences of positive integers $\{Q_\alpha\}_{\alpha=1}^M$, such that $Q_\alpha =\left\lbrace  i_1^{(\alpha)},i_2^{(\alpha)},\cdots,i_{K_\alpha}^{(\alpha)}\right\rbrace $ with $i_j^{(\alpha)}\in \{1,\cdots,N\}$ and $Q_\alpha\cap Q_{\alpha'}=\varnothing$ for $\alpha\neq \alpha'$, where a given sequence $Q_\alpha$ represents precisely the qubits belonging to the $\alpha$-th subsystem.
\end{itemize}
In the following we will say that $\{Q_\alpha\}_{\alpha=1}^M$ identifies a generic $M$-partition of an $N$-qubit system.

The set of $N$-qubit separable states $\mathcal{S}_{\{Q_\alpha\}_{\alpha=1}^M}$ with respect to the $M$-partition $\{Q_\alpha\}_{\alpha=1}^M$ contains all, and only, states $\varsigma$ of the form
\begin{equation}\label{Eq:SeparableStates}
\varsigma = \sum_{i} p_{i} \tau_{i}^{(1)} \otimes \tau_{i}^{(2)} \otimes \ldots \otimes \tau_{i}^{(M)},
\end{equation}
where $\{p_{i}\}$ forms a probability distribution and $\tau_{i}^{(\alpha)}$ are arbitrary states of the  $\alpha$-th subsystem. In other words, any $\{Q_\alpha\}_{\alpha=1}^M$-separable state can be written as a convex combination of product states that are all factorised with respect to the same partition $\{Q_\alpha\}_{\alpha=1}^M$. On the other hand, the set of $N$-qubit $M$-separable states $\mathcal{S}_{M}$ contains all, and only, states that can be written as convex combinations of product states, each of which is factorised with respect to an $M$-partition that need not be the same. One can easily see that the set of $M$-separable states is the convex hull of the union of all the sets of $\{Q_\alpha\}_{\alpha=1}^M$-separable states obtained by considering all the possible $M$-partitions $\{Q_\alpha\}_{\alpha=1}^M$.

Any valid measure of multiparticle entanglement must be zero on the relevant set of separable states and monotonically non-increasing under LOCC. In the partition-dependent setting, a LOCC with respect to a particular $\{Q_\alpha\}_{\alpha=1}^M$-partition amounts to allowing each of the $M$ parties to perform local operations on their qubits, and communicate with any other party via a classical channel~\cite{Horodecki2009}. Conversely, in the partition-independent setting, one considers operations that are LOCC with respect to all of the $M$-partitions, which can be shown to be all and only the single-particle LOCC. 
A convex combination of single-particle local unitaries acting on a state $\varrho$, given by
\begin{equation}
\sum_{i} p_{i} U_{i}^{(1)} \otimes U_{i}^{(2)} \otimes \ldots \otimes U_{i}^{(N)} \varrho U_{i}^{(1) \dagger} \otimes U_{i}^{(2) \dagger} \otimes \ldots \otimes U_{i}^{(N) \dagger},
\end{equation}
is a particular type of single-particle LOCC (requiring only one-way communication). It can be physically achieved by allowing one of the subsystems $\alpha$ to randomly select a local unitary $U_{i}^{(\alpha)}$ by using the probability distribution $\{p_{i}\}$ and then to communicate the result to all the other subsystems.

One can also impose that a  measure of $N$-particle entanglement is convex under convex combinations of quantum states, i.e.
\begin{eqnarray}
E_{\{Q_\alpha\}_{\alpha=1}^M}(q \varrho + (1-q) \varrho') &\leq & q E_{\{Q_\alpha\}_{\alpha=1}^M}(\varrho) + (1-q) E_{\{Q_\alpha\}_{\alpha=1}^M}(\varrho'), \nonumber \\
E_{M}(q \varrho + (1-q) \varrho') &\leq & q E_{M}(\varrho) + (1-q) E_{M}(\varrho')
\end{eqnarray}
for some probability $q$ and quantum states $\varrho$ and $\varrho'$, which ensures that classical mixing of quantum states cannot lead to an increasing of entanglement. Any measure obeying these properties is referred to as a convex entanglement monotone. In this work we adopt geometric measures of multiparticle entanglement, defined in terms of the distance to the relevant set of separable states. For a given distance $D$, generic distance-based measures of the multiparticle entanglement of an $N$-qubit state $\varrho$, quantifying how much $\varrho$ is not $\{Q_\alpha\}_{\alpha=1}^M$-separable (resp., $M$-separable), are given by, respectively,
\begin{eqnarray}
E_{\{Q_\alpha\}_{\alpha=1}^M}^{D}(\varrho) &\equiv & \inf _{\varsigma \in \mathcal{S}_{\{Q_\alpha\}_{\alpha=1}^M}} D(\varrho, \varsigma), \label{esupd}\\
E_{M}^{D}(\varrho) &\equiv & \inf _{\varsigma \in \mathcal{S}_{M}} D(\varrho, \varsigma), \label{esupi}
\end{eqnarray}
where Eq.~(\ref{esupd}) refers to the partition-dependent setting, and Eq.~(\ref{esupi}) to the partition-independent one.
It is sufficient for the distance $D$ to obey contractivity and joint convexity (see Methods in the main text) for $E_{\{Q_\alpha\}_{\alpha=1}^M}^{D}$ and $E_{M}^{D}$ to be convex entanglement monotones \cite{PlenioVirmani2007}.

\section{\sf \bfseries The set of $\mathcal{M}_{N}^{3}$ states}



In this appendix we show some relevant properties of the subclass of $N$-qubit states with all maximally mixed marginals that we refer to as $\mathcal{M}_N^3$ states. Their matrix representation in the computational basis is the following:
\begin{equation}\label{eq:NqubitBellState}
\varpi = \frac{1}{2^{N}} \left( \mathbb{I}^{\otimes N} + \sum_{i=1}^{3} c_{i} \sigma_{i}^{\otimes N}\right),
\end{equation}
where $\mathbb{I}$ is the $2 \times 2$ identity matrix, $\sigma_{i}$ is the $i$-th Pauli matrix, $c_{i} = {\rm Tr}\left[\varpi \sigma_{i}^{\otimes N}\right]\in[-1,1]$ and $N>1$. These states are denoted by the triple $\{c_{1},c_{2},c_{3}\}$.

The characterisation of the $\mathcal{M}_{N}^{3}$ states is manifestly different between the even and odd $N$ case. For even $N$, the eigenvectors and eigenvalues are given by, respectively,
\begin{equation}\label{eq:EigenvectorsM3NstatesNeven}
|\beta_i^{\pm}\rangle=\frac{1}{\sqrt{2}}\left(\mathbb{I}^{\otimes N} \pm \sigma_1^{\otimes N} \right) |i\rangle,
\end{equation}
and
\begin{equation}\label{eq:EigenvaluesNBellStates}
\lambda_p^{\pm}=\frac{1}{2^N}\left[1\pm c_1 \pm (-1)^{N/2}(-1)^p c_2 +(-1)^p c_3 \right],
\end{equation}
where $i\in \{1,\cdots,2^{N-1}\}$, $\{|i\rangle\}_{i=1}^{2^N}$ is the binary ordered $N$-qubit computational basis and finally $p$ is the parity of $|\beta_i^{\pm}\rangle$ with respect to the parity operator along the $z$-axis $\Pi_3 = \sigma_3^{\otimes N}$, i.e.
\begin{equation}\label{eq:definitionofparity}
\Pi_3|\beta_i^{\pm}\rangle = (-1)^p |\beta_i^{\pm}\rangle.
\end{equation}

In the $\{c_{1},c_{2},c_{3}\}$-space, the set of $\mathcal{M}_{N}^{3}$ states with even $N$ is represented by the tetrahedron ${\cal T}_{(-1)^{N/2}}$ with vertices $\{1,(-1)^{N/2},1\}$, $\{-1,-(-1)^{N/2},1\}$, $\{1,-(-1)^{N/2},-1\}$ and $\{-1,(-1)^{N/2},-1\}$, as illustrated in Fig.~2(b) in the main text. This tetrahedron is constructed simply by imposing the non-negativity of the four eigenvalues (\ref{eq:EigenvaluesNBellStates}) of such $\mathcal{M}_{N}^{3}$ states.

For odd $N$, the eigenvectors and eigenvalues of the $\mathcal{M}_N^3$ states can be easily written in spherical coordinates as
\begin{eqnarray}\label{eq:EigenvectorsM3NstatesNodd}
|\alpha_i^{\pm}\rangle&=&\cos\left[\frac{\theta}{2} +(1\mp(-1)^p)\frac{\pi}{4}\right]|i\rangle  \\
&&+ (-1)^{p} e^{i(-1)^{p}(-1)^{\frac{N-1}{2}}\phi} \sin\left[\frac{\theta}{2} +(1\mp(-1)^p)\frac{\pi}{4}\right]\sigma_1^{\otimes N}|i\rangle , \nonumber
\end{eqnarray}
and
\begin{equation}\label{eq:EigenvaluesNBellStatesNodd}
\lambda_{\pm}=\frac{1}{2^N}\left(1\pm r\right),
\end{equation}
where $i\in \{1,\cdots,2^{N-1}\}$, $\{|i\rangle\}_{i=1}^{2^N}$ is again the binary ordered $N$-qubit computational basis, $p$ is the parity of $|i\rangle$ with respect to the parity operator $\Pi_3=\sigma_3^{\otimes N}$, $c_1= r \sin\theta \cos\phi$, $c_2= r \sin\theta \sin\phi$ and $c_3=r \cos\theta$, with $r=\sqrt{c_1^2+c_2^2+c_3^2}$, $\theta\in[0,\pi]$ and $\phi\in[0,2\pi[$.

Consequently, thanks again to the semi-positivity constraint, the set of $\mathcal{M}_N^3$ states with odd $N$ is represented in the $\{c_{1},c_{2},c_{3}\}$-space by the unit ball ${\cal B}_{1}$ centred into the origin, as shown in Fig.~2(c) in the main text.

\section{\sf \bfseries $\mathcal{M}^{3}_{N}$-fication}\label{sec:M3ificationsection}
The following Theorem is crucial for providing a lower bound to any multiparticle entanglement monotone of any state $\varrho$ and for analytically computing the multiparticle geometric entanglement of any $\mathcal{M}_{N}^{3}$ state $\varpi$.
\begin{theorem}\label{Theorem:MCubification}
Any $N$-qubit state $\varrho$ can be transformed into a corresponding $\mathcal{M}^{3}_{N}$ state $\varrho_{\mathcal{M}^{3}_{N}}$ through a fixed operation, $\Theta$, that is a single-qubit LOCC and such that
\begin{equation}
\Theta(\varrho) = \varrho_{\mathcal{M}^{3}_{N}} = \frac{1}{2^{N}}\left( \mathbb{I}^{\otimes N} + \sum_{i=1}^{3} c_{i} \sigma_{i}^{\otimes N}\right),
\end{equation}
where $c_{i} = \mbox{{\rm Tr}}(\varrho \sigma_{i}^{\otimes N})$.
\end{theorem}

\noindent \textit{Proof} The first part of the proof was sketched in the Methods section and is repeated here for completeness.

We will give the form of $\Theta(\varrho)$, show that $\Theta$ is a single-qubit LOCC, and finally prove that it transforms any $N$-qubit state $\varrho$ into $\varrho_{\mathcal{M}^{3}_{N}}$.

To define $\Theta(\varrho)$, we begin by setting $2(N-1)$ single-qubit local unitaries
\begin{eqnarray}
\{U_{j}\}_{j=1}^{2(N-1)}=\{(\sigma_{1} \otimes \sigma_{1} \otimes I^{\otimes N-2}),
(I \otimes \sigma_{1} \otimes \sigma_{1} \otimes I^{\otimes N-3}),\nonumber \\
\ldots
(I^{\otimes N-3} \otimes \sigma_{1} \otimes \sigma_{1} \otimes I),
(I^{\otimes N-2} \otimes \sigma_{1} \otimes \sigma_{1}) \nonumber \\,
(\sigma_{2} \otimes \sigma_{2} \otimes I^{\otimes N-2}),
(I \otimes \sigma_{2} \otimes \sigma_{2} \otimes I^{\otimes N-3}),\nonumber \\
\ldots
(I^{\otimes N-3} \otimes \sigma_{2} \otimes \sigma_{2} \otimes I),
(I^{\otimes N-2} \otimes \sigma_{2} \otimes \sigma_{2})
\}. \nonumber \\
\end{eqnarray}
Then, we fix a sequence of states $\{\varrho_{0},\varrho_{1}, \ldots \varrho_{2(N-1)}\}$ defined by
\begin{equation}\label{Eq:ConvexIteration}
\varrho_{j} \equiv \frac{1}{2}\left( \varrho_{j-1}+U_{j}\varrho_{j-1}U_{j}^{\dagger} \right)
\end{equation}
for $j \in \{1,2, \ldots 2(N-1)\}$. By setting $\varrho_{0} = \varrho$ and $\varrho_{2(N-1)}=\Theta(\varrho)$, we define the required LOCC channel, i.e.~$\Theta(\varrho)= \frac{1}{2^{2(N-1)}}\sum_{i=1}^{2^{2(N-1)}} U_{i}' \varrho U_{i}'^{\dagger}$ where $U_{i}'$ are the following unitaries
\begin{equation}
\{U_{i}'\}_{i=1}^{2^{2(N-1)}}=
\left \{
  \begin{tabular}{c}
  $\mathbb{I}^{\otimes N}$ \\
  $\{U_{i_{1}}\}_{i_{1}=1}^{2(N-1)}$ \\
  $\{U_{i_{2}}U_{i_{1}}\}_{i_{2}>i_{1}=1}^{2(N-1)}$ \\
  $\cdots$ \\
  $\{U_{i_{2(N-1)}} \ldots U_{i_{2}}U_{i_{1}}\}_{i_{2(N-1)}>\ldots>i_{2}>i_{1}=1}^{2(N-1)}$ \\
  \end{tabular}
\right \}.
\end{equation}
It is clear that $\{U_{i}'\}_{i=1}^{2^{2(N-1)}}$ are unitaries that still act locally on individual qubits. Since $\Theta$ is a convex mixture of such local unitaries, we conclude that $\Theta$ is a single-qubit LOCC.

Now we will show that $\Theta(\varrho) = \varrho_{\mathcal{M}^{3}_{N}}$. Consider the arbitrary $N$-qubit state $\varrho$ written in the form
\begin{equation}\label{Eq:GeneralNQubitState}
\varrho=\frac{1}{2^{N}} \sum_{i_{1} i_{2} \ldots i_{N}=0}^{3} R_{i_{1}i_{2} \ldots i_{N}}^{\varrho} \sigma_{i_{1}} \otimes \sigma_{i_{2}} \ldots \otimes \sigma_{i_{N}},
\end{equation}
where the $R_{i_{1},i_{2}, \ldots i_{N}}^{\varrho}={\rm Tr}\left[\varrho\  \sigma_{i_{1}} \otimes \sigma_{i_{2}} \ldots \otimes \sigma_{i_{N}} \right] \in [-1,1]$ are the correlation tensor elements of $\varrho$ with  $\sigma_{0}=\mathbb{I}$. Convex combination of two arbitrary $N$-qubit states $\varrho$ and $\varrho'$ gives
\begin{equation}\label{Eq:ConvexCombinationRForm}
q \varrho + (1-q) \varrho' = \frac{1}{2^{N}} \sum_{i_{1} i_{2} \ldots i_{N}=0}^{3} R_{i_{1}i_{2} \ldots i_{N}}^{q \varrho+(1-q)\varrho '} \sigma_{i_{1}} \otimes \sigma_{i_{2}} \ldots \otimes \sigma_{i_{N}}
\end{equation}
where $R_{i_{1}i_{2} \ldots i_{N}}^{q \varrho+(1-q)\varrho '} = q R_{i_{1}i_{2} \ldots i_{N}}^{\varrho}+(1-q)R_{i_{1}i_{2} \ldots i_{N}}^{\varrho'}$.

We will now understand the evolution of the $R_{i_{1}i_{2} \ldots i_{N}}^{\varrho_{j}}$ for each step $j$ in Eq.~(\ref{Eq:ConvexIteration}). The action of $U_{1}$ on $\varrho$ is
\begin{eqnarray}
U_{1} \varrho U_{1}^{\dagger}= \frac{1}{2^{N}} \sum_{i_{1}i_{2} \ldots i_{N}=0}^{3} R_{i_{1}i_{2} \ldots i_{N}}^{\varrho} \sigma_{1}\sigma_{i_{1}}\sigma_{1} \otimes \sigma_{1}\sigma_{i_{2}}\sigma_{1} \nonumber \\
\otimes \sigma_{i_{3}} \ldots \otimes \sigma_{i_{N}}.
\end{eqnarray}
From $\sigma_{1} \sigma_{i} \sigma_{1} = -(-1)^{\delta_{0i}+\delta_{1i}}\sigma_{i}$ we have that the correlation tensor elements of $U_{1} \varrho U_{1}^{\dagger}$ are $R_{i_{1}i_{2} \ldots i_{N}}^{U_{1} \varrho U_{1}^{\dagger}} = (-1)^{\delta_{0i_1}+\delta_{1i_1}+\delta_{0i_2}+\delta_{1i_2}} R_{i_{1}i_{2} \ldots i_{N}}^{\varrho}$. By using Eq.~(\ref{Eq:ConvexIteration}) and Eq.~(\ref{Eq:ConvexCombinationRForm}), it is clear that the $R_{i_{1}i_{2} \ldots i_{N}}^{\varrho_{1}}$ of $\varrho_{1}$ are $R_{i_{1}i_{2} \ldots i_{N}}^{\varrho}$ if $i_{1}$ and $i_{2}$ are (i) any combination of only $1$ and $0$ or (ii) any combination of only $2$ and $3$, and zero otherwise.

Generally, for $j \in [1,N-1]$, the $R_{i_{1}i_{2} \ldots i_{N}}^{\varrho_{j}}$ of $\varrho_{j}$ are $R_{i_{1}i_{2} \ldots i_{N}}^{\varrho_{j-1}}$ if $i_{j}$ and $i_{j+1}$ are (i) any combination of only $1$ and $0$ or (ii) any combination of only $2$ and $3$, and zero otherwise. For $j \in [N,2(N-1)]$ the conditions are analogous, where the $R_{i_{1}i_{2} \ldots i_{N}}^{\varrho_{j}}$ of $\varrho_{j}$ are $R_{i_{1}i_{2} \ldots i_{N}}^{\varrho_{j-1}}$ if $i_{j}$ and $i_{j+1}$ are (i) any combination of only $2$ and $0$ or (ii) any combination of only $1$ and $3$, and zero otherwise. For the final state $\varrho_{2(N-1)}$, the only nonzero $R_{i_{1}i_{2} \ldots i_{N}}^{\varrho_{2(N-1)}}$ are those for which $\{i_{1}i_{2}\ldots i_{N}\}$ consists of only $0$, $1$, $2$, or $3$, and that for these elements $R_{i_{1}i_{2} \ldots i_{N}}^{\varrho_{2(N-1)}}=R_{i_{1}i_{2} \ldots i_{N}}^{\varrho}$. Therefore
\begin{eqnarray}
\Theta(\varrho) =\varrho_{2(N-1)} &=& \frac{1}{2^{N}} \sum_{i=0}^{3} R_{i i \ldots i}^{\varrho} \sigma_{i} \otimes \sigma_{i} \ldots \otimes \sigma_{i} \nonumber \\
&\equiv & \frac{1}{2^{N}}\left( \mathbb{I}^{\otimes N} + \sum_{i=1}^{3} c_{i} \sigma_{i}^{\otimes N}\right) = \varrho_{\mathcal{M}^{3}_{N}}
\end{eqnarray}
where we have used $R_{ii \ldots i}^{\varrho} = \mbox{{\rm Tr}}(\varrho \sigma_{i}^{\otimes N}) \equiv c_{i}$ for $i \in \{1,2,3\}$ and $R_{00 \ldots 0}^{\varrho} = \mbox{{\rm Tr}}(\varrho) =1$.
\begin{flushright}
$\blacksquare$
\end{flushright}


Herein, we will refer to $\varrho_{\mathcal{M}^{3}_{N}}= \Theta(\varrho)$ as the $\mathcal{M}^3_N$-fication of the state $\varrho$. Theorem \ref{Theorem:MCubification} has two major implications. The first implication applies to any multiparticle entanglement monotone, be it partition-dependent or independent. We have that
\begin{eqnarray}
E_{\{Q_\alpha\}_{\alpha=1}^M}(\varrho_{\mathcal{M}^{3}_{N}}) &=& E_{\{Q_\alpha\}_{\alpha=1}^M}(\Theta (\varrho)) \leq E_{\{Q_\alpha\}_{\alpha=1}^M}(\varrho),\\
E_{M}(\varrho_{\mathcal{M}^{3}_{N}}) &=& E_{M}(\Theta (\varrho)) \leq E_{M}(\varrho),
\end{eqnarray}
where in the first equality we use $\varrho_{\mathcal{M}^{3}_{N}} = \Theta (\varrho)$ and in the inequality we use the monotonicity under single-qubit LOCC of any measure of multiparticle entanglement and the fact that $\Theta$ is a single-qubit LOCC. In other words, the multiparticle entanglement of the $\mathcal{M}^3_N$-fication $\varrho_{\mathcal{M}^{3}_{N}}$ of any state $\varrho$ provides us with a lower bound of the multiparticle entanglement of $\varrho$.

The second implication applies specifically to distance-based measures of multiparticle entanglement, although regardless of whether such a measure is partition-dependent or independent. We have that, for any $\mathcal{M}_{N}^{3}$ state $\varpi$ and any  separable state $\varsigma$,
\begin{equation}\label{Eq:M3NCloser}
D(\varpi,\varsigma_{\mathcal{M}_{N}^{3}}) = D(\Theta(\varpi),\Theta(\varsigma)) \leq D(\varpi,\varsigma),
\end{equation}
where in the first equality we use the invariance of any $\mathcal{M}_{N}^{3}$ state through $\Theta$ and that $\Theta(\varsigma) \equiv \varsigma_{\mathcal{M}_{N}^{3}}$ is the $\mathcal{M}^{3}$-fication of $\varsigma$, and in the inequality we use the contractivity of the distance through any completely positive trace-preserving channel. Moreover, the $\mathcal{M}^{3}_{N}$-fication $\varsigma_{\mathcal{M}_{N}^{3}}$ of any separable state $\varsigma$, regardless of whether $\varsigma$ is $\{Q_\alpha\}_{\alpha=1}^M$-separable or $M$-separable, is a separable $\mathcal{M}_{N}^{3}$ state of the same kind as $\varsigma$, since $\Theta$ is a single-qubit LOCC and thus leaves any set of separable states invariant. Therefore, both the sets $\mathcal{S}_{\{Q_\alpha\}_{\alpha=1}^M}^{\mathcal{M}_{N}^{3}}$ and $\mathcal{S}_{M}^{\mathcal{M}_{N}^{3}}$ of, respectively, $\{Q_\alpha\}_{\alpha=1}^M$-separable and $M$-separable $\mathcal{M}_{N}^{3}$ states will be crucial to identify (see Appendix \ref{Appendix:SetOfSeperableM3N}), since they allow us to use Eq.~(\ref{Eq:M3NCloser}) to say that for any  distance-based  measure of multiparticle entanglement of an $\mathcal{M}_{N}^{3}$ state $\varpi$,
\begin{eqnarray}
E_{\{Q_\alpha\}_{\alpha=1}^M}^{D} (\varpi)&\equiv& \inf_{\varsigma \in \mathcal{S}_{\{Q_\alpha\}_{\alpha=1}^M}} D(\varpi,\varsigma) = \inf_{\varsigma_{\mathcal{M}_{N}^{3}}\in \mathcal{S}_{\{Q_\alpha\}_{\alpha=1}^M}^{\mathcal{M}_{N}^{3}}} D(\varpi,\varsigma_{\mathcal{M}_{N}^{3}}), \nonumber \\
E_{M}^{D} (\varpi)&\equiv& \inf_{\varsigma \in \mathcal{S}_{M}} D(\varpi,\varsigma) = \inf_{\varsigma_{\mathcal{M}_{N}^{3}}\in \mathcal{S}_{M}^{\mathcal{M}_{N}^{3}}} D(\varpi,\varsigma_{\mathcal{M}_{N}^{3}}),
\end{eqnarray}
i.e.~that one of the closest $\{Q_\alpha\}_{\alpha=1}^M$-separable (resp., $M$-separable) states $\varsigma_{\varpi}$ to an $\mathcal{M}_{N}^{3}$ state $\varpi$ is itself an $\mathcal{M}_{N}^{3}$ state. We now formalise these two results as corollaries.

\begin{corollary}\label{Corollary:EntanglementInequality}
For any $N$-qubit state $\varrho$, the multiparticle entanglement of the corresponding $\mathcal{M}^{3}_{N}$-fied state $\varrho_{\mathcal{M}^{3}_{N}}$ is always less than or equal to the multiparticle entanglement of $\varrho$, i.e.
\begin{eqnarray}
E_{\{Q_\alpha\}_{\alpha=1}^M}(\varrho_{\mathcal{M}^{3}_{N}}) &\leq& E_{\{Q_\alpha\}_{\alpha=1}^M}(\varrho),\\
E_{M}(\varrho_{\mathcal{M}^{3}_{N}}) &\leq& E_{M}(\varrho),
\end{eqnarray}
for any $\{Q_\alpha\}_{\alpha=1}^M$-partition of the $N$-qubit system and any $2\leq M \leq N$.
\end{corollary}

\begin{corollary}
For any contractive distance $D$ and any $\mathcal{M}_{N}^{3}$ state $\varpi$, one of the closest $\{Q_\alpha\}_{\alpha=1}^M$-separable (resp., $M$-separable) states $\varsigma_{\varpi}$ to $\varpi$ is itself an $\mathcal{M}_{N}^{3}$ state, i.e.
\begin{equation}
\varsigma_{\varpi} = \frac{1}{2^{N}} \left( \mathbb{I}^{\otimes N} + \sum_{i} s_{i} \sigma_{i}^{\otimes N}\right),
\end{equation}
for any $\{Q_\alpha\}_{\alpha=1}^M$-partition of the $N$-qubit system and any $2\leq M \leq N$.
\end{corollary}

Theorem \ref{Theorem:MCubification} allows for another result which will be useful to characterise the set of separable $\mathcal{M}_{N}^{3}$ states.
\begin{corollary}\label{Corollary:MCubifiedPositivity}
The set of the triples $\{c_{1},c_{2},c_{3}\}$, with $c_{i} = \mbox{{\rm Tr}}(\varrho \sigma_{i}^{\otimes N})$, obtained by considering any possible $N$-qubit state $\varrho$ is
\begin{itemize}
\item{the unit ball ${\cal B}_{1}$, when $N$ is odd;}
\item{the tetrahedron ${\cal T}_{(-1)^{N/2}}$, when $N$ is even.}
\end{itemize}
\end{corollary}

This is because the set of $\mathcal{M}^{3}_{N}$-fications of all the states coincides exactly with the set of $\mathcal{M}_{N}^{3}$ states. Indeed, the $\mathcal{M}^{3}_{N}$-fication channel $\Theta$ makes the entire set of states collapse into the set of $\mathcal{M}^3_N$ states, whereas it leaves the set of $\mathcal{M}^3_N$ states invariant. Herein, we shall refer to the triple $\{c_{1},c_{2},c_{3}\}$, with $c_{i} = \mbox{{\rm Tr}}(\varrho \sigma_{i}^{\otimes N})$, as the Pauli correlation vector corresponding to the state $\varrho$.


\section{\sf \bfseries The set of separable $\mathcal{M}^{3}_{N}$ states}\label{Appendix:SetOfSeperableM3N}

We are now ready to characterise the sets $\mathcal{S}_{\{Q_\alpha\}_{\alpha=1}^M}^{\mathcal{M}_{N}^{3}}$ and $\mathcal{S}_{M}^{\mathcal{M}_{N}^{3}}$ of, respectively, $\{Q_\alpha\}_{\alpha=1}^M$-separable and $M$-separable $\mathcal{M}_N^3$ states. 
The first ingredient is to note that $\mathcal{S}_{\{Q_\alpha\}_{\alpha=1}^M}^{\mathcal{M}_{N}^{3}}$ coincides exactly with the set $\Theta \left[ \mathcal{S}_{\{Q_\alpha\}_{\alpha=1}^M} \right]$ of the $\mathcal{M}^{3}_{N}$-fications of any $\{Q_\alpha\}_{\alpha=1}^M$-separable state. Furthermore, we note that since any $\mathcal{M}_{N}^{3}$ state is invariant under  any permutation of the $N$ qubits, then the set of $\{Q_\alpha\}_{\alpha=1}^M$-separable $\mathcal{M}_{N}^{3}$ states $\mathcal{S}_{\{Q_\alpha\}_{\alpha=1}^M}^{\mathcal{M}_{N}^{3}}$ does not depend on which qubits belong to each of the subsystems. Therefore we need only to specify the cardinalities $\{K_\alpha\}_{\alpha=1}^M$ to completely characterise $\mathcal{S}_{\{Q_\alpha\}_{\alpha=1}^M}^{\mathcal{M}_{N}^{3}}$, and we will herein refer to the latter as the set of $\{K_\alpha\}_{\alpha=1}^M$-separable $\mathcal{M}_{N}^{3}$ states $\mathcal{S}_{\{K_\alpha\}_{\alpha=1}^M}^{\mathcal{M}_{N}^{3}}$.

\begin{theorem}\label{Theorem:SeparableM3NStates}
For any $N$, the set of separable $\mathcal{M}_{N}^{3}$ states $\mathcal{S}_{\{K_\alpha\}_{\alpha=1}^M}^{\mathcal{M}_{N}^{3}}$ is either
\begin{itemize}
\item{the set of all $\mathcal{M}_{N}^{3}$ states, for any allowed $\{K_\alpha\}_{\alpha=1}^M$ partition such that $K_{\alpha}$ is odd for at most one value of $\alpha$;}
\item{the set of $\mathcal{M}_{N}^{3}$ states 
represented in the $\{c_{1},c_{2},c_{3}\}$-space by the unit octahedron ${\cal O}_{1}$ with vertices $\{\pm 1, 0, 0\}$, $\{0, \pm 1, 0\}$ and $\{0, 0, \pm 1\}$, for any allowed $\{K_\alpha\}_{\alpha=1}^M$ partition such that $K_{\alpha}$ is odd for more than one value of $\alpha$.}
\end{itemize}
\end{theorem}
\noindent \textit{Proof}

In order to characterise the set of $\{K_\alpha\}_{\alpha=1}^M$-separable $\mathcal{M}_{N}^{3}$ states, $\mathcal{S}_{\{K_\alpha\}_{\alpha=1}^M}^{\mathcal{M}_{N}^{3}}$, we simply need to identify its representation in the $\{c_{1},c_{2},c_{3}\}$-space. Since $\mathcal{S}_{\{K_\alpha\}_{\alpha=1}^M}^{\mathcal{M}_{N}^{3}}=\Theta\left[\mathcal{S}_{\{Q_\alpha\}_{\alpha=1}^M}\right]$, we know that such a representation is the set of Pauli correlation vectors corresponding to all the elements of $\mathcal{S}_{\{Q_\alpha\}_{\alpha=1}^M}$.



Due to Eq.~(\ref{Eq:SeparableStates}), the Pauli correlation vector of any $\varsigma \in \mathcal{S}_{\{Q_\alpha\}_{\alpha=1}^M}$ is given by
\begin{eqnarray}\label{Eq:SeperableM3ification}
s_{j} &=& \mbox{{\rm Tr}}\left( \varsigma \sigma_{j}^{\otimes N}\right) = \mbox{{\rm Tr}} \left[ \left( \sum_{i} p_{i} \tau_{i}^{(1)} \otimes \tau_{i}^{(2)} \otimes \ldots \otimes \tau_{i}^{(M)} \right) \sigma_{j}^{\otimes N}\right] \nonumber \\
&=&\sum_{i} p_{i} \mbox{{\rm Tr}}\left[ \tau_{i}^{(1)} \sigma_{j}^{\otimes K_{1}} \otimes \tau_{i}^{(2)} \sigma_{j}^{\otimes K_{2}} \otimes \ldots \otimes \tau_{i}^{(M)} \sigma_{j}^{\otimes K_{M}} \right] \nonumber \\
&=&\sum_{i} p_{i} \prod_{\alpha =1}^{M} \mbox{{\rm Tr}} \left( \tau_{i}^{(\alpha)} \sigma_{j}^{\otimes K_{\alpha}} \right) =\sum_{i} p_{i} \prod_{\alpha =1}^{M} c_{i,j}^{(\alpha)}
\end{eqnarray}
where in the final equality we denote $c_{i,j}^{(\alpha)} = \mbox{{\rm Tr}} \left(\tau_{i}^{(\alpha)} \sigma_{j}^{\otimes K_{\alpha}}\right)$ as the $j$-th component of the Pauli correlation vector $\vec{c}_{i}^{(\alpha)}=\{c_{i,1}^{(\alpha)},c_{i,2}^{(\alpha)},c_{i,3}^{(\alpha)}\}$ corresponding to the arbitrary state $\tau_{i}^{(\alpha)}$ of subsystem $\alpha$. Eq.~(\ref{Eq:SeperableM3ification}) can be simplified further by introducing the Hadamard product as the componentwise multiplication of vectors, i.e.~for $\vec{u}=\{u_{1},u_{2},u_{3}\}$ and $\vec{v}=\{v_{1},v_{2},v_{3}\}$ the Hadamard product is $\vec{u} \circ \vec{v}= \{u_{1}v_{1},u_{2}v_{2},u_{3}v_{3}\}$. Using the Hadamard product gives
Eq.~(\ref{Eq:SeperableM3ification}) as
\begin{equation}\label{Eq:SVector}
\vec{s} = \sum_{i} p_{i} \vec{c}_{i}^{(1)} \circ \vec{c}_{i}^{(2)} \circ \ldots \circ \vec{c}_{i}^{(M)},
\end{equation}
i.e., that the Pauli correlation vector of any $\{Q_\alpha\}_{\alpha=1}^M$-separable state is a convex combination of Hadamard products of Pauli correlation vectors corresponding to subsystem states. Due to Corollary \ref{Corollary:MCubifiedPositivity}, we know that $\vec{c}_{i}^{(\alpha)} \in {\cal B}_{1}$ when $K_{\alpha}$ is odd and $\vec{c}_{i}^{(\alpha)} \in {\cal T}_{(-1)^{K_{\alpha}/2}}$ when $K_{\alpha}$ is even, and so $\mathcal{S}_{\{K_\alpha\}_{\alpha=1}^M}^{\mathcal{M}_{N}^{3}}$ is represented by the following set
\begin{equation}\label{Eq:SeparableM3NHadamard}
\mathcal{S}_{\{K_\alpha\}_{\alpha=1}^M}^{\mathcal{M}_{N}^{3}} = conv \left( A^{(1)} \circ A^{(2)} \circ \ldots \circ A^{(M)} \right),
\end{equation}
with
\begin{equation}
A^{(\alpha)} =\left\{
  \begin{array}{lr}
    {\cal B}_{1} \,\,\,\,\,\,\,\,\,\,\,\,\,\,\, \mbox{if } K_{\alpha} \mbox{ is odd,}\\
    {\cal T}_{(-1)^{K_{\alpha}/2}} \,\,\, \mbox{if } K_{\alpha} \mbox{ is even,}\\
  \end{array}
\right.
\end{equation}
where we define the Hadamard product between any two sets $A$ and $B$ as $A \circ B = \{\vec{a} \circ \vec{b} \, | \, \vec{a}\in A \, , \, \vec{b} \in B\}$ and the convex hull $conv (A)$ is the set of all possible convex combinations of elements in $A$. The commutativity and associativity of the Hadamard product allow us to rearrange the ordering in Eq.~(\ref{Eq:SeparableM3NHadamard}) in the following way
\begin{equation}\label{Eq:SeparableM3NSimpleHadamard}
\mathcal{S}_{\{K_\alpha\}_{\alpha=1}^M}^{\mathcal{M}_{N}^{3}} = conv \left[ \left( \mathop{\mathlarger{\mathlarger{\mathlarger{\bigcirc}}}}_{\mathsmaller{\mu: K_{\mu} even}}{\cal T}_{(-1)^{K_{\mu}/2}}\right) \circ \left( \mathop{\mathlarger{\mathlarger{\mathlarger{\bigcirc}}}}_{\mathsmaller{\nu: K_{\nu} odd}}{\cal B}_{1}\right) \right],
\end{equation}
where $\bigcirc_{\alpha=1}^{n} A^{(\alpha)} = A^{(1)} \circ A^{(2)} \circ \ldots \circ A^{(n)}$.

By writing any vector in ${\cal T}_{\pm 1}$ as a convex combination of the vertices of ${\cal T}_{\pm 1}$, one can easily show that
\begin{eqnarray}\label{Eq:HadamardTetrahedra}
{\cal T}_{-1} \circ {\cal T}_{-1}  &=& {\cal T}_{1}, \nonumber \\
{\cal T}_{1} \circ {\cal T}_{1}    &=& {\cal T}_{1}, \nonumber \\
{\cal T}_{1} \circ {\cal T}_{-1}   &=& {\cal T}_{-1},
\end{eqnarray}
so that
\begin{equation}\label{Eq:HadamardTetrahedraResult}
\mathop{\mathlarger{\mathlarger{\mathlarger{\bigcirc}}}}_{\mathsmaller{\mu: K_{\mu} even}}{\cal T}_{(-1)^{K_{\mu}/2}} = {\cal T}_{(-1)^{\mathcal{M}_{-}}},
\end{equation}
where $\mathcal{M}_{-}$ is the number of $K_{\mu}$ with odd $K_{\mu}/2$. Similarly, one can see that
\begin{equation}\label{Eq:HadamardTetrahedraAndBall}
{\cal T}_{\pm 1} \circ {\cal B}_{1} = {\cal B}_{1}.
\end{equation}

Finally, we have that
\begin{equation}\label{Eq:hadamardproductofballs}
conv \left( \bigcirc_{\mathsmaller{i=1}}^{n} {\cal B}_{1} \right) = {\cal O}_{1}\ \ \forall n \geq 2.
\end{equation}

Indeed, since $\{\{\pm 1,0,0\},\{0,\pm 1,0\},\{0,0,\pm 1\}\} \subset \bigcirc_{\mathsmaller{i=1}}^{n} {\cal B}_{1}$ and $conv \{\{\pm 1,0,0\},\{0,\pm 1,0\},\{0,0,\pm 1\}\} = {\cal O}_{1}$, we know that ${\cal O}_{1} \subseteq conv \left( \bigcirc_{\mathsmaller{i=1}}^{n} {\cal B}_{1} \right)$. Now we will show that ${\cal O}_{1} \supseteq conv \left(\bigcirc_{\mathsmaller{i=1}}^{n} {\cal B}_{1} \right)$. To do so, it is sufficient to see that
\begin{equation}\label{Eq:BallBallOctahedronVector}
\vec{b} \circ \vec{b}' \in {\cal O}_{1}
\end{equation}
for any $\vec{b},\vec{b}'\in {\cal B}_{1}$, which trivially implies that $\bigcirc_{\mathsmaller{i=1}}^{n} {\cal B}_{1} \subseteq {\cal O}_{1}$, and so $conv \left(\bigcirc_{\mathsmaller{i=1}}^{n} {\cal B}_{1}\right) \subseteq conv \left( {\cal O}_{1} \right)= {\cal O}_{1}$. Equation (\ref{Eq:BallBallOctahedronVector}) holds since
\begin{eqnarray}
\left| b_{1} b_{1}'\right|+\left| b_{2} b_{2}'\right|+\left| b_{3} b_{3}'\right| = \left| b_{1} \right| \left| b_{1}'\right|+\left| b_{2} \right| \left| b_{2}'\right|+\left| b_{3} \right| \left| b_{3}'\right| \nonumber \\
= \vec{n}\cdot \vec{n}' = \left| \left| \vec{n} \right|\right|  \left| \left| \vec{n}' \right|\right| \cos \theta \leq 1, \nonumber \\
\end{eqnarray}
where we define $\vec{n}=\{|b_{1}|,|b_{2}|,|b_{3}|\}$ and $\vec{n}'=\{|b_{1}'|,|b_{2}'|,|b_{3}'|\}$, respectively, as the vectors corresponding to $\vec{b}$ and $\vec{b}'$ in the positive octant of the unit ball, and $\theta$ as the angle between these vectors.


Now, due to Eqs.~(\ref{Eq:SeparableM3NSimpleHadamard}), (\ref{Eq:HadamardTetrahedraResult}), (\ref{Eq:HadamardTetrahedraAndBall}) and (\ref{Eq:hadamardproductofballs}), and the fact that $conv(A)=A$ for any convex set $A$, we identify four cases:
\begin{enumerate}
\item{if $K_{\alpha}$ is even for any $\alpha$ then
\begin{eqnarray}
\mathcal{S}_{\{K_\alpha\}_{\alpha=1}^M}^{\mathcal{M}_{N}^{3}} &=& conv \left( \mathop{\mathlarger{\mathlarger{\mathlarger{\bigcirc}}}}_{\mathsmaller{\mu: K_{\mu} even}}{\cal T}_{(-1)^{K_{\mu}/2}} \right) \nonumber \\ &=& conv \left( {\cal T}_{(-1)^{\mathcal{M}_{-}}}\right) \nonumber \\ &=& {\cal T}_{(-1)^{\mathcal{M}_{-}}},
\end{eqnarray}
where $\mathcal{M}_{-}$ is the number of $K_{\mu}$ with odd $K_{\mu}/2$;
\label{Condition:AllEven}
}
\item{if $K_{\alpha}$ is odd for just one value of $\alpha$ then
\begin{eqnarray}
\mathcal{S}_{\{K_\alpha\}_{\alpha=1}^M}^{\mathcal{M}_{N}^{3}} &=& conv \left[\left( \mathop{\mathlarger{\mathlarger{\mathlarger{\bigcirc}}}}_{\mathsmaller{\mu: K_{\mu} even}}{\cal T}_{(-1)^{K_{\mu}/2}} \right) \circ {\cal B}_{1}\right] \nonumber \\ &=& conv \left( {\cal T}_{\pm 1} \circ {\cal B}_{1}\right) \nonumber \\ &=& {\cal B}_{1};
\end{eqnarray}
\label{Condition:OnlyOneOdd}
}
\item{if $K_{\alpha}$ is odd for all values of $\alpha$ then
\begin{eqnarray}
\mathcal{S}_{\{K_\alpha\}_{\alpha=1}^M}^{\mathcal{M}_{N}^{3}} &=& conv \left( \mathop{\mathlarger{\mathlarger{\mathlarger{\bigcirc}}}}_{\mathsmaller{\nu: K_{\nu} odd}} {\cal B}_{1} \right) \nonumber \\ &=& {\cal O}_{1};
\end{eqnarray}
\label{Condition:AllOdd}
}
\item{otherwise,
\begin{eqnarray}
\mathcal{S}_{\{K_\alpha\}_{\alpha=1}^M}^{\mathcal{M}_{N}^{3}} &=& conv \left[ \left( \mathop{\mathlarger{\mathlarger{\mathlarger{\bigcirc}}}}_{\mathsmaller{\mu: K_{\mu} even}}{\cal T}_{(-1)^{K_{\mu}/2}}\right) \circ \left( \mathop{\mathlarger{\mathlarger{\mathlarger{\bigcirc}}}}_{\mathsmaller{\nu: K_{\nu} odd}}{\cal B}_{1}\right) \right] \nonumber \\
&=& conv \left[ {\cal T}_{\pm 1} \circ \left( \mathop{\mathlarger{\mathlarger{\mathlarger{\bigcirc}}}}_{\mathsmaller{\nu: K_{\nu} odd}}{\cal B}_{1}\right) \right] \nonumber \\
&=& conv \left[ {\cal T}_{\pm 1}  \circ {\cal B}_{1} \circ \ldots \circ {\cal B}_{1} \right] \nonumber \\
&=& conv \left( \mathop{\mathlarger{\mathlarger{\mathlarger{\bigcirc}}}}_{\mathsmaller{\nu: K_{\nu} odd}} {\cal B}_{1} \right) \nonumber \\
&=& {\cal O}_{1}.
\end{eqnarray}
\label{Condition:Otherwise}
}
\end{enumerate}

For any even $N$-qubit system, only a $\{K_\alpha\}_{\alpha=1}^M$ partitioning within cases \ref{Condition:AllEven}, \ref{Condition:AllOdd} and \ref{Condition:Otherwise} may be realised. In case \ref{Condition:AllEven}, i.e.~when $K_{\alpha}$ is even for any $\alpha$, we have $\mathcal{S}_{\{K_\alpha\}_{\alpha=1}^M}^{\mathcal{M}_{N}^{3}} = {\cal T}_{(-1)^{\mathcal{M}_{-}}}$, where $\mathcal{M}_{-}$ is the number of $K_{\alpha}$ with odd $K_{\alpha}/2$. However, one can simply see that $(-1)^{\mathcal{M}_{-}}=(-1)^{N/2}$, and thus $\mathcal{S}_{\{K_\alpha\}_{\alpha=1}^M}^{\mathcal{M}_{N}^{3}}$ is the set ${\cal T}_{(-1)^{N/2}}$ of all $\mathcal{M}_{N}^{3}$ states. Otherwise, in cases \ref{Condition:AllOdd} and \ref{Condition:Otherwise}, we have $\mathcal{S}_{\{K_\alpha\}_{\alpha=1}^M}^{\mathcal{M}_{N}^{3}} = {\cal O}_{1}$.

For any odd $N$-qubit system, only a $\{K_\alpha\}_{\alpha=1}^M$ partitioning within cases \ref{Condition:OnlyOneOdd}, \ref{Condition:AllOdd} and \ref{Condition:Otherwise} may be realised. In case \ref{Condition:OnlyOneOdd}, i.e.~when $K_{\alpha}$ is odd for only one $\alpha$, we have $\mathcal{S}_{\{K_\alpha\}_{\alpha=1}^M}^{\mathcal{M}_{N}^{3}} = {\cal B}_{1}$, and thus $\mathcal{S}_{\{K_\alpha\}_{\alpha=1}^M}^{\mathcal{M}_{N}^{3}}$ is the set ${\cal B}_{1}$ of all $\mathcal{M}_{N}^{3}$ states. Otherwise, in cases \ref{Condition:AllOdd} and \ref{Condition:Otherwise}, we have $\mathcal{S}_{\{K_\alpha\}_{\alpha=1}^M}^{\mathcal{M}_{N}^{3}} = {\cal O}_{1}$.

\begin{flushright}
$\blacksquare$
\end{flushright}

By identifying the set of separable $\mathcal{M}_{N}^{3}$ states $\mathcal{S}_{\{Q_\alpha\}_{\alpha=1}^M}^{\mathcal{M}_{N}^{3}}$, Theorem \ref{Theorem:SeparableM3NStates} implies the following corollary.
\begin{corollary}
For any multiparticle entanglement monotone $E_{\{Q_\alpha\}_{\alpha=1}^M}$ and any $\mathcal{M}_{N}^{3}$ state $\varpi$ partitioned along any given $\{Q_\alpha\}_{\alpha=1}^M$-partition, $E_{\{Q_\alpha\}_{\alpha=1}^M} (\varpi) = 0$ if
\begin{enumerate}
\item{$K_{\alpha}$ is odd for at most one value of $\alpha$;}
\item{$K_{\alpha}$ is odd for more than one value of $\alpha$ and $\left| c_{1}\right|+\left| c_{2}\right|+\left| c_{3}\right| \leq 1$ for $c_{i} = \mbox{{\rm Tr}}(\varpi \sigma_{i}^{\otimes N})$.}
\end{enumerate}
\end{corollary}

Now we are ready to characterise also the set of $M$-separable $\mathcal{M}_{N}^{3}$ states $\mathcal{S}_{M}^{\mathcal{M}_{N}^{3}}$. Indeed we know that $\mathcal{S}_{M}^{\mathcal{M}_{N}^{3}}$ is just the convex hull of the union of all the sets of  $\{Q_\alpha\}_{\alpha=1}^M$-separable $\mathcal{M}_{N}^{3}$ states $\mathcal{S}_{\{Q_\alpha\}_{\alpha=1}^M}^{\mathcal{M}_{N}^{3}}$ obtained by considering all the possible $M$-partitions  $\{Q_\alpha\}_{\alpha=1}^M$. Furthermore, one can easily see that for any $M\leq \left \lceil{N/2}\right \rceil$ one can always find an $M$-partition $\{Q_\alpha\}_{\alpha=1}^M$ such that $K_{\alpha}$ is odd for at most one value of $\alpha$ and thus $\mathcal{S}_{\{Q_\alpha\}_{\alpha=1}^M}^{\mathcal{M}_{N}^{3}} = \mathcal{M}_{N}^{3}$, whereas for any $M> \left \lceil{N/2}\right \rceil$ this is impossible and thus $\mathcal{S}_{\{Q_\alpha\}_{\alpha=1}^M}^{\mathcal{M}_{N}^{3}} ={\cal O}_{1}$ for any possible $M$-partition $\{Q_\alpha\}_{\alpha=1}^M$. This immediately implies the following two Corollaries.

\begin{corollary}\label{Corollary:MSeparableM3NStates}
For any $N$, the set of $M$-separable $\mathcal{M}_{N}^{3}$ states $\mathcal{S}_{M}^{\mathcal{M}_{N}^{3}}$ is either
\begin{itemize}
\item{the set of all $\mathcal{M}_{N}^{3}$ states, for any $M\leq \left \lceil{N/2}\right \rceil$;}
\item{the set of $\mathcal{M}_{N}^{3}$ states 
represented in the $\{c_{1},c_{2},c_{3}\}$-space by the unit octahedron ${\cal O}_{1}$ with vertices $\{\pm 1, 0, 0\}$, $\{0, \pm 1, 0\}$ and $\{0, 0, \pm 1\}$, for any $M>\left \lceil{N/2}\right \rceil$.}
\end{itemize}
\end{corollary}

\begin{corollary}
For any multiparticle entanglement monotone $E_{M}$ and any $\mathcal{M}_{N}^{3}$ state $\varpi$, $E_{M} (\varpi) = 0$ if
\begin{enumerate}
\item{$M\leq \left \lceil{N/2}\right \rceil$;}
\item{$M > \left \lceil{N/2}\right \rceil$ and $\left| c_{1}\right|+\left| c_{2}\right|+\left| c_{3}\right| \leq 1$ for $c_{i} = \mbox{{\rm Tr}}(\varpi \sigma_{i}^{\otimes N})$.}
\end{enumerate}
\end{corollary}

\section{\sf \bfseries Multiparticle entanglement of $\mathcal{M}_{N}^{3}$ states}\label{Appendix:M3NEntanglement}

We now provide the analytical expressions for both the partition-dependent and partition-independent geometric measures of multiparticle entanglement $E_{\{Q_\alpha\}_{\alpha=1}^M}^{D}(\varpi)$ and $E_{M}^{D}(\varpi)$ of any $\mathcal{M}_{N}^{3}$ state $\varpi$. Within the partition-dependent setting we will restrict to any nontrivial partition $\{K^{'}_\alpha\}_{\alpha=1}^M$, i.e. such that $K^{'}_\alpha$ is odd for at least two values of $\alpha$, whereas within the partition-independent setting we will restrict to any non trivial number of parties $M'$, i.e. such that $M'>\left \lceil{N/2}\right \rceil$. According to Appendix \ref{sec:M3ificationsection} and \ref{Appendix:SetOfSeperableM3N}, in both cases we simply need to find the minimal distance from $\varpi$ to the set of $\mathcal{M}_{N}^{3}$ states inside the unit octahedron $\mathcal{O}_{1}$. In the even $N$ case, the closest state is the same for any convex and contractive distance (note that every jointly convex distance is convex), while in the odd $N$ case this is not true.

\subsection{\sf \bfseries Even $N$ case}

For even $N$, both the $\{K^{'}_\alpha\}_{\alpha=1}^M$- and $M'$-inseparable $\mathcal{M}^3_N$ states belong to the four corners obtained by removing the unit octahedron $\mathcal{O}_{1}$ from the whole tetrahedron $\mathcal{T}_{(-1)^{N/2}}$ of $\mathcal{M}^3_N$ states. In the following we will focus only on the corner containing the vertex $\{-1,(-1)^{N/2},-1\}$, since all the $\mathcal{M}^3_N$ states belonging to the other three corners can be obtained from this by simply applying a single-qubit local unitary $\sigma_{i} \otimes \mathbb{I}^{\otimes N-1}$, $i \in \{1,2,3\}$, under which any sort of multiparticle entanglement is invariant.

In order to characterise all the $\mathcal{M}^3_N$ states with even $N$ belonging to the $\{-1,(-1)^{N/2},-1\}$-corner, it will be convenient to move from the coordinate system $\{c_1,c_2,c_3\}$ to a new coordinate system $(p,q,h)$, where we assign the coordinates $\left( \frac{1}{3},\frac{1}{3},1\right)$ to the vertex $\{-1,(-1)^{N/2},-1\}$ and the coordinates
\begin{eqnarray}\label{eq:RoundBracketCoordinateSystem}
p&=&\frac{1+c_1-(-1)^{N/2}c_2-c_3}{3+c_1-(-1)^{N/2}c_2+c_3},\\
q&=&\frac{1+c_1+(-1)^{N/2}c_2+c_3}{3+c_1-(-1)^{N/2}c_2+c_3},\\
h&=& (-1-(c_1 -(-1)^{N/2} c_2 + c_3))/2,
\end{eqnarray}
to any other point in the corner. In order to avoid confusion between the above two coordinate systems, we will denote an $\mathcal{M}^3_N$ state $\varpi$ with curly brackets when representing it in the $\{c_1,c_2,c_3\}$ coordinate system, whereas we will denote $\varpi$ with round brackets when representing it in the $(p,q,h)$ coordinate system. Specifically, the $\mathcal{M}^3_N$ states represented by the triples $(p,q,h)$, with a fixed value of $h\in[0,1[$, correspond in the $\{c_1,c_2,c_3\}$-space to all, and only, the $\mathcal{M}^3_N$ states belonging to the triangle with the following vertices:
\begin{eqnarray}
V_1(h)&=&\left \lbrace -h,(-1)^{N/2} h,-1 \right \rbrace,\nonumber\\
V_2(h)&=&\left \lbrace -h,(-1)^{N/2},-h\right \rbrace,\label{eq:verticesoftheisoentangletriangle}\\
V_3(h)&=&\left \lbrace -1,(-1)^{N/2}h,-h\right \rbrace,\nonumber
\end{eqnarray}
in such a way that
\begin{equation}\label{eq:triangulation}
(p,q,h)= p V_1(h) + q V_2(h) + (1-p-q)V_3(h).
\end{equation}
These triangles corresponding to constant values of $h$ will play a crucial role, as they represent the sets of $\mathcal{M}^3_N$ states with constant $\{K^{'}_\alpha\}_{\alpha=1}^M$- and $M'$-inseparable multiparticle entanglement for even $N$. In particular, for $h=0$ we get one of the faces of the octahedron of $\{K^{'}_\alpha\}_{\alpha=1}^M$- and $M'$-separable states, whereas with increasing  $h$, we will prove that both the $\{K^{'}_\alpha\}_{\alpha=1}^M$- and $M'$-inseparable multiparticle entanglement of the $\mathcal{M}^3_N$ states belonging to the corresponding triangle will increase monotonically. We will now show that the $\{K^{'}_\alpha\}_{\alpha=1}^M$-separable (resp., $M'$-separable) state represented by the triple $(p,q,0)$ is one of the closest $\{K^{'}_\alpha\}_{\alpha=1}^M$-separable (resp., $M'$-separable) states to the $\mathcal{M}^3_N$ state $(p,q,h)$.

\begin{lemma}\label{th:theclosestseparablestateisonthefaceoftheoctahedron}
For every even $N$, according to any convex and contractive distance, one of the closest $\{K^{'}_\alpha\}_{\alpha=1}^M$-separable (resp., $M'$-separable) states $\varsigma_\varpi$ to any $\mathcal{M}^3_N$ state $\varpi$ belonging to the $\{-1,(-1)^{N/2},-1\}$-corner is always an $\mathcal{M}^3_N$ state of the form $(p',q',0)$ for some $p',q' \in [0,1]$, $p'+q' \leq 1$.
\end{lemma}


\noindent \textit{Proof}.
Let $\varpi$ and $\varsigma$ be, respectively, any $\mathcal{M}^3_N$ state belonging to the $\{-1,(-1)^{N/2},-1\}$-corner and any $\{K^{'}_\alpha\}_{\alpha=1}^M$-separable $\mathcal{M}^3_N$ state, i.e.~any $\mathcal{M}^3_N$ state contained in the unit octahedron $\mathcal{O}_{1}$. There will always be a $\{K^{'}_\alpha\}_{\alpha=1}^M$-separable $\mathcal{M}^3_N$ state $\varsigma'$, belonging to the octahedron face whose vertices are $V_1(0)$, $V_2(0)$, and $V_3(0)$ given in Eqs.~(\ref{eq:verticesoftheisoentangletriangle}), such that
$\varsigma' = \lambda \varpi + (1-\lambda) \varsigma$ for some $\lambda\in[0,1]$. Now, for any convex distance, the following holds
\begin{eqnarray}
&&D(\varpi,\varsigma' ) \\\nonumber
&=&D(\varpi, \lambda \varpi + (1-\lambda) \varsigma ) \\\nonumber
&\leq& \lambda D(\varpi, \varpi ) + (1-\lambda) D(\varpi,\varsigma) \\\nonumber
&=& (1-\lambda) D(\varpi,\varsigma) \\\nonumber
&\leq& D(\varpi,\varsigma).
\end{eqnarray}
As one of the closest $\{K^{'}_\alpha\}_{\alpha=1}^M$-separable states $\varsigma_\varpi$ to any $\mathcal{M}^3_N$ state $\varpi$ is always a $\{K^{'}_\alpha\}_{\alpha=1}^M$-separable $\mathcal{M}^3_N$ state, then the above inequality implies that, for any $\mathcal{M}^3_N$ state $\varpi$ belonging to the  $\{-1,(-1)^{N/2},-1\}$-corner, $\varsigma_\varpi$ always belongs to the triangle with vertices $V_1(0)$, $V_2(0)$, and $V_3(0)$  i.e.~ $\varsigma_\varpi=(p',q',0)$ for some $p',q'\in[0,1]$, $p'+q'\leq 1$. Exactly the same proof holds when substituting $\{K^{'}_\alpha\}_{\alpha=1}^M$-separability with $M'$-separability.

\begin{flushright}
$\blacksquare$
\end{flushright}

\begin{lemma}\label{th:anycontractivedistanceismagictranslationalinvariant}
For every even $N$, any contractive distance satisfies the following translational invariance property:
\begin{eqnarray}\label{eq:magictranslationalinvariance1}
D\left(( p,q,h) ,( p,q,0) \right)= D\left(\left( \frac{1}{3},\frac{1}{3},h\right),\left(\frac{1}{3},\frac{1}{3},0\right)\right),
\end{eqnarray}
for any $p,q\in[0,1]$ with $p+q\leq 1$ and $h\in[0,1[$.
\end{lemma}

\noindent \textit{Proof}.
First of all, by considering the following single-qubit LOCC,
\begin{equation}\label{eq:LOCCforthetriangulation}
\Lambda_{\{p,q\}}(\varrho) = p \varrho + q U_1 \varrho U_1^\dagger + (1-p-q)U_2 \varrho U_2^\dagger
\end{equation}
where $p,q\in[0,1]$, $p+q\leq 1$ and
\begin{eqnarray}
U_1 &=&   S_2^{\otimes N} S_1^{\otimes N} F_N,\\
U_2 &=&    S_1^{\otimes N} F_N S_2^{\otimes N},
\end{eqnarray}
with $S_i = \frac{1}{\sqrt{2}}\left(\mathbb{I} + i \sigma_i\right) $ and $F_N=\sigma_1^{\otimes (N/2 + 1)}\otimes \mathbb{I}^{\otimes (N/2-1)}$,
we have the following inequality,
\begin{eqnarray} \label{eq:inequalityduetodephasing1}
&&D\left(\left(\frac{1}{3},\frac{1}{3},h\right),\left(\frac{1}{3},\frac{1}{3},0\right)\right) \nonumber \\
&=& D\left(\Lambda_{\left\lbrace\frac{1}{3},\frac{1}{3}\right\rbrace}\left( p,q,h\right),\Lambda_{\left\lbrace\frac{1}{3},\frac{1}{3}\right\rbrace}(p,q,0)\right)  \nonumber \\
&\leq& D(\left( p,q,h\right),\left( p,q,0\right)),
\end{eqnarray}
where the final inequality is due to the contractivity of the distance $D$, whereas the first equality is due to the fact that
\begin{eqnarray}
\left(\frac{1}{3},\frac{1}{3},h\right) &=& \Lambda_{\left\lbrace\frac{1}{3},\frac{1}{3}\right\rbrace}\left( p,q,h\right),
\end{eqnarray}
which in turn is due to Eqs.~(\ref{eq:verticesoftheisoentangletriangle}), (\ref{eq:triangulation}), and:
\begin{eqnarray}
U_1\{c_1,c_2,c_3\}U_1^\dagger&=&\{-(-1)^{N/2}c_2,-(-1)^{N/2}c_3,c_1\},\\
U_2\{c_1,c_2,c_3\}U_2^\dagger&=&\{c_3,-(-1)^{N/2}c_1,-(-1)^{N/2}c_2\}.
\end{eqnarray}

In order to prove the opposite inequality and thus Eq.~(\ref{eq:magictranslationalinvariance1}), we now introduce a global $N$-qubit  channel $\Omega$ with operator-sum representation
\begin{equation}\label{Eq:KrausRepresentation}
\Omega(\varrho) = \sum_{i=1}^{2^{N}} A_{i}\varrho A_{i}^{\dagger},
\end{equation}
where
\begin{eqnarray}\label{Eq:KrausOperators}
\{A_i\}_{i=1}^{2^N} = \left\lbrace \{|\Psi_j^+\rangle\langle \Phi_j^+|\}_{j=1}^{2^{N-2}}, \{|\Psi_j^+\rangle\langle \Phi_j^-|\}_{j=1}^{2^{N-2}}, \right. \nonumber\\
\left. \{|\Psi_j^+\rangle\langle \Psi_j^+|\}_{j=1}^{2^{N-2}},\{|\Psi_j^-\rangle\langle \Psi_j^-|\}_{j=1}^{2^{N-2}} \right\rbrace
\end{eqnarray}
with the $2^{N}$ Kraus operators satisfying $\sum_{i} A_{i}^{\dagger} A_{i} = \mathbb{I}^{\otimes N}$, where $\{|\Phi_j^{\pm}\rangle\}$ and $\{ |\Psi_j^{\pm}\rangle\}$ constitute the binary ordered $N$-qubit eigenvectors $\{|\beta_i^\pm\rangle\}$ with even and odd parity, respectively, i.e. they are such that
\begin{eqnarray}
\Pi_3 |\Phi_j^{\pm}\rangle&=&|\Phi_j^{\pm}\rangle,\nonumber\\
\Pi_3 |\Psi_j^{\pm}\rangle&=&-|\Psi_j^{\pm}\rangle, \label{eq:PhiandPsi}
\end{eqnarray}
where $j\in\{1,\cdots,2^{N-2}\}$. It will be crucial in the following to see that the effect of $\Omega$ on an $\mathcal{M}_N^3$ state represented by the triple $\left(\frac{1}{3},\frac{1}{3},h\right)$ is given by
\begin{equation}\label{eq:ActionoftheglobalNqubitrephasingchannel}
\Omega\left(\left(\frac{1}{3},\frac{1}{3},h\right)\right)=\left( 1,0,h\right).
\end{equation}

Thanks to Eqs. (\ref{eq:EigenvaluesNBellStates}) and (\ref{eq:PhiandPsi}), one gets that the spectral decomposition of an $\mathcal{M}_N^3$ state with even $N$ can be written as follows:
\begin{eqnarray}\label{eq:spectraldecompositionofM3NstateswithevenN}
\{c_1,c_2,c_3\}&& \\ \nonumber
&=& \frac{1}{2^N}\left[1 + c_1 + (-1)^{N/2} c_2 + c_3 \right]\sum_j|\Phi_j^{+}\rangle\langle\Phi_j^{+}|\\ \nonumber
&+& \frac{1}{2^N}\left[1 - c_1 - (-1)^{N/2} c_2 + c_3 \right]\sum_j|\Phi_j^{-}\rangle\langle\Phi_j^{-}| \\ \nonumber
&+&\frac{1}{2^N}\left[1 + c_1 - (-1)^{N/2} c_2 - c_3 \right]\sum_j|\Psi_j^{+}\rangle\langle\Psi_j^{+}|\\ \nonumber
&+& \frac{1}{2^N}\left[1 - c_1 + (-1)^{N/2}c_2 - c_3 \right] \sum_j|\Psi_j^{-}\rangle\langle\Psi_j^{-}|. \\ \nonumber
\end{eqnarray}

Consequently, the spectral decompositions of the $\mathcal{M}_N^3$ states represented by the triples $\left(\frac{1}{3},\frac{1}{3},h\right)$ and $(1,0,h)$ are, respectively,
\begin{eqnarray}\label{eq:SpectralDecompositionOfNoisySmolinState}
\left(\frac{1}{3},\frac{1}{3},h\right)&=&\left\lbrace-\frac{2h+1}{3},(-1)^{N/2}\frac{2h+1}{3},-\frac{2h+1}{3}\right\rbrace  \nonumber \\ \nonumber
&=& \frac{1}{2^{N-1}}\left(\frac{1-h}{3} \right)\sum_j|\Phi_j^{+}\rangle\langle\Phi_j^{+}|\\
&+& \frac{1}{2^{N-1}}\left(\frac{1-h}{3} \right)\sum_j|\Phi_j^{-}\rangle\langle\Phi_j^{-}|\\ \nonumber
&+&\frac{1}{2^{N-1}}\left(\frac{1-h}{3} \right)\sum_j|\Psi_j^{+}\rangle\langle\Psi_j^{+}|\\ \nonumber
&+&\frac{1}{2^{N-1}}\left(1 + h \right)\sum_j|\Psi_j^{-}\rangle\langle\Psi_j^{-}|,\\ \nonumber
\end{eqnarray}
and
\begin{eqnarray}
(1,0,h)&=&\left\lbrace -h,(-1)^{N/2} h,-1\right\rbrace  \nonumber \\
&=&\frac{1}{2^{N-1}}\left(1 - h \right)\sum_j|\Psi_j^{+}\rangle\langle\Psi_j^{+}|\\ \nonumber
&+&\frac{1}{2^{N-1}}\left(1 + h \right)\sum_j|\Psi_j^{-}\rangle\langle\Psi_j^{-}|.\\ \nonumber
\end{eqnarray}
By exploiting the following equalities
\begin{eqnarray}
\Omega(|\Phi_j^{+}\rangle\langle\Phi_j^{+}|)=|\Psi_j^{+}\rangle\langle\Psi_j^{+}|,\\ \nonumber
\Omega(|\Phi_j^{-}\rangle\langle\Phi_j^{-}|)=|\Psi_j^{+}\rangle\langle\Psi_j^{+}|,\\ \nonumber
\Omega(|\Psi_j^{+}\rangle\langle\Psi_j^{+}|)=|\Psi_j^{+}\rangle\langle\Psi_j^{+}|,\\ \nonumber
\Omega(|\Psi_j^{-}\rangle\langle\Psi_j^{-}|)=|\Psi_j^{-}\rangle\langle\Psi_j^{-}|,\\ \nonumber
\end{eqnarray}
and the linearity of the channel $\Omega$, we immediately get Eq.~(\ref{eq:ActionoftheglobalNqubitrephasingchannel}). We then have the inequality
\begin{eqnarray}\label{eq:inequalityduetorephasing1}
&&D(\left( p,q,h\right),\left( p,q,0\right))\nonumber \\ \nonumber
&=& D\left(\Lambda_{\{p,q\}} \left( \Omega \left(\frac{1}{3},\frac{1}{3},h\right) \right),\Lambda_{\{p,q\}} \left( \Omega \left(\frac{1}{3},\frac{1}{3},0 \right) \right) \right)  \nonumber \\
&\leq& D\left(\left(\frac{1}{3},\frac{1}{3},h\right),\left(\frac{1}{3},\frac{1}{3},0\right)\right),
\end{eqnarray}
where the final inequality is again due to the contractivity of the distance $D$, whereas the first equality is due to the fact that
\begin{eqnarray}
\left( p,q,h\right) &=&\Lambda_{\{p,q\}} \left( \Omega \left(\frac{1}{3},\frac{1}{3},h\right) \right),\nonumber
\end{eqnarray}
which in turn is due to Eqs.~(\ref{eq:ActionoftheglobalNqubitrephasingchannel}), (\ref{eq:LOCCforthetriangulation}), (\ref{eq:triangulation}) and (\ref{eq:verticesoftheisoentangletriangle}) . By putting together the two opposite inequalities (\ref{eq:inequalityduetodephasing1}) and (\ref{eq:inequalityduetorephasing1}), we immediately get the invariance of  Eq.~(\ref{eq:magictranslationalinvariance1}) for any contractive distance.

\begin{flushright}
$\blacksquare$
\end{flushright}

Now we are ready to find out the analytical expression of one of the closest $\{K^{'}_\alpha\}_{\alpha=1}^M$-separable (resp., $M'$-separable) states $\varsigma_\varpi$ to any $\mathcal{M}^3_N$ state $\varpi$ belonging to the $\{-1,(-1)^{N/2},-1\}$-corner.

\begin{theorem}\label{lem:nearestseparablestatetoanM3Nstate}
For any even $N$, according to any convex and contractive distance, the $\mathcal{M}^3_N$ state $(p,q,0)$ is one of the closest $\{K^{'}_\alpha\}_{\alpha=1}^M$-separable (resp., $M'$-separable) states to the $\mathcal{M}_N^3$ state $(p,q,h)$.
\end{theorem}
\noindent \textit{Proof}.

Thanks to Lemma \ref{th:theclosestseparablestateisonthefaceoftheoctahedron}, which holds for any convex and contractive distance and any even $N$, we just need to prove that for any $p',q'\in[0,1]$, $p'+q'\leq 1$,
\[D(\left( p,q,h\right),\left( p,q,0\right))
\leq D(\left( p,q,h\right),\left( p',q',0\right)).\]
In fact
\begin{eqnarray} \nonumber
&&D(\left( p,q,h\right),\left( p,q,0\right))  \\ \nonumber
&=& D\left(\left(\frac{1}{3},\frac{1}{3},h\right),\left(\frac{1}{3},\frac{1}{3},0\right)\right)  \\ \nonumber
&=& D\left(\Lambda_{\left\lbrace\frac{1}{3},\frac{1}{3}\right\rbrace}\left( p,q,h\right),\Lambda_{\left\lbrace\frac{1}{3},\frac{1}{3}\right\rbrace}(p',q',0)\right)  \nonumber \\
&\leq& D(\left( p,q,h\right),\left( p',q',0\right)), \nonumber
\end{eqnarray}
where the first equality is due to Lemma \ref{th:anycontractivedistanceismagictranslationalinvariant}, which holds for any contractive distance and any even $N$, the second equality is due to the fact that
\begin{eqnarray}
\left(\frac{1}{3},\frac{1}{3},h\right) &=& \Lambda_{\left\lbrace\frac{1}{3},\frac{1}{3}\right\rbrace}\left( p,q,h\right),\\
\left(\frac{1}{3},\frac{1}{3},0\right) &=& \Lambda_{\left\lbrace\frac{1}{3},\frac{1}{3}\right\rbrace}(p',q',0),
\end{eqnarray}
with $\Lambda_{\left\lbrace\frac{1}{3},\frac{1}{3}\right\rbrace}$ representing the LOCC expressed by Eq.~(\ref{eq:LOCCforthetriangulation}), and finally the inequality is due to the contractivity of the distance $D$.

\begin{flushright}
$\blacksquare$
\end{flushright}

Now that we know the analytical expression of one of the closest $\{K^{'}_\alpha\}_{\alpha=1}^M$- and $M'$-separable states to any $\mathcal{M}^3_N$  state with even $N$ according to any convex and contractive distance, we can unveil the general hierarchy of both the $\{K^{'}_\alpha\}_{\alpha=1}^M$- and $M'$-inseparable multiparticle entanglement of these $\mathcal{M}^3_N$ states with respect to any geometric entanglement monotone  $E^D_{\{K^{'}_\alpha\}_{\alpha=1}^M}$ and $E^D_{M'}$, respectively.

\begin{corollary}\label{th:hierarchyofthetriangles}
For every even $N$ and according to any valid geometric measure of $\{K^{'}_\alpha\}_{\alpha=1}^M$- and $M'$-inseparable multiparticle entanglement $E^D_{\{K^{'}_\alpha\}_{\alpha=1}^M}$ and $E^D_{M'}$, the following holds:
\begin{eqnarray}
E^D_{\{K^{'}_\alpha\}_{\alpha=1}^M}((p,q,h))&=& E^D_{\{K^{'}_\alpha\}_{\alpha=1}^M}((p',q',h))\label{eq:isoentangledM3Nstates} \\
E^D_{M'}((p,q,h))&=& E^D_{M'}((p',q',h))\label{eq:isoMprimeentangledM3Nstates} \\
E^D_{\{K^{'}_\alpha\}_{\alpha=1}^M}((p,q,h))&\leq& E^D_{\{K^{'}_\alpha\}_{\alpha=1}^M}((p',q',h')),\label{eq:hierarchyofentanglementasafunctionofc}\\
E^D_{M'}((p,q,h))&\leq& E^D_{M'}((p',q',h')),\label{eq:hierarchyofMprimeentanglementasafunctionofc}
\end{eqnarray}
for any $h\leq h'$.
\end{corollary}

\noindent \textit{Proof}.

Let us start by proving Eq.~(\ref{eq:isoentangledM3Nstates}). By using Theorem \ref{lem:nearestseparablestatetoanM3Nstate} and Lemma \ref{th:anycontractivedistanceismagictranslationalinvariant}, we obtain
\begin{eqnarray}
E^D_{\{K^{'}_\alpha\}_{\alpha=1}^M}((p,q,h))&=&  D((p,q,h),(p,q,0))\\\nonumber
&=&  D\left(\left(\frac{1}{3},\frac{1}{3},h\right),\left(\frac{1}{3},\frac{1}{3},0\right)\right)\\\nonumber
&=&  D((p',q',h),(p',q',0))\\\nonumber
&=&  E^D_{\{K^{'}_\alpha\}_{\alpha=1}^M}((p',q',h)),\nonumber
\end{eqnarray}
for any $p,q,p',q'\in[0,1]$, $p+q\leq 1$, $p'+q'\leq 1$ and $h\in[0,1[$.

In order to prove Eq.~(\ref{eq:hierarchyofentanglementasafunctionofc}), let us consider the $\mathcal{M}^3_N$ states $\varpi=(p,q,h)$, $\varpi'=(p,q,h')$ such that $h\leq h'$, and $\varsigma=\varsigma_\varpi=\varsigma_{\varpi'}=(p,q,0)$, which is one of the closest $\{K^{'}_\alpha\}_{\alpha=1}^M$-separable states to both $\varpi$ and $\varpi'$ according to Theorem \ref{lem:nearestseparablestatetoanM3Nstate}.  We can write $\varpi = \lambda \varpi' + (1-\lambda) \varsigma,$ for some $\lambda\in[0,1]$. Now, by using the convexity of the distance and Eq.~(\ref{eq:isoentangledM3Nstates}), we get
\begin{eqnarray}
E^D_{\{K^{'}_\alpha\}_{\alpha=1}^M}((p,q,h))&=&D(\varpi,\varsigma ) \\\nonumber
&=&D(\lambda \varpi' + (1-\lambda) \varsigma, \varsigma ) \\\nonumber
&\leq& \lambda D(\varpi', \varsigma ) + (1-\lambda) D(\varsigma,\varsigma) \\\nonumber
&=& \lambda D(\varpi', \varsigma )\\\nonumber
&\leq& D(\varpi',\varsigma).\\\nonumber
&=& E^D_{\{K^{'}_\alpha\}_{\alpha=1}^M}((p,q,h'))\\\nonumber
&=& E^D_{\{K^{'}_\alpha\}_{\alpha=1}^M}((p',q',h'))\nonumber .
\end{eqnarray}

In order to prove Eqs. (\ref{eq:isoMprimeentangledM3Nstates}) and (\ref{eq:hierarchyofMprimeentanglementasafunctionofc}), we just remark that exactly the same proof holds when substituting $\{K^{'}_\alpha\}_{\alpha=1}^M$-separability with $M'$-separability.

\begin{flushright}
$\blacksquare$
\end{flushright}

We are now ready to apply the above general results to calculate the geometric multiparticle entanglement $E_{\{K_\alpha^{'}\}_{\alpha=1}^M}^{D}(\varpi)$ and $E_{M'}^{D}(\varpi)$ of any $\mathcal{M}_{N}^{3}$ state $\varpi$ for particular instances of $D$. As we have just shown, for $\mathcal{M}_{N}^{3}$ states in the $\{-1,(-1)^{N/2},-1\}$-corner, $E_{\{K_\alpha^{'}\}_{\alpha=1}^M}^{D}(\varpi)=E_{M'}^{D}(\varpi)=D\left( \left( \frac{1}{3},\frac{1}{3}, h\right) ,\left( \frac{1}{3},\frac{1}{3},0\right) \right) = f_{D}(h)$, where $f_{D}(h)$ is some monotonically increasing function of $h$ only, which depends on the chosen distance $D$. By local unitary equivalence, this is true indeed for any $\mathcal{M}_{N}^{3}$ state in any of the four corners if we define the generalised $h$ to be $h_{\varpi} = \frac{1}{2}(\sum_{j=1}^{3} |c_{j}|-1)$ . Table I in the main text shows $f_D(h_{\varpi})$ for the relative entropy, trace, infidelity, squared Bures, and squared Hellinger distance.

Here we show the derivation of the given expressions for $f_{D}(h)$. Since the two $\mathcal{M}_N^3$ states $\left( \frac{1}{3},\frac{1}{3}, h\right)$ and $\left( \frac{1}{3},\frac{1}{3},0\right)$ are diagonal in the same basis, we have that their distance reduces to the corresponding classical distance between the probability distributions formed by their eigenvalues, denoted by $P_h$ and $P_0$ respectively. We recall that the classical relative entropy, trace, infidelity, squared Bures, and squared Hellinger distance between two probability distributions $P=\{p_i\}$ and $Q=\{q_i\}$ are given by, respectively $D_{\rm RE}(P,Q)=\sum_{i}p_{i} \log_2 (p_{i}/q_{i})$, $D_{\rm Tr}(P,Q)=\sum_{i} \left|p_{i}-q_{i} \right|/2$, $D_{\rm F}(P,Q)=1-F(P,Q)$, and $D_{\rm B}(P,Q)=D_{\rm H}(P,Q)=2\left(1-\sqrt{F(P,Q)}\right)$, where $F(P,Q)=\left(\sum_{i} \sqrt{p_{i}q_{i}}\right)^{2}$ is the classical fidelity. Consequently, by using Eq.~(\ref{eq:SpectralDecompositionOfNoisySmolinState}) to get both $P_h$ and $P_0$, we obtain the desired expressions for $f_{D}(h)$.

\subsection{\sf \bfseries Odd $N$ case}

Let us now turn our attention to the evaluation of the geometric multiparticle entanglement $E^D_{\{K^{'}_\alpha\}_{\alpha=1}^M}(\varpi)$ and $E^D_{M'}(\varpi)$ of an $\mathcal{M}^3_N$ state $\varpi$  in the case of odd $N$. Unlike the even $N$ case, where all the convex and contractive distances $D$ concur on what is one of the closest separable states to an $\mathcal{M}^3_N$ state, there is unfortunately no such agreement in the odd $N$ case. In the following we will focus in particular on the trace distance-based geometric measures of multiparticle entanglement $E^{D_{\rm Tr}}_{\{K^{'}_\alpha\}_{\alpha=1}^M}(\varpi)$ and $E^{D_{\rm Tr}}_{M'}(\varpi)$ of any $\mathcal{M}^3_N$ state with odd $N$, when considering as usual a partition ${\{K^{'}_\alpha\}_{\alpha=1}^M}$ such that $K^{'}_\alpha$ is odd for more than one value of $\alpha$ and a number of parties $M'>\left \lceil{N/2}\right \rceil$, respectively.

We know that the trace distance-based multiparticle entanglement of an $\mathcal{M}_N^3$ state $\varpi$ with odd $N$ is the minimal distance from $\varpi$ to the unit octahedron ${\cal O}_1$. Due to convexity of the trace distance and of the unit octahedron $\mathcal{O}_1$, we get that one of the closest $\{K^{'}_\alpha\}_{\alpha=1}^M$-separable (resp., $M'$-separable) $\mathcal{M}^3_N$ states to an entangled $\mathcal{M}^3_N$ state $\varpi$ must belong necessarily to the boundary of the octahedron, i.e. either to one of its faces or edges. We can easily see that the trace distance between two arbitrary odd $N$ $\mathcal{M}^3_N$ states $\varpi_1=\left\lbrace c_1^{(1)},c_2^{(1)},c_3^{(1)}\right\rbrace$ and $\varpi_2=\left\lbrace c_1^{(2)},c_2^{(2)},c_3^{(2)}\right\rbrace$ is nothing but one half of the Euclidean distance between their representing triples, i.e.
\begin{equation}\label{eq:TraceDistanceIsEuclidean}
D_{\rm Tr}(\varpi_1,\varpi_2)=\frac{1}{2}\sqrt{\sum_{i=1}^3 \left(c_i^{(1)}-c_i^{(2)}\right)^2}.
\end{equation}
This is proven as follows. In order to evaluate the trace distance between any two $\mathcal{M}^3_N$ states with odd $N$, we just need to calculate the eigenvalues of their difference, because
\begin{equation}\label{eq:tracedistane}
D_{\rm Tr}(\varpi_1,\varpi_2)= \frac{1}{2} \text{Tr}(|\varpi_1 - \varpi_2|) = \frac{1}{2} \sum_i |\lambda_i|,
\end{equation}
where $\{\lambda_i\}$ are just the eigenvalues of $\varpi_1 - \varpi_2$. Since $\varpi_1 - \varpi_2 = \frac{1}{2^N} \sum_i d_i \sigma_i^{\otimes N}$, with $d_i=c_i^{(1)}-c_i^{(2)}$, one can easily see that its eigenvectors are exactly the ones expressed in Eq.~(\ref{eq:EigenvectorsM3NstatesNodd}), with $d_1=r\sin\theta\cos\phi$, $d_2=r\sin\theta\sin\phi$ and $d_3=r \cos\theta$  while its eigenvalues are given by either $\frac{1}{2^N}r$ or $-\frac{1}{2^N}r$, with $r=\sqrt{d_1^2 + d_2^2+d_3^2}$. By putting these eigenvalues into Eq.~(\ref{eq:tracedistane}) one immediately gets Eq.~(\ref{eq:TraceDistanceIsEuclidean}).

An immediate consequence of Eq.~(\ref{eq:TraceDistanceIsEuclidean}) is that the closest $\{K^{'}_\alpha\}_{\alpha=1}^M$-separable (resp., $M'$-separable) $\mathcal{M}^3_N$ state $\varsigma_\varpi=\{s_1,s_2,s_3\}$ to an entangled $\mathcal{M}^3_N$ state $\varpi=\{c_1,c_2,c_3\}$ is just its Euclidean orthogonal projection onto the boundary of the unit octahedron $\mathcal{O}_1$.
We can thus distinguish between the following two cases:
\begin{enumerate}
\item the Euclidean projection of $\{c_1,c_2,c_3\}$ onto the boundary of the unit octahedron $\mathcal{O}_1$ falls onto one of its faces. This case happens if, and only if, $0\leq \mbox{sign}(c_i)s_i \leq 1$ for any $i$, where $\{s_i = \mbox{sign}(c_i)(1 - |c_1| - |c_2| - |c_3| + 3 |c_i|)/3\}$ is exactly the triple representing such Euclidean projection;\label{item:ProjectionOntoTheFaces}
\item the Euclidean projection of $\{c_1,c_2,c_3\}$ onto the boundary of the unit octahedron $\mathcal{O}_1$ falls onto one of its edges. This case happens when any of the conditions listed in case \ref{item:ProjectionOntoTheFaces} do not hold. Moreover, the triple $\{s_1,s_2,s_3\}$ representing such Euclidean projection is given by $s_k=0$ and $s_i=\mbox{sign}(c_i)(1-\sum_{j\neq k}|c_j| + 2|c_i|)/2$, where $k$ is set by $f_k = \min\{f_1,f_2,f_3\}$ with $f_i = \sqrt{c_i^2 + (1-\sum_{j\neq i}|c_j|)^2/2}$.\label{item:ProjectionOntoTheEdges}
\end{enumerate}
This provides the explicit expression reported in Eq.~(3) in the main text.

\section{\sf \bfseries Genuine geometric multiparticle entanglement of the GHZ-diagonal states}

In this appendix we will adopt our approach to evaluate exactly the geometric genuine multiparticle entanglement $E_2^D$ of any $N$-qubit GHZ-diagonal state $\xi$ with respect to any contractive and jointly convex distance $D$.

Recall that any $N$-qubit state $\varrho$ can be transformed via a single-qubit LOCC $\Gamma$ into a GHZ-diagonal state $\varrho_{\text{GHZ}}\equiv \Gamma(\varrho)$ with eigenvalues given by $p_i^\pm=\langle\beta_i^\pm|\varrho|\beta_i^\pm\rangle$~\cite{Hofmann2014}, a procedure referred to as GHZ-diagonalisation of $\varrho$ in the main text.  The entanglement quantification is then based on the following two arguments.

First, we have that one of the closest $2$-separable states to a GHZ-diagonal state is itself GHZ-diagonal. Indeed, for any GHZ-diagonal state $\xi$ and any $2$-separable state $\varsigma$, we have that
\begin{equation}\label{Eq:GHZCloser}
D(\xi,\varsigma_{\text{GHZ}}) = D(\Gamma(\xi),\Gamma(\varsigma)) \leq D(\xi,\varsigma),
\end{equation}
where in the first equality we use the invariance of any GHZ-diagonal state through $\Gamma$ and that $\Gamma(\varsigma) \equiv \varsigma_{\text{GHZ}}$ is the GHZ-diagonalisation of $\varsigma$, and in the inequality we use the contractivity of the distance through any completely positive trace-preserving channel. Moreover, the GHZ-diagonalisation $\varsigma_{\text{GHZ}}$ of any $2$-separable state $\varsigma$ is a $2$-separable GHZ-diagonal state since $\Gamma$ is a single-qubit LOCC.

Therefore, the set $\mathcal{S}_{2}^{\mathcal{G}}$ of $2$-separable GHZ-diagonal states turns out to be the relevant one in order to compute exactly any distance-based measure of genuine multiparticle entanglement of a GHZ-diagonal state $\xi$, thus dramatically simplifying the ensuing optimisation as follows:
\begin{equation}\label{Eq:GenuineGeometricGHZ}
E_{2}^{D} (\xi)\equiv \inf_{\varsigma \in \mathcal{S}_{2}} D(\xi,\varsigma) = \inf_{\varsigma_{\text{GHZ}}\in \mathcal{S}_{2}^{\mathcal{G}}} D(\xi,\varsigma_{\text{GHZ}}).
\end{equation}

Now, let us consider an arbitrary GHZ-diagonal state $\xi$ and rearrange its GHZ eigenstates $\{|\beta_{i}\rangle\}_{i=1}^{2^{N}}$ in such a way that the corresponding eigenvalues $\{p^{\xi}_{i}\}_{i=1}^{2^{N}}$ are in non-increasing order. It is well known that $\xi$ is $2$-separable if, and only if, $p^{\xi}_{1} \leq 1/2$ \cite{Guhne2010}. For $p^{\xi}_{1} > 1/2$, we will show that one of the closest $2$-separable GHZ-diagonal states $\varsigma_{\xi}$ has eigenvalues $\{p^{\varsigma_{\xi}}_{i}\}_{i=1}^{2^{N}}$ such that $p^{\varsigma_{\xi}}_{1} = 1/2$, with $\{p^{\varsigma_{\xi}}_{i}\}_{i=1}^{2^{N}}$ corresponding again to the ordering of GHZ eigenstates $\{|\beta_{i}\rangle\}_{i=1}^{2^{N}}$ set by $\xi$. This result further simplifies the  optimisation in Eq. \ref{Eq:GenuineGeometricGHZ}.

Consider any $2$-separable GHZ-diagonal state $\varsigma$, it holds that there will always be a $2$-separable GHZ-diagonal state $\varsigma'$ with eigenvalues $\{p^{\varsigma'}_{i}\}_{i=1}^{2^{N}}$ and $p^{\varsigma'}_{1} = 1/2$ such that $\varsigma' = \lambda \xi + (1-\lambda) \varsigma$ for some $\lambda\in[0,1]$. Now, for any convex distance, the following holds
\begin{eqnarray}
&&D(\xi,\varsigma' ) \\\nonumber
&=&D(\xi, \lambda \xi + (1-\lambda) \varsigma ) \\\nonumber
&\leq& \lambda D(\xi, \xi ) + (1-\lambda) D(\xi,\varsigma) \\\nonumber
&=& (1-\lambda) D(\xi,\varsigma) \\\nonumber
&\leq& D(\xi,\varsigma).
\end{eqnarray}
This inequality immediately implies that one of the closest $2$-separable GHZ-diagonal states $\varsigma_{\xi}$ to a $2$-inseparable GHZ-diagonal state $\xi$ is of the form $\varsigma'$, which is formalised as a corollary below.
\begin{corollary}\label{th:ClosestGHZDiagForm}
For any convex and contractive distance $D$ and any fully inseparable GHZ-diagonal state $\xi$, whose GHZ eigenstates $\{|\beta_{i}\rangle\}_{i=1}^{2^{N}}$ are arranged such that the corresponding eigenvalues $\{p^{\xi}_{i}\}_{i=1}^{2^{N}}$ are in non-increasing order, one of the closest $2$-separable states $\varsigma_{\xi}$ to $\xi$ is itself a GHZ-diagonal state with eigenvalues $\{p^{\varsigma_{\xi}}_{i}\}_{i=1}^{2^{N}}$ such that $p^{\varsigma_{\xi}}_{1} = 1/2$, with $\{p^{\varsigma_{\xi}}_{i}\}_{i=1}^{2^{N}}$  corresponding again to the ordering of GHZ eigenstates $\{|\beta_{i}\rangle\}_{i=1}^{2^{N}}$ set by $\xi$.
\end{corollary}

We can now apply this Corollary to calculate the geometric genuine multiparticle entanglement $E_{2}^{D}(\xi)$ of any GHZ-diagonal state $\xi$ for particular instances of $D$. Since the closest $2$-separable state $\varsigma_\xi$ to a GHZ-diagonal state $\xi$ is also a GHZ-diagonal state, they are diagonal in the same basis and their distance reduces to the corresponding classical distance between the probability distributions formed by their eigenvalues, denoted by $P_\xi$ and $P_{\varsigma_\xi}$ respectively. By using the expressions given earlier in Appendix \ref{Appendix:M3NEntanglement} of the classical relative entropy, trace, infidelity, squared Bures, and squared Hellinger distance between two probability distributions $P_\xi$ and $P_{\varsigma_\xi}$, and minimising it over all probability distributions $P_{\varsigma_\xi}$ such that $p^{\varsigma_{\xi}}_{1}=1/2$, one easily obtains the desired expressions for $E_{2}^{D}(\xi)$ expressed in the main text.

\end{document}